\documentclass[aps,10pt, pra,twocolumn,showpacs,superscriptaddress,amsfonts,amsmath,floatfix]{revtex4-1}
\usepackage[T1]{fontenc}
\usepackage{graphicx}
\usepackage{color}
\newcommand{\ket}[1]{| #1 \rangle}
\newcommand{\bra}[1]{\langle #1 |}
\newcommand{\braket}[2]{\langle #1 | #2 \rangle}
\newcommand{\ketbra}[2]{| #1 \rangle \langle #2 |}
\newcommand{\meanv}[1]{\langle #1 \rangle}
\newcommand{\re}{\mathrm{Re}\, }
\newcommand{\im}{\mathrm{Im}\,}
\newcommand{\dss}{\displaystyle}
\newcommand{\non}{\nonumber}
\newcommand{\I}{{\rm{i}}}
\newcommand{\D}{{\rm{d}}}
\newcommand{\E}{e}
\newcommand{\tr}{\operatorname{tr}}

\newcommand{\nn}{\nonumber}
\newcommand{\Heff}{\hat{H}_{\rm eff}}
\newcommand{\nopone}{\hat{n}_1}
\newcommand{\noptwo}{\hat{n}_2}
\newcommand{\nopi}{\hat{n}_i}

\newcommand{\Nop}{\hat{N}}
\newcommand{\aopone}{\hat{a}_1}
\newcommand{\aoptwo}{\hat{a}_2}
\newcommand{\aopi}{\hat{a}_i}

\newcommand{\lt}{\left(}
\newcommand{\rt}{\right)}
\newcommand{\bbr}[1]{\left(  #1 \right)}

\newcommand{\be}{\begin{equation}}
\newcommand{\ee}{\end{equation}}
\newcommand{\ba}{\begin{eqnarray}}
\newcommand{\ea}{\end{eqnarray}}

\newcommand{\fr}{\frac}

\def\proba{{\rm I\kern -.18em P}}

\newcommand{\sinc}{\operatorname{sinc}}

\newcommand{\ie}{i.e.\;}

\newcommand{\mv}{{\boldsymbol{m}}}

\newcommand{\rv}{{\boldsymbol{r}}}

\newcommand{\sv}{{\boldsymbol{s}}}

\newcommand{\vv}{{\boldsymbol{v}}}

\newcommand{\Ee}{{\cal E}}

\newcommand{\Ll}{{\cal L}}

\begin{document}

\title {Effect of one-, two-, and three-body atom loss processes  on superpositions of phase states  in Bose-Josephson junctions}
\author{D. Spehner}
\email{Dominique.Spehner@ujf-grenoble.fr}
\affiliation{Univ. Grenoble Alpes and CNRS, Institut Fourier, F-38000 Grenoble, France}
\affiliation{Univ. Grenoble Alpes, LPMMC, F-38000 Grenoble, France}
\affiliation{CNRS, LPMMC, F-38000 Grenoble, France}
\author{K. Pawlowski}
\affiliation{Center for Theoretical Physics PAN, 02-668 Warsaw, Poland}
\affiliation{5.\,Physikalisches Institut, Universit\"at Stuttgart, D-70569 Stuttgart, Germany}
\affiliation{Laboratoire Kastler Brossel, Ecole Normale Sup\'erieure, F-75231 Paris, France}
\author{G. Ferrini}
\affiliation{Laboratoire Kastler Brossel, Universit\'e Pierre et Marie Curie, 
F-75000 Paris, France}
\author{A. Minguzzi}
\affiliation{Univ. Grenoble Alpes, LPMMC, F-38000 Grenoble, France}
\affiliation{CNRS, LPMMC, F-38000 Grenoble, France}
\date{\today}


\begin{abstract}

In a two-mode Bose-Josephson junction formed by a binary mixture of
ultracold atoms, macroscopic superpositions  of phase states are
produced during the time evolution after a sudden quench to zero of
the coupling amplitude. Using  quantum trajectories and an
exact diagonalization of the master equation, we study
the effect of one-, two-, and three-body atom losses on the
superpositions by analyzing separately the amount of quantum
correlations in each subspace with fixed atom number. The  
quantum correlations  useful for atom interferometry are estimated 
using the quantum Fisher information. We identify the choice of
parameters leading to the largest Fisher information,  thereby
showing that, for all kinds of loss processes, quantum correlations can be
partially protected  from decoherence when the losses are strongly asymmetric in the two modes.
\end{abstract}

\maketitle

\section{Introduction}

Non-classical states such as squeezed states and macroscopic superpositions  
of coherent states are particularly 
interesting for high-precision interferometry since they allow for phase resolution beyond the standard quantum limit.  
One of the systems where such states may be engineered is a  Bose-Einstein condensate (BEC) made of metastable vapors of ultracold atoms. 
This system displays a  wide tunability of parameters: the interaction between atoms can be controlled by Feshbach 
resonances \cite{fano1961, feshbach1958}, and,  
by using  optical lattices, the BEC can be coherently split into  up to few thousands sub-systems with controlled 
tunneling between them \cite{Bloch2005,Bloch2008,Weitenberg2011}. 
When the condensed atoms are trapped in a double-well potential, they realize an external Bose-Josephson junction (BJJ).
The spatial wave functions localized inside a single well constitute the two modes of the BJJ and
the tunneling between the wells leads to an inter-mode coupling. An internal BJJ is formed by
condensed  atoms in two hyperfine states  resonantly coupled by 
a microwave radio-frequency field, trapped in a single harmonic well. 
In both cases, when inter-mode coupling dominates interactions, the ground state of the BJJ 
is a spin coherent state (CS), that is, a product state in which all atoms are in the same superposition of the two modes.
After a sudden quench to zero of the coupling, the dynamical evolution builds up entangled states because of the
 interactions between atoms. In the absence of decoherence mechanisms, the system evolves first into squeezed states 
\cite{kitagawa1993, soerensen2001a, soerensen2001}, then to multi-component superpositions of CSs~\cite{yurke1986, stoler1971}, 
and then has a revival in the initial CS.

To date, only squeezed states, which appear  at times much shorter than the revival time, 
have been realized experimentally \cite{esteve2008, riedel2010,gross2010}. At longer times, 
recombination and collision processes leading to losses of atoms in the BEC give rise to strong decoherence effects and eventually to
the disappearance of the BEC. Particle losses also   destroy  partially the coherence of the squeezed states, as analyzed quantitatively 
in~\cite{sinatra1998, yun2008, yun2009, pawlowskiBackground}. 
The phase noise produced by magnetic fluctuations in internal BJJs is another important source of decoherence~\cite{riedel2010,ferrini2010}.
The superpositions of CSs appear later in the evolution  and are expected to be more fragile than squeezed states.
The main theoretical studies on decoherence effects on such superpositions have focused on the influence of the coupling 
of the atoms with  the electromagnetic vacuum  \cite{huang2006} and the impact of phase noise \cite{ferrini2010a}.
In particular, it has been shown in \cite{ferrini2010,ferrini2010a} that the coherences of the superpositions are not 
strongly degraded by phase noise, and this degradation  does not increase with the number of atoms in the BJJ. 
Under current experimental conditions, photon scattering is typically negligible and phase noise  can be decreased by
using a  spin-echo technique \cite{gross2010}. In such conditions, the most important source of decoherence is particle losses. 
Three kinds of loss processes  may play a role: one-body losses, due to inelastic collisions between trapped atoms and the background 
gas; two-body losses, resulting from scattering of two atoms in the magnetic trap, which changes their spin and gives them enough
 kinetic energy to be ejected from the trap; and three-body losses, where a three-body collision event produces a molecule and ejects
 a third atom out of the trap. 

In a previous work \cite{pawlowski2013},
we have analyzed the impact of two-body losses on the superpositions of coherent states produced in internal BJJs.  
In this paper, we extend this analysis and study the combined effect of one-body, two-body,  and three-body losses on the formation
of the superposition states. By using a quantum trajectory approach we find,
in agreement with   Ref.~\cite{sinatra2012}, that  for all types of losses the fluctuations in the atomic interaction energy produced by the random loss events
give rise to  an effective phase noise. 
We show that for weak loss rates this noise is responsible for the strongest decoherence effect. 
The tunability of the scattering lengths by Feshbach resonances makes it possible to switch the effective phase noise off
in the mode loosing more atoms, without changing the interaction strength  in the unitary dynamics, \ie keeping 
the formation times of the superpositions fixed.   One may in this way partially protect the coherences of the superpositions
for strongly asymmetric losses in the two modes, as it has been already pointed out in  Ref.~\cite{pawlowski2013} in the case of two-body losses. 
We show in this work that this   result applies to all loss processes and that for moderate loss rates   the corresponding states
are more useful for high-precision atom interferometry than the squeezed states.
This usefulness for interferometry is quantified by the quantum Fisher information 
$F$, which is related to the best  phase  precision achievable in one measurement 
according to $(\Delta \varphi)_{\text{best}}= 1/\sqrt{F}$~\cite{braunstein1994}.
We calculate the Fisher information as a function of time in the lossy BJJ 
by using an exact diagonalization of the master equation.


The paper is organized as follows.   In Sec.\ \ref{sec-madel_and_methods} we recall the 
dynamical evolution in a BJJ  in the absence of tunneling and present the theoretical tools used  to analyze it.
We first  introduce  the Bose-Hubbard model and
the Markovian master equation  describing the dynamics in the presence of 
particle losses (Sec.~\ref{sec-no_loss}).  In the remaining part of the section, we give 
a brief account on atom interferometry (Sec.~\ref{sec-def_Fisher}) and
on the quantum trajectory method for solving master equations (Sec.~\ref{eq-general_QJ}).
Our main results on the time evolution of the quantum Fisher information  in 
a lossy BJJ are presented in Sec.~\ref{quantum_correlations}.  
These results are explained in Sec.~\ref{sec-phys_explan} with the help of the quantum trajectory approach. 
We analyze separately the contributions to the total atomic density  matrix
of quantum trajectories which do not experience any loss (Sec.~\ref{eq-subspace_N_0_atoms}) and 
of trajectories having a single or several loss events (Sec.~\ref{sec-subspace_N_0-2_atoms}). 
The various physical effects leading to an increase or a decrease of the Fisher information at the formation times of the macroscopic superpositions are 
described in detail  (Sec.~\ref{sec-reduced_phase_noise}).
Section~\ref{conclusions} contains a summary and conclusive remarks.
Four appendices offer some additional technical details.

\section{  Model and methods} \label{sec-madel_and_methods}
\subsection{Quenched dynamics of a Bose-Josephson Junction} \label{sec-no_loss}

In this subsection  we first recall the main features of the dynamics of a
two-mode Bose-Josephson junction (BJJ) in the quantum regime
after a sudden quench of the inter-mode coupling to zero. We then introduce the Markovian 
master equation describing atom losses in the BJJ   and the conditional density matrices with fixed numbers of atoms.

\subsubsection{Initial coherent state and Husimi distribution}

We denote by  
$\aopi$, $\aopi^\dagger$, and $\nopi= \aopi^\dagger \aopi$ the bosonic annihilation,
creation, and number operators in mode $i=1,2$. The total number of atoms in the BJJ is given by the 
operator $\Nop =\nopone + \noptwo$.
The Fock states $\ket{n_1,n_2}$ are the joint eigenstates of $\nopone$ and $\noptwo$ with 
eigenvalues $n_1$ and $n_2$, respectively.
Initially, the BJJ is in its
ground state in the regime where inter-mode coupling dominates interactions. 
This initial state is well approximated  by
\begin{equation} \label{eq-initial_state}
\ket{\psi (0)} = \ket{N_0; \phi=0} \equiv \ket{N_0;\theta=\frac{\pi}{2}, \phi=0} 
\, ,
\end{equation}
where $N_0$ is the initial number of atoms and
\begin{eqnarray} \label{eq-def_coherent_state}
\non
\ket{N;\theta,\phi} 
& = & 
\sum_{n_1=0}^N \lt \begin{array}{c} N \\ n_1 \end{array} \rt^{1/2}
  \frac{(\tan (\theta/2))^{n_1}}{[1+\tan^2 (\theta/2)]^{\frac{N}{2}}}
\\
& &  e^{-\I n_1 \phi} \ket{n_1,n_2=N-n_1}
\end{eqnarray}
are the SU(2)-coherent states (CSs) for $N$ atoms \cite{Zhang1990}.

An arbitrary (pure or mixed) state $\hat{\rho}$ with $N$ atoms can be represented by its
Husimi distribution on the Bloch sphere of radius $N/2$,
\begin{displaymath}
 Q_{N} (\theta,\phi) = \frac{1}{\pi}\bra{N ; \theta , \phi}\hat{\rho}\ket{N ; \theta , \phi}\,.
\end{displaymath}
This distribution provides a useful information on the  phase content of $\hat{\rho}$.
The initial CS \eqref{eq-initial_state} has a Husimi distribution with a single  
peak at $(\theta , \phi) = (\frac{\pi}{2}, 0)$ of width $\approx  1/\sqrt{N_0}$, 
as shown in the panel (a) of Fig.~\ref{fig1}.

\subsubsection{Dynamics in the absence of atom losses}
\label{sec-dynamics_no_loss}

\begin{figure}
\begin{minipage}[t]{4.2cm}%
\centering
(a)
\includegraphics[width=\textwidth]{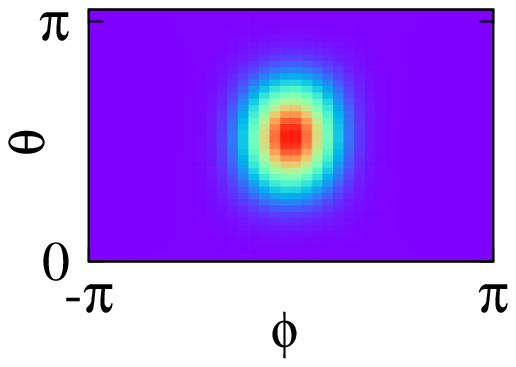}
\end{minipage}\hspace{0.1cm}
\begin{minipage}[t]{4.2cm}%
\centering
(b)
\includegraphics[width=\textwidth]{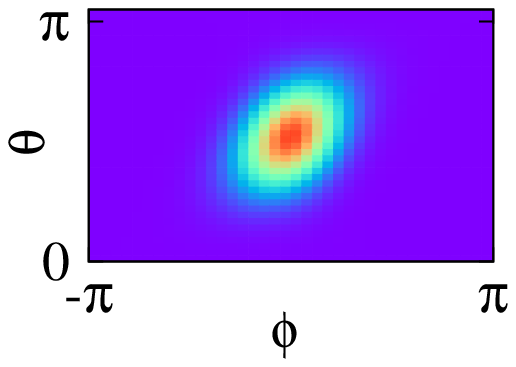}
\end{minipage}

\begin{minipage}[b]{4.2cm}%
\centering
(c)
\includegraphics[width=\textwidth]{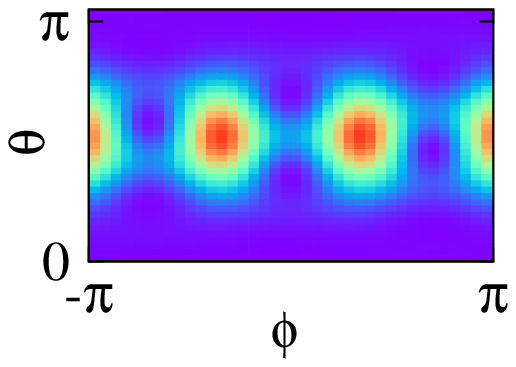}
\end{minipage}\hspace{0.1cm}
\begin{minipage}[b]{4.2cm}%
\centering
(d)
\includegraphics[width=\textwidth]{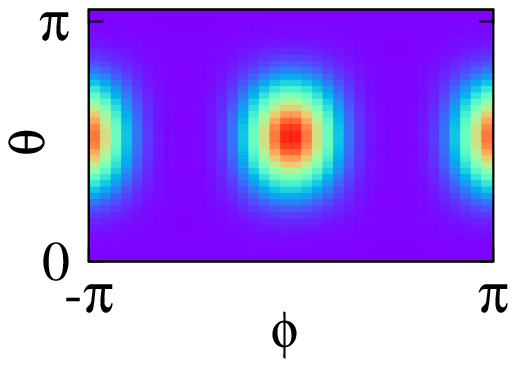}
\end{minipage}
\caption{(Color online) Husimi functions in the absence of losses in the BJJ 
at some specific times: 
(a) $t=0$ (coherent state);
(b) $t=T/40$ (spin squeezed state) , (c) $t=T/6$ (3-component superposition of phase states), (d) $t=T/4$
(2-component superposition). 
Other parameters: $U_1=U_2=\frac{2\pi}{T}$, $U_{12}=0$,   $E_1=E_2=0$, and $N_0=10$. 
}
\label{fig1}
\end{figure}

After a sudden quench to zero of the inter-mode coupling at time $t=0$, 
the two-mode Bose-Hubbard Hamiltonian of the atoms reads~\cite{Milburn97}
\begin{equation} \label{eq-H_0}
\hat{H}_0 = \sum_{i=1,2} \lt E_i \nopi + \frac{U_i}{2} \nopi (\nopi -1) \rt
+ U_{12} \nopone \noptwo ,
\end{equation}
where $E_i$ is the energy of the mode $i$, $U_i$ the interaction energy
between atoms in the same mode $i$, and $U_{12}$ the interaction energy 
between atoms in different modes ($U_{12}=0$ for external BJJs). 
For a fixed total number of atoms $N_0=\nopone+\noptwo$, the Hamiltonian (\ref{eq-H_0}) has a quadratic term in
the relative  number operator
$\nopone-\noptwo$ of the form $\chi (\nopone-\noptwo)^2/4$, with the effective 
interaction energy
\begin{equation} \label{eq-def_chi}  
\chi = \frac{U_1+U_2-2 U_{12}}{2}\;.
\end{equation}
The atomic state $|\psi^{(0)} (t)\rangle = e^{-\I t \hat{H}_0} \ket{\psi (0)}$ 
displays a periodic evolution with period
$T=2 \pi/\chi $ if $N_0$ is even and $T/2$ if $N_0$ is odd.  Before the revival, the dynamics
drives the system first into squeezed states at times $t \approx T N_0^{-\frac{2}{3}}$~\cite{kitagawa1993} 
(see panel (b) in Fig.~\ref{fig1}). At the later times
\begin{equation}
t_q=\frac{\pi}{\chi  q} = \frac{T}{2q}
\quad , \quad  q=2,3,\ldots ,
\end{equation}
the atoms are  in macroscopic superpositions of coherent states,
\be
\label{eq-q-component_cat_state}
| \psi^{(0)} (t_q) \rangle
 = \sum_{k=0}^{q-1} c_{k,q} \bigl| N_0; \phi_{k,q}  \bigr\rangle
\, ,
\ee
with coefficients $c_{k,q}$ of equal moduli $q^{-1/2}$ and
phases $\theta = \pi/2$ and $\phi_{k,q}=\phi_{0,q}+2 \pi k/q$, 
where $\phi_{0,q}$ depends on $q$, $N_0$, and the energies $E_i$ and $U_i$~\cite{yurke1986, stoler1971}.
 In particular, at time $t=t_2$ the BJJ is in the superposition 
$(\ket{N_0; \phi_{0,2} } - \ket{N_0; \phi_{1,2} } )/\sqrt{2}$ of two CSs located
on the equator of the Bloch sphere at diametrically opposite points.
Panels (c) and (d) of Fig. \ref{fig1} show the Husimi distributions of the states (\ref{eq-q-component_cat_state}) for 
 $q=2$ and $q=3$.

It is easy to determine the matrix elements of the density matrix $\hat{\rho}^{(0)} (t)
 =  |\psi^{(0)} (t)\rangle\langle\psi^{(0)} (t)|$ in the Fock basis.
They have time-independent moduli
\begin{equation} \label{eq-density_matrix_no_losses}
 | \bra{n_1,n_2} \hat{\rho}^{(0)} (t) \ket{n_1',n_2'} | 
 =\frac{1}{2^{N_0}} 
\left( \begin{array}{c} N_0\\ n_1 \end{array} \right)^{1/2}  
 \left( \begin{array}{c} N_0\\ n_1' \end{array} \right)^{1/2}
\end{equation}
behaving  in the limit $N_0\gg 1$ like
\begin{equation}
\label{eq-density_matrix_no_losses_asymptotic}
\sqrt{\frac{2}{\pi N_0}} \exp \Bigl\{ - \frac{1}{N_0}\Bigl(  \bigl( n_1-\frac{N_0}{2} \bigr)^2 
+ \bigl( n_1'-\frac{N_0}{2} \bigr)^2 \Bigr) \Bigr\} \;,
\end{equation}
where we have set $n_2 =N_0-n_1$ and $n_2'=N_0-n_1'$.

At the time $t_q$ of formation of the superposition (\ref{eq-q-component_cat_state}),
it is convenient to decompose $\hat{\rho}^{(0)} (t_q)$ 
as a sum of a ``diagonal part''  $[\hat{\rho}^{(0)} (t_q)]_{\rm d}$,
corresponding to the statistical mixture of the CSs in the superposition, 
and an ``off-diagonal part'' $[\hat{\rho}^{(0)}(t_q)]_{\rm od}$ describing 
the coherences between these CSs. Defining $[\hat{\rho}^{(0)}(t_q)]_{kk'}=c_{k,q} c_{k',q}^\ast
|N_0; \phi_{k,q}\rangle \langle N_0;  \phi_{k',q}|$, one has~\cite{ferrini2010}
\begin{eqnarray} \label{eq-density_matrix_no_losses_1}
\nonumber
[\hat{\rho}^{(0)} (t_q)]_{\rm d}
& = & \sum_{k=0}^{q-1} [\hat{\rho}^{(0)}(t_q)]_{k k} 
\\
{[ \hat{\rho}^{(0)}(t_q)]_{\rm od}}
& = & 
 \sum_{k \not= k' =0}^{q-1} [\hat{\rho}^{(0)}(t_q)]_{k k'}\;.
\end{eqnarray}
These  diagonal and off-diagonal parts exhibit remarkable structures in the 
Fock basis, which allow to read them easily from the total density matrix~\cite{ferrini2010}:
\begin{equation} \label{eq-diag_and_off_diag_rho}
\begin{array}{lcl} 
\bra{n_1,n_2} [ \hat{\rho}^{(0)} (t_q)]_{\rm d} \ket{n_1',n_2'} & =  & 
0 \text{ if $n_1' \not= n_1$ modulo $q$}
\\[2ex]
\bra{n_1,n_2} [ \hat{\rho}^{(0)} (t_q) ]_{\rm od} \ket{n_1',n_2'} & = & 
0 \text{ if $n_1'=n_1$ modulo $q$.}
\end{array}
\end{equation}
The off-diagonal part does almost not contribute to the Husimi distribution. 
The Husimi plots  in  Fig.~\ref{fig1} (c,d) thus essentially  show the diagonal parts only.
On the other hand, the quantum correlations useful for interferometry (\ie giving rise to high values of the Fisher
information, see below)
are contained in the off-diagonal part~\cite{ferrini2010}.

\subsubsection{Master equation in the presence of atom losses}
\label{master_eq}

We account for loss processes  in the BJJ by considering the
Markovian master equation~\cite{anglin1997, jack2002, jack2003}
\begin{equation} \label{eq-master_equation}
\frac{\D \hat{\rho}}{\D t} 
 =  
- \I \bigl[ \hat{H}_0 , \hat{\rho} (t) \bigr] 
+ ( {\Ll}_{\text{1-body}} + {\Ll}_{\text{2-body}}  
+ {\Ll}_{\text{3-body}}) (\hat{\rho}(t)) 
\end{equation}
where we have set $\hbar=1$, $\hat{\rho} (t)$ is the atomic density matrix, and
the superoperators ${\Ll}_{\text{1-body}}$, ${\Ll}_{\text{2-body}}$, and ${\Ll}_{\text{3-body}}$ describe
one-body, two-body, and three-body losses, respectively. 
They are given by
\begin{eqnarray} \label{eq-generator_2_body}
\nn
{\Ll}_{\text{1-body}} ( \hat{\rho} )
&  = &
 \sum_{i=1,2} \alpha_i 
  \Bigl( 
   \aopi \, \hat{\rho} \, {\aopi}^\dagger
   - \frac{1}{2} \bigl\{ \nopi , \hat{\rho}  \bigr\}
  \Bigr)
\\
\nn
{\Ll}_{\text{2-body}} ( \hat{\rho} )
&  = &
 \sum_{1 \leq i \leq j \leq 2} \gamma_{ij} 
  \Bigl(
   \aopi \hat{a}_j \, \hat{\rho} \, {\aopi}^\dagger \hat{a}_j^\dagger 
   - \frac{1}{2} \bigl\{ {\aopi}^\dagger \hat{a}_j^\dagger \aopi \hat{a}_j  , \hat{\rho}  \bigr\}
  \Bigr)
\\
\nn
{\Ll}_{\text{3-body}} ( \hat{\rho} )
&  = &
 \sum_{1\leq i \leq j \leq k \leq 2} \kappa_{ijk} 
  \Bigl( 
   \aopi \hat{a}_j \hat{a}_k \, \hat{\rho} \, \aopi^\dagger \hat{a}_j^\dagger \hat{a}_k^\dagger
\\
& &
\hspace*{5mm}   - \frac{1}{2} \bigl\{ \aopi^\dagger \hat{a}_j^\dagger \hat{a}_k^\dagger \aopi \hat{a}_j \hat{a}_k
 , \hat{\rho}  \bigr\}
  \Bigr)\;,
\end{eqnarray}
where  the rates $\alpha_i$, $\gamma_{ij}$, and $\kappa_{ijk}$
correspond to the loss of one atom in the mode $i$,
of two atoms in the modes $i$ and $j$, and  
 of three atoms in the modes $i$, $j$, and $k$ (with $i,j,k =1,2$), respectively,
and $\{ \cdot , \cdot\}$ denotes the anti-commutator.
To shorten notation  we write the loss rate of two (three) atoms in the same mode $i$ as  $\gamma_{i} = \gamma_{ii}$  
($\kappa_i = \kappa_{iii}$) and set $\kappa_{12}=\kappa_{112}$ and $\kappa_{21}=\kappa_{122}$.
Note that the inter-mode rates $\gamma_{12}$, $\kappa_{12}$, and $\kappa_{21}$ vanish
for external BJJs. 
The loss rates depend on the macroscopic  wave function of the condensate and thus on the number of atoms
and interaction energies $U_i$. As far as the number of lost atoms at the revival time $T$ 
remains small with respect to the initial atom number $N_0$, one may, however, assume that these rates are 
time-independent in the time interval $[0,T]$. Hereafter we always  assume that this is the case.

\subsubsection{Conditional states}

The master equation (\ref{eq-master_equation})
does not couple sectors with different numbers of atoms $N$. As a result, if the 
density matrix $\hat{\rho} (t)$ 
has initially no coherences between states with different $N$'s
then such coherences are absent at all times $t \geq 0$. 
Hence  
\begin{equation} \label{eq-block-structure}
\hat{\rho} (t ) = \sum_{N=0}^{N_0} \widetilde{\rho}_N (t) 
\quad ,\quad  \widetilde{\rho}_N (t) = w_N (t) \hat{\rho}_N (t)\;,
\end{equation}
where $\widetilde{\rho}_N (t)$ ($\hat{\rho}_N (t)$) is the unnormalized (normalized) density matrix 
with a well-defined atom number  $N$ (that is, 
$\bra{n_1,n_2} \widetilde{\rho}_N (t) \ket{n_1',n_2'} = 0$ for $n_1 + n_2 \not= N$ or $n_1' + n_2' \not= N$) and
$w_N(t) \geq 0$ is the probability of finding $N$ atoms in the BJJ   at time $t$ (thus $\sum_N w_N (t) = 1$).
The matrix $\hat{\rho}_N (t)$ is the conditional state following a measurement of $\hat{N}$.
More precisely, it describes the state of the BJJ  when one
selects among many single-run experiments
 those for which the measured atom number at time $t$ is equal to $N$ and one averages
all experimental results over these ``post-selected'' single-run experiments, disregarding all the others.
In this sense, $\hat{\rho}_N (t)$ contains a more precise physical information  than the total
density matrix $\hat{\rho}(t)$. To have access to this information,  one must be able to       
extract samples with a well-defined  number of atoms initially (since we assumed an initial state
with $N_0$ atoms) and after the evolution time $t$. 
Even though the precise measurement of $\hat{N}$
is still an experimental challenge,
the precision  has increased by orders of magnitude during the last years~\cite{itah2010, gross2011, hume2013}.

\subsection{Quantum correlations useful for interferometry} \label{sec-def_Fisher}

A useful quantity characterizing quantum correlations (QCs) between   particles in systems involving  many atoms is the
quantum Fisher information. 
Let us recall briefly its definition and its link with phase estimation in
atom interferometry (see~\cite{Klauder1986, ferrini2010, pezze09} for more detail).
In a  Mach-Zehnder atom interferometer, an input state $\hat{\rho}_{\text{in}}$ 
is first transformed into a
superposition of two modes, analogous to the two arms of an optical interferometer. These
modes acquire distinct phases $\varphi_1$ and $\varphi_2$ during the subsequent quantum
evolution and are finally recombined to read out interference fringes, from which
the phase shift $\varphi=\varphi_1-\varphi_2$ is inferred.
  We assume in the whole paper that during  this interferometric sequence one can neglect inter-particle interactions (nonlinear terms in the Hamiltonian
(\ref{eq-H_0})) and 
loss processes. This is well justified in the experiments of Ref.~\cite{gross2010}.
The dependence of the phase sensitivity on inter-particle interactions has been studied in~\cite{Grond11, tikhonenkov10}.
Under this assumption, the  
 output state of the interferometer is
$\hat{\rho}_{\text{out}}(\varphi) = e^{-\I \varphi \hat J_{\vec{n}}} \hat{\rho}_{\text{in}}
e^{\I \varphi \hat{J}_{\vec{n}}}$,
where  $\hat{J}_{\vec{n}}= n_x \hat{J}_x + n_y \hat{J}_y + n_z \hat{J}_z$ is the angular momentum 
generating a rotation on the Bloch sphere   along the axis defined by the unit vector $\vec{n}$, with 
 $\hat J_x=(\hat a^\dagger_1 \hat a_2+ \hat a^\dagger_2 \hat a_1)/2$,
$\hat J_y=-i (\hat a^\dagger_1 \hat a_2-\hat a^\dagger_2 \hat a_1)/2$, and
$\hat J_z = (\hat a^\dagger_1 \hat a_1- \hat a^\dagger_2 \hat a_2)/2$.

The phase shift $\varphi$ is determined by means of a statistical estimator depending on the results of measurements
on the output state $\hat{\rho}_{\text{out}}(\varphi)$. The best precision on $\varphi$ 
that can be achieved (that is, optimizing over  all possible  estimators and measurements)
is given   by~\cite{braunstein1994} 
\be
\label{eq:Cramer_Rao}
(\Delta \varphi)_{\text{best}}
= \fr{1}{\sqrt{ {\cal{M}} \,F ( \hat{\rho}_{\rm in}, \hat{J}_{\vec{n}} ) }}\;,
\ee
where ${\cal{M}}$ is the number of measurements and 
\begin{equation} \label{eq_Fisher_N}
F  ( \hat{\rho} ,\hat{J}_{\vec{n}} ) 
 = 2\sum_{k,l,p_k+p_l >0} \frac{(p_k-p_l)^2}{p_k+p_l} 
 \bigl| \langle k | \hat{J}_{\vec{n}}  | l \rangle \bigr|^2
\end{equation}
is the quantum Fisher information. Here, $\{ | l \rangle \}$ is an orthonormal basis diagonalizing $\hat{\rho}$, 
$\hat{\rho} | l \rangle  =  p_l | l \rangle$.
The quantum Fisher information thus measures the amount of QCs in the input state 
that can be used to enhance phase sensitivity  with respect to the shot noise limit
$ (\Delta \varphi)_{\rm SN} = 1/\sqrt{ {\cal{M}} \,\langle \hat{N} \rangle}$, that is, to the sensitivity obtained
by using $\langle \hat{N} \rangle$ independent atoms. 
Since $\hat{J}_{\vec{n}}$ does not couple
subspaces with different $N$'s, 
it follows from Eq.(\ref{eq_Fisher_N}) and from the block structure (\ref{eq-block-structure}) of
$\hat{\rho}$ that
\begin{equation} \label{eq-total_Fisher}
F ( \hat{\rho} ,\hat{J}_{\vec{n}} ) = \sum_{N=0}^{N_0} w_N  F ( \hat{\rho}_N  ,\hat{J}_{\vec{n}})\;,
\end{equation}
where $ F_N ( \hat{\rho}_N,\hat{J}_{\vec{n}})$ is the Fisher information   of 
the conditional state $\hat{\rho}_N$ with $N$ atoms and $w_N$ is the corresponding probability.

It is shown in~\cite{HyllusPRL2010} that 
if $F ( \hat{\rho},\hat{J}_{\vec{n}} )$ is larger than the average number of atoms $\langle \hat{N} \rangle$
then the atoms are entangled.
According to Eq.(\ref{eq:Cramer_Rao}), the condition  
$F ( \hat{\rho} ,\hat{J}_{\vec{n}} )> \langle \hat{N} \rangle$ is a necessary and 
sufficient condition for sub-shot noise sensitivity 
$(\Delta \varphi)_{\text{best}}< (\Delta \varphi)_{\rm SN}$.

In order to obtain a measure of QCs independent of the  
direction $\vec{n}$ of the interferometer,
we optimize the Fisher information over all unit vectors $\vec{n}$ and define~\cite{Hyllus2010},  
\begin{equation} \label{eq-F_N(t)}
 F ( \hat{\rho} )  = \max_{\| \vec{n} \|=1} F ( \hat{\rho} ,\hat{J}_{\vec{n}} ) = 4 C_{\rm max} \;.
\end{equation}
Here, $C_{\rm max}$ is the largest eigenvalue of the 
$3\times 3$ real symmetric covariance  matrix
\begin{equation}
\label{eq:covariance_general}
C_{ a b} 
  = 
\fr{1}{2} \sum_{k,l,p_k+p_l>0} \frac{(p_k-p_l)^2}{p_k+p_l}
\re  \bigl\{ \langle k | \hat{J}_a| l \rangle 
\langle l | \hat{J}_b| k  \rangle \bigr\} \; ,
\end{equation}
with $a,b = 1,2,3$.
For simplicity we  write $F_{\rm tot} (t) \equiv F ( \hat{\rho}(t) )$ for the optimized Fisher information   of the total
atomic density matrix $\hat{\rho}(t)$ at time $t$  (note that the direction $\vec{n}$ maximizing $ F ( \hat{\rho} ,\hat{J}_{\vec{n}} )$ depends on $t$). 
When studying the QCs in the conditional states we  
optimize over $\vec{n}$ independently in each subspace and 
define $F_N(t) $   as  in (\ref{eq-F_N(t)}), by replacing $\hat{\rho}$ by $\hat{\rho}_N (t)$ in this formula.
Note that $F_{\rm tot} (t)$ is not equal to $\sum_N w_N (t) F_N (t)$, because the optimal 
directions may be different in each subspace. 

\begin{figure*}

\begin{minipage}[t]{8.5cm}%
\centering
(a) symmetric case 
\includegraphics[width=\textwidth]{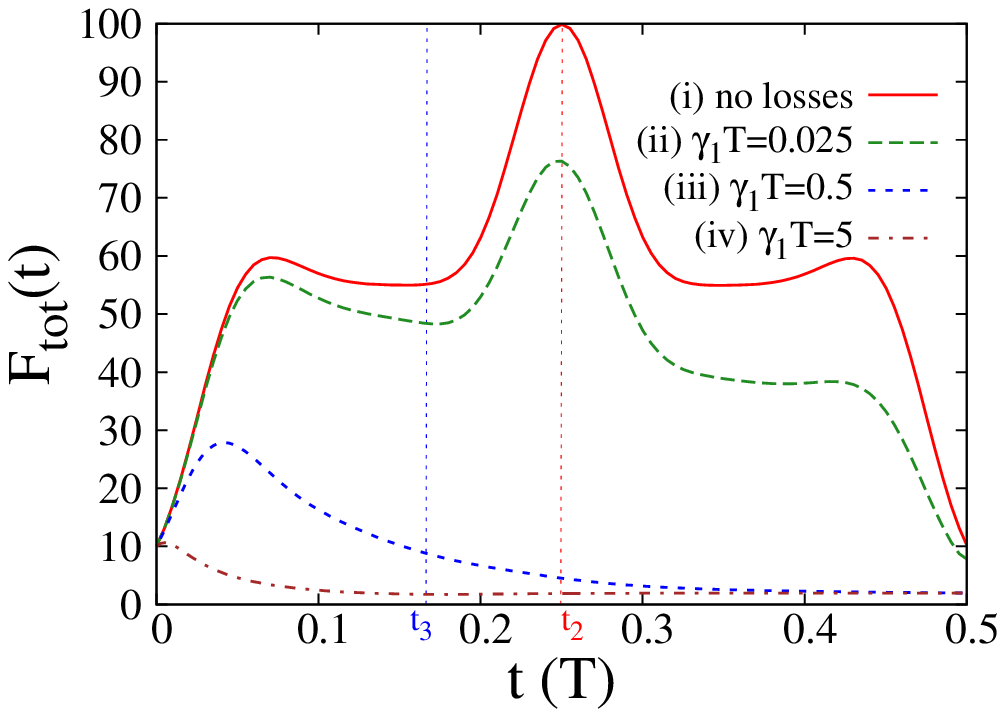}
\end{minipage}\hspace{1cm}
\begin{minipage}[t]{7.5cm}%
\vspace{0.5cm}
\centering
(b) $\gamma_1 T= \gamma_2 T = 0.5$ \\
\includegraphics[width=\textwidth]{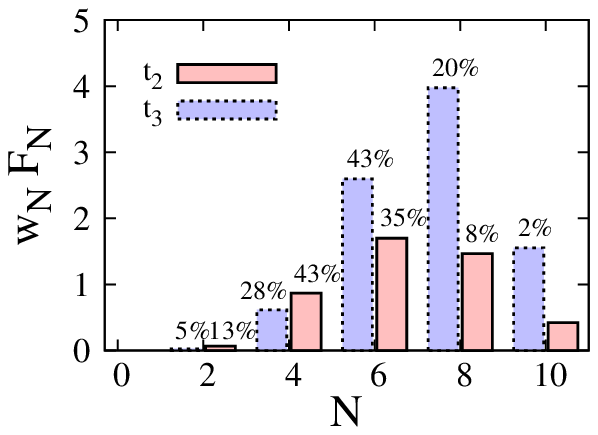} 
\end{minipage}
\begin{minipage}[t]{8.5cm}%
(c) all cases\\
\includegraphics[width=\textwidth]{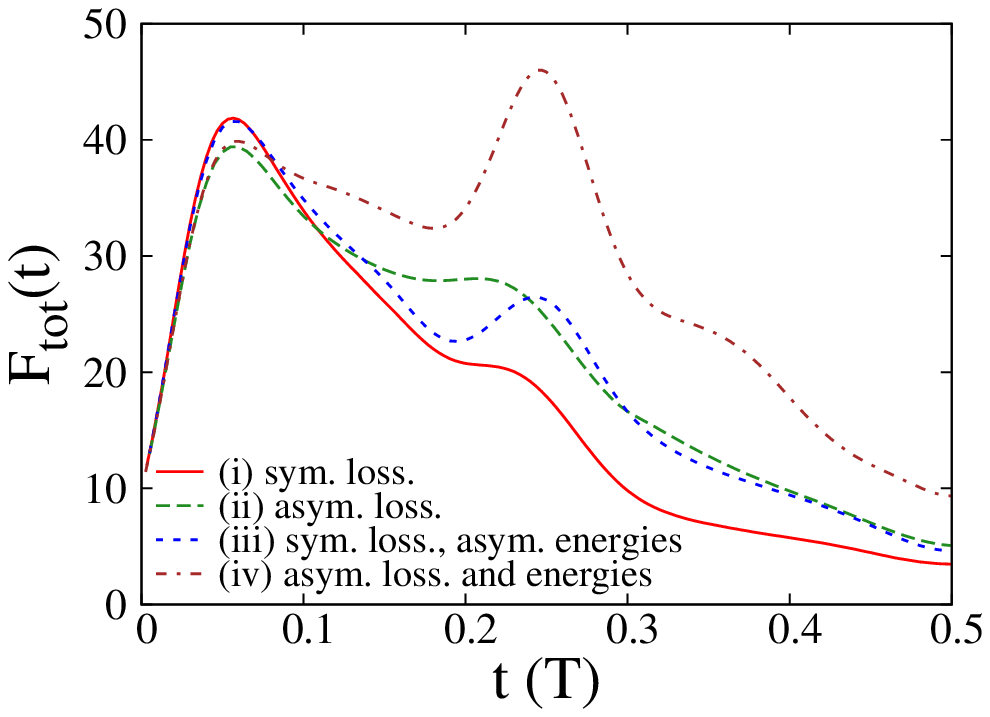}
\end{minipage}
\hspace{1cm}
\begin{minipage}[t]{7.5cm}
\vspace{0.5cm}
\centering
(d) $t=t_2=T/4$\\
\includegraphics[width=\textwidth]{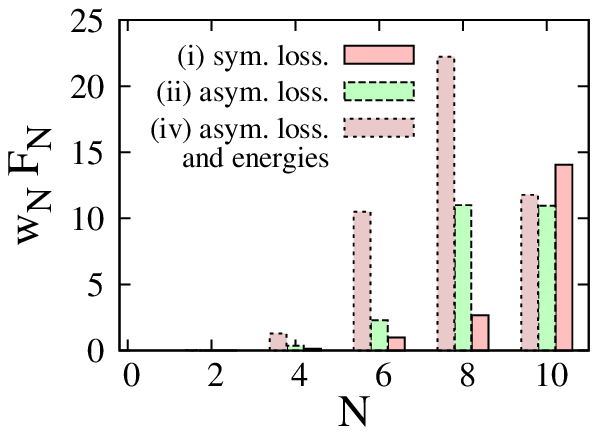}
\end{minipage}
\caption{(Color online) (a) Total quantum Fisher information $F_{\rm tot} (t)$ versus time $t$ (in units of 
$T=2\pi/\chi$) for symmetric 
two-body loss rates $\gamma_1=\gamma_2$ in each mode and $\gamma_{12}=\alpha_i = \kappa_i = \kappa_{ij} = 0$.
The different curves correspond to  (from top to bottom) $\gamma_1 T=0$, $0.025$, $0.5$, and $5$. 
The dotted vertical lines indicate the formation times $t_2=T/4$ and $t_3=T/6$ of the 
2- and 3-component superpositions.
The histogram (b) shows the contributions $w_N F_N (\hat{\rho}_N ,J_{\vec{n}_{\text{opt}}})$ to $F_{\rm tot}$ of the subspaces with 
different atom numbers $N$  [see Eq.\eqref{eq-total_Fisher}] for  
two different times, $t_2$ (pink boxes) and  $t_3$ (blue boxes),
and for the loss rates indicated above the histogram.
The percentages on top of each boxes are the probabilities $w_N$ of finding  $N$ atoms at these times
(weights smaller than 1\% are not indicated). (c) Same as in (a) for
(i)~symmetric losses ($\gamma_1=\gamma_2 = 0.177/T$) and energies ($U_1 = U_2$);
(ii)~asymmetric losses ($\gamma_1 = 0.6/T$, $\gamma_2 = 0$) and symmetric energies 
($U_1 =U_2$);
(iii)~symmetric losses ($\gamma_1=\gamma_2 = 0.177/T$) and asymmetric energies   
($U_2=U_{12}< U_1$);
(iv)~asymmetric losses ($\gamma_1 = 0$, $\gamma_2 = 0.6/T$) and energies ($U_2 =U_{12}<U_1$).
The loss rates are chosen in such a way that the number of lost atoms at time $t_2$ is the same and equal to about 3 in all cases.
The interaction energies $U_i$ are such that $T=4\pi/(U_1+U_2- 2 U_{12})$ is the same in all cases. 
(d)~Histogram of the contributions of the subspaces with $N$ atoms to $F_{\rm tot} (t_2)$ for the same values 
of $\gamma_i$ and $U_i$ as in (c) in the cases (i) (right pink boxes), (ii) (middle green boxes), and (iv) (left purple boxes).
In all panels $N_0=10$, $\gamma_{12}=0$,   and  one- and three-body losses are absent. All results are obtained from the exact diagonalization method 
(see Appendix~\ref{app:exact_diagonalization}).
}
\label{fig2}
\end{figure*}

In the absence of losses, the two-component superposition of CSs
has  the highest possible Fisher information $F [\hat{\rho}^{(0)} (t_2)]= N_0^2$, which is for $N_0\gg 1$
approximately twice larger than  that of the  superpositions with $q$ components, $3 \leq q\lesssim N_0^{1/2}$~\cite{pezze09,ferrini2010}. 
The upper solid curve in Fig.~\ref{fig2}(a) shows 
$F_{\rm tot} (t)$ in the absence of losses as a function of time for $N_0=10$ atoms in the BJJ.

\subsection{Quantum trajectories} \label{eq-general_QJ}

We solve the master equation \eqref{eq-master_equation} using two methods:
the quantum jump approach and an exact diagonalization. 
We outline in this section   the first approach, which yields a tractable analytical solution in the case of few loss events and
gives physical intuition on the various decoherence mechanisms.
This approach will be used  to explain the results provided by the exact diagonalization method, which
offers the exact solution for the whole density matrix when inter-mode losses are absent, 
i.e. $\gamma_{12} = \kappa_{21} = \kappa_{12}  =0 $.
 The  exact diagonalization method is described in  Appendix \ref{app:exact_diagonalization}.
We use it mostly to compute numerically the Fisher information.

In the quantum jump description, the state of the atoms is
a pure state $\ket{\psi (t)}$ which evolves randomly in time as 
follows~\cite{carmichael1991,moelmer1993,belavkin1990,barchielli1991,Knight98,Haroche}.
At random times $s$ quantum jumps occur and the atomic state is transformed as
\begin{equation} \label{eq-jump_dyn}
\ket{\psi(s-)} 
 \longrightarrow 
 \ket{\psi (s+)} = \frac{\hat{M}_{m} \ket{\psi(s-)}}{\| \hat{M}_{m} \ket{\psi(s-)}\| }\;,
\end{equation}
where the index $m$ labels the type of 
jump and $\hat{M}_{m}$ is the corresponding jump operator. In our case,
restricting for the moment our attention to two-body losses, one has three types of jumps: 
the  loss of two atoms in the first mode, with  $\hat{M}_{2,0} = \aopone^2$,  the 
loss of two atoms in the second mode, with  $\hat{M}_{0,2} = \aoptwo^2$, and 
the loss of one atom in each mode, with $\hat{M}_{1,1} = \aopone \aoptwo$.
The probability that a jump $m$ occurs in  
the infinitesimal time interval $[s, s+\D s]$ is 
$\D p_m (s) = \Gamma_m \| \hat{M}_m \ket{\psi(s)}\|^2 \D s$, where $\Gamma_m$ is the jump rate in
the loss channel $m$. Using the notation of Sec.~\ref{master_eq}, one has 
$\Gamma_{2,0} = \gamma_1$, $\Gamma_{0,2} = \gamma_2$, and $\Gamma_{1,1} = \gamma_{12}$. 
Between jumps, the wave function $\ket{\psi(t)}$ evolves
according to the effective non self-adjoint Hamiltonian 
$\Heff = \hat{H_0} - \I \hat{D}_{\rm 2-body}$ with
\begin{eqnarray} \label{eq-K_2_body}
\hat{D}_{\rm 2-body}  & = &  
\frac{1}{2} \sum_{m} \Gamma_m \hat{M}_m^{\dagger} \hat{M}_m 
\\ \nn
& = &
\frac{1}{2} \sum_{i=1,2} \gamma_i  \nopi ( \nopi -1 ) 
+  \frac{\gamma_{12}}{2}  \nopone \noptwo \,.
\end{eqnarray}
The physical origin of the damping term
  comes from the gain of information acquired on the atomic state by conditioning the system to 
have no loss in a given time interval~\cite{moelmer1993,Haroche}: 
the longer the time interval, the smaller must be the number
of atoms left in the BJJ in the mode losing atoms. 

The random wave function at time $t$ reads 
\begin{eqnarray}  \label{eq-no_jump}
\nn
\ket{\psi_J (t)} & = & \frac{\ket{\widetilde{\psi}_J(t)}}{\| {\widetilde{\psi}_J(t)}\|}
\\
\nn
\ket{\widetilde{\psi}_J(t)} 
&  = &  
  e^{-\I(t-s_J)  \Heff } \hat{M}_{m_J} e^{-\I (s_J-s_{J-1})  \Heff} \hat{M}_{m_{J-1}} \cdots 
\\
& & 
\cdots e^{-\I \Heff (s_2-s_1) } \hat{M}_{m_1} e^{-\I s_1 \Heff } \ket{\psi (0)}\;,
\end{eqnarray}
where $J$ is the number of loss events in the time interval $[0,t]$, 
$0 \leq s_1 \leq \cdots \leq s_J \leq t$ are the random loss times, and $m_1,\ldots, m_J$ the random loss types.
The time evolution of the wave function $t \mapsto \ket{\psi_J (t)}$ for a fixed realization of the
jump process is called a quantum trajectory.

The probability to have no atom loss between times $0$ and $t$ is given by $\|e^{-\I t \Heff} \ket{\psi(0)}\|^2$.
The probability to have 
$J$ loss events in $[0,t]$, with the $\nu$th event   of type $m_\nu$ occurring in the time interval
$[s_\nu,s_\nu + \D s_\nu]$, $\nu=1,\ldots , J$, is 
\begin{eqnarray} \label{eq-distrib_jump_times}
\nn
& & \D p_{m_1,\ldots,m_J}^{(t)} (s_1,\ldots , s_J;J) 
\\ 
& & \hspace{1cm} = 
  \Gamma_{m_1} \ldots \Gamma_{m_J} \| {\widetilde{\psi}_J(t)} \|^2 \D s_1 \ldots \D s_J \,.
\end{eqnarray}

The link of this approach with the master equation description is that  
the average over
all quantum trajectories  
(that is, over the number of jumps $J$, the jump times $s_\nu$, and the jump types $m_\nu$)
 of the rank-one projector 
$\ketbra{\psi_J (t)}{\psi_J (t)}$ yields the density matrix $\hat{\rho}(t)$  solution of  
the master equation (\ref{eq-master_equation})~\cite{moelmer1993}. 
We thus recover the block structure (\ref{eq-block-structure})
of the atomic density matrix, with
\begin{eqnarray} \label{eq-density_matrix_subspace_N_0-j}
\nn 
\widetilde{\rho}_{N_J}(t)  
& = & \sum_{m_1,\ldots,m_J} \Gamma_{m_1} \ldots \Gamma_{m_J}
\int_{0\leq s_1 \leq \cdots \leq s_J \leq t} \D s_1 \ldots \D s_J 
\\
& & 
\ketbra{\widetilde{\psi}_J (t)}{\widetilde{\psi}_J(t)}\;,
\end{eqnarray}
  where we have set $N_J = N_0 - 2 J$.
Therefore, quantum trajectories provide a natural and    efficient tool
to study  the conditional states $\hat{\rho}_{N_J} (t)$ with $N_J$
atoms, which only depend on 
quantum trajectories having $J$ two-body loss events.

It is straightforward to extend the above description to include also one- and three-body losses. 
This is achieved by 
adding new types of jumps with jump operators 
$\hat{M}_{1,0}= \hat{a}_1$,  $\hat{M}_{0,1}= \hat{a}_2$ (for one-body losses) and    
$\hat{M}_{3,0}= \hat{a}_1^3$, $\hat{M}_{0,3}= \hat{a}_2^3$,
$\hat{M}_{2,1} = \hat{a}_1^2 \hat{a}_2$, and $\hat{M}_{1,2} = \hat{a}_1 \hat{a}_2^2$
(for three-body losses). The corresponding jump rates are
$\Gamma_{1,0}=\alpha_1$, $\Gamma_{0,1}=\alpha_2$, $\Gamma_{3,0}=\kappa_1$, $\Gamma_{0,3}=\kappa_2$,
$\Gamma_{2,1}=\kappa_{12}$, and $\Gamma_{1,2}=\kappa_{21}$.
The conditional state $\hat{\rho}_{N} (t)$ is obtained 
 by summing the right-hand side of Eq.(\ref{eq-density_matrix_subspace_N_0-j})
over all $J$ and all $r_1,\ldots,r_J  \in \{1,2,3\}$ such that $N_0-\sum_{\nu=1}^J r_\nu = N$,
$r_\nu$ being the number of atoms lost in the $\nu$th loss event.   
The effective Hamiltonian becomes 
$\Heff= \hat{H}_0 -\I  \hat{D}$ with $\hat{D}=\hat{D}_{\rm 1-body} + \hat{D}_{\rm 2-body} + \hat{D}_{\rm 3-body}$ and
\begin{eqnarray} \label{eq-D_1body}
\hat{D}_{\rm 1-body}  
& = &
 \frac{1}{2}\sum_{i=1,2} \alpha_i \nopi 
\\
\nn
\hat{D}_{\rm 3-body} 
&  = &
\frac{1}{2}  \sum_{i=1,2} \kappa_i \nopi ( \nopi-1) (\nopi -2) 
\\  \label{eq-D_3body}
& & + \frac{1}{2}  \sum_{i \not= j} \kappa_{ij} \nopi ( \nopi -1 ) \hat{n}_j \,.
\end{eqnarray}

\section{Main results}\label{quantum_correlations}

We present in this section our main results on the   time evolution of the QCs  in the atomic state under the quenched 
dynamics in the presence of atom losses. The amount of QCs is estimated by the total quantum Fisher information 
$F_{\rm tot} (t)$.

Before investigating the combined effect of the various loss processes, we start by a detailed analysis 
of a small atomic sample with $N_0=10$ atoms subject to 
two-body losses only and without inter-mode losses
(the latter losses cannot be addressed by our exact diagonalization and will be discussed in 
Sec.~\ref{sec-phys_explan}).  
  Figure \ref{fig2} shows the effect of increasing the loss rates in a symmetric model with
$\gamma_1=\gamma_2$ and  $U_1=U_2$  (panel (a)). 
The Fisher information, which in the absence of losses is characterized by a broad peak at the time $t_2=T/4$ of
 formation of the two-component superposition, rapidly decreases once the loss rate increases.  
If, however,  an  asymmetric model is chosen 
-- with parameters yielding the same  average number of lost atoms at time $t_2$ and the same revival time $T$ as in the symmetric case --
  we find  that the Fisher information  is considerably increased (panel (c) in Fig.~\ref{fig2}). The most favorable situation 
turns out to be the one 
with  asymmetric energies $U_2=U_{12} < U_1$ and vanishing loss rate $\gamma_1=0$ in the   mode with largest interactions  
 (a similar result would be obtained for $U_1=U_{12} < U_2$ and $\gamma_2 =0$). 
The histograms shown in panels (b) and (d) in Fig.~\ref{fig2} give  the  contributions   
$w_N ( t ) F ( \hat{\rho}_N ( t ),\hat{J}_{\vec{n}})$ to $F_{\rm tot} (t )$ 
of the various subspaces with fixed atom numbers $N$ (see~(\ref{eq-total_Fisher})),
evaluated in the direction $\vec{n} = \vec{n}_{\rm opt}$ optimizing the Fisher information of the total state, for $t=t_2,t_3$. 
One infers from these histograms  
that the aforementioned effect is non-trivial, namely, the large Fisher informations at times $t_2$ and $t_3$ 
for asymmetric rates and energies
do not come from the contribution of the subspace with $N=N_0$.

\begin{figure}[ht]
\includegraphics[width=8.5cm]{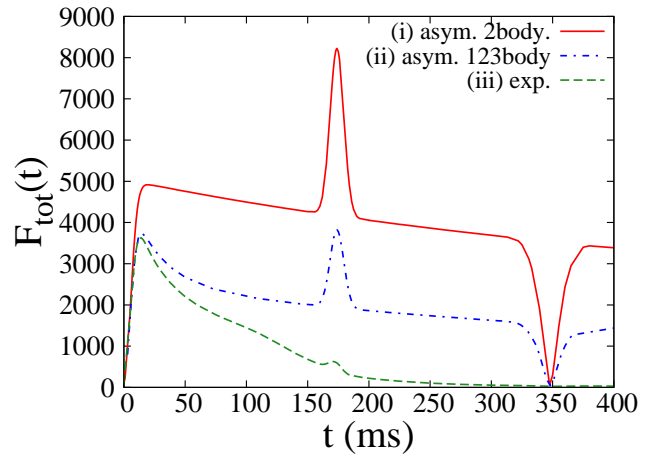}
\caption{(Color online) Total quantum Fisher information $F_{\rm tot} (t)$ versus time $t$ (in units of 
$T$) from exact diagonalization for   more  realistic experimental conditions (see Appendix~\ref{app-parameters})
with  $N_0 = 100$, $U_2  =  U_{12}$, $U_1-U_{12}  = 18.056$Hz, and 
(i)~asymmetric  two-body losses $\gamma_2 = 0.0127$Hz and $\gamma_1 = 0$ without one- and three-body losses; 
(ii) one-, two-, and three-body losses in the second mode with rates $\alpha_2 = 0.4$Hz, 
$\gamma_2 =  0.0127$Hz,
 $\kappa_2 =  1.08\times 10^{-6} $Hz, and no losses in the first mode;
(iii) symmetric one- and three-body losses and asymmetric two-body losses, 
$\alpha_1 = \alpha_2 = 0.2$Hz, $\gamma_2 =  0.0127$Hz, $\gamma_1=0$, and 
$\kappa_1 = \kappa_2 = 0.54\times 10^{-6} $Hz. The case (iii) roughly corresponds to the experimental conditions in Refs.~\cite{riedel2010,gross2010}.
}
\label{fig3}
\end{figure}

We study now  an atomic sample with $N_0=100$ atoms initially in the case where several loss processes are combined 
together.
Figure \ref{fig3} shows the total quantum Fisher information for experimentally relevant parameters extracted 
from Refs.~\cite{sinatra1998, riedel2010, yun2009}   (we explain  how these parameters are obtained in Appendix~\ref{app-parameters}).
As can be seen in this figure,   in the presence of two-body asymmetric losses only, the QCs of the superpositions  are 
well preserved as in the small atomic sample discussed above.
   This asymmetric situation is realized in the experiment of Ref.~\cite{gross2010},
the two-body losses occurring mainly in the upper internal level~\cite{gross2010dissJPB}.    
When one- and three-body losses -- which are also present in this experiment -- are added,  the 
coherences of the superpositions are still preserved provided that all losses occur   in the second mode (upper energy level).
In this case, the QCs   can be protected against atom  losses by 
tuning the interaction energy $U_2$ such that $U_2 = U_{12}$. This shows that the results of Ref.~\cite{pawlowski2013}
concerning two-body losses hold for one- and three-body losses as well.
However, when symmetric one-body or three-body losses are added, the QCs are destroyed on a much shorter 
time scale and the peak in the Fisher information at time $t_2$ disappears. 
In the experiments of Refs.~\cite{riedel2010,gross2010},  the one-body losses are symmetric   since 
they are due to collisions 
with atoms from the background gas, which are equally likely for the two internal atomic states.
These one-body symmetric losses  are therefore much more detrimental to the QCs than asymmetric two-body losses.

\section{Quantum correlations in the subspaces with fixed atom numbers} \label{sec-phys_explan}
\subsection{Overview of the results from the quantum jump method}

In order to explain the behavior of the total Fisher information
observed in the exact diagonalization approach,
we  analyze separately the contributions of each subspace with a fixed atom number to the total
Fisher information.   This is done by using the quantum jump approach of Sec.~\ref{eq-general_QJ}.
We argue in what follows that the stronger decoherence for  symmetric two-body loss rates and energies 
in Fig.~\ref{fig2}(c) originates from a ``destructive interference''  
(exact cancellation) when adding the contributions of the two loss channels
at time $t_2$. A similar cancellation occurs at time $t_3$ for symmetric three-body losses, but it 
is absent for one-body losses.
Moreover, the weaker decoherence for completely asymmetric losses obtained by tuning the interaction energies as described above
comes from the absence of dephasing in the mode $i$ losing atoms.
This effect is somehow trivial for external BJJs: there  this absence of dephasing   occurs for a
vanishing interaction energy $U_i=U_{12}=0$; for such $U_i$ 
the collision processes responsible of two-body losses in the mode $i$ are suppressed (moreover,
our assumption that the loss rates are independent of the energies is not justified anymore). In contrast,
for internal BJJs decoherence is reduced when $U_i$ is equal to the inter-mode 
interaction $U_{12} \not= 0$ and the effect is non-trivial. We shall explain it by invoking the effective phase noise 
produced by atom losses in the presence of
interactions~\cite{sinatra2012, pawlowski2013}.

\subsection{Conditional state in the subspace with the initial number of atoms}
\label{eq-subspace_N_0_atoms}

\begin{figure*}
\begin{minipage}[t]{0.2\textwidth}
\centering $\quad\gamma_1 T=0$
\includegraphics[width=\textwidth]{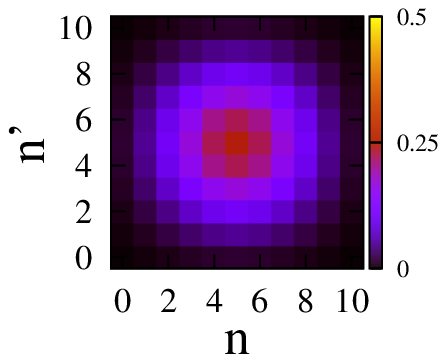}
\end{minipage}
\hspace{0.01\textwidth}
\begin{minipage}[t]{0.2\textwidth}
\centering $\quad\gamma_1 T=2$
\includegraphics[width=\textwidth]{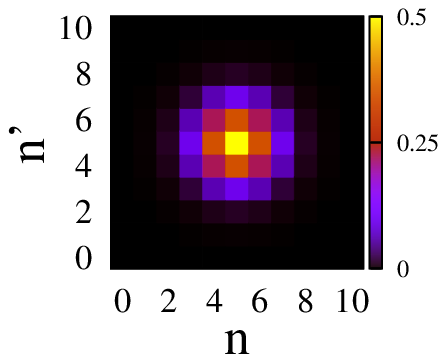}
\end{minipage}
\hspace{0.01\textwidth}
\begin{minipage}[t]{0.2\textwidth}
\centering $\quad\gamma_1 T=10$
\includegraphics[width=\textwidth]{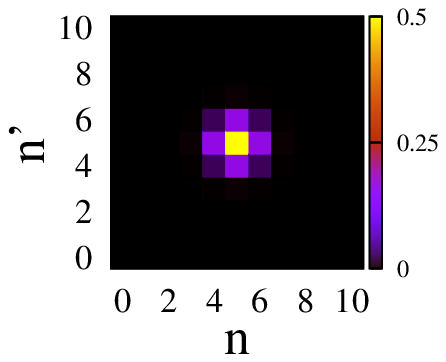}
\end{minipage}
\hspace{0.01\textwidth}
\begin{minipage}[t]{0.2\textwidth}
\centering $\quad\gamma_1 T=20$
\includegraphics[width=\textwidth]{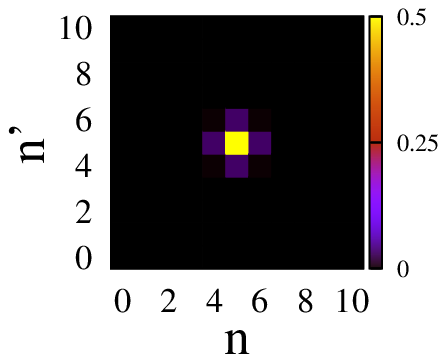}
\end{minipage}

\begin{minipage}[t]{0.2\textwidth}
\centering
 $\gamma_1 T=0$
\includegraphics[width=\textwidth]{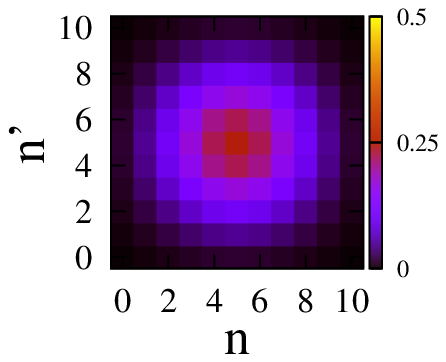}
\end{minipage}
\hspace{0.01\textwidth}
\begin{minipage}[t]{0.2\textwidth}
\centering
 $\gamma_1 T=4$
\includegraphics[width=\textwidth]{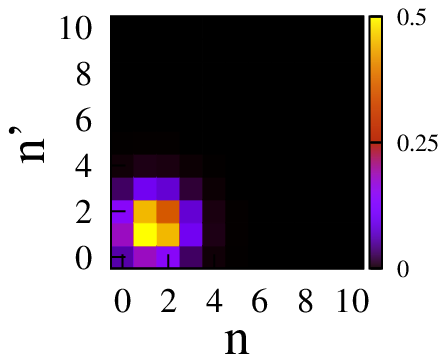}
\end{minipage}
\hspace{0.01\textwidth}
\begin{minipage}[t]{0.2\textwidth}
\centering
 $\gamma_1 T=20$
\includegraphics[width=\textwidth]{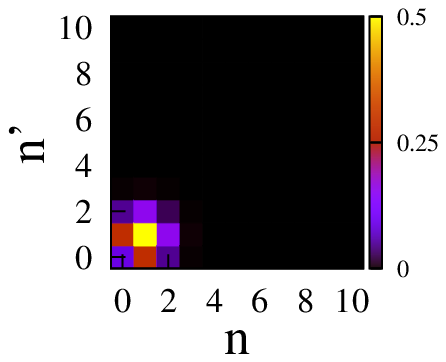}
\end{minipage}
\hspace{0.01\textwidth}
\begin{minipage}[t]{0.2\textwidth}
\centering
 $\gamma_1 T=40$
\includegraphics[width=\textwidth]{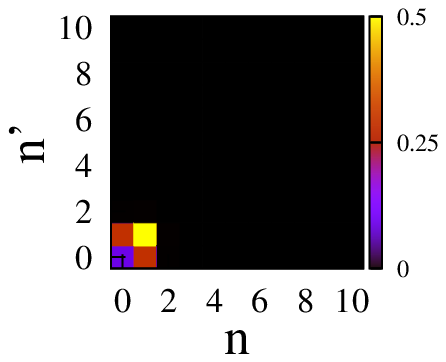}
\end{minipage}
\caption{
(Color online) Moduli of the matrix elements in the Fock basis 
of the state $\hat{\rho}_{N_0} (t_3)$ in the subspace with $N_0$ atoms 
at the time of formation $t_3 = T/6$ of the 3-component superposition, from the exact diagonalization
method.
Upper panels: symmetric two-body losses 
($\gamma_1=\gamma_2$).
Lower panels:  completely asymmetric two-body losses ($\gamma_2=0$). The values of $\gamma_1$ 
are indicated on the top of each panel. Other parameters:
$\alpha_i=\gamma_{12}= \kappa_i =\kappa_{ij}=0$ and $N_0=10$.
 }
\label{fig4}
\end{figure*}

\begin{figure}
\includegraphics[width=8.6cm]{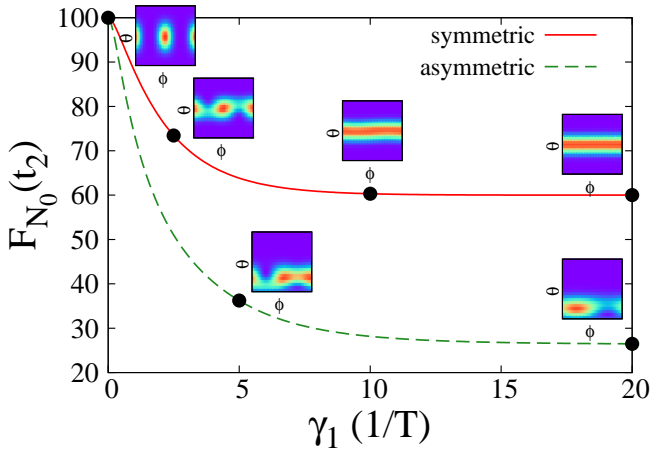} 
\caption{(Color online)
Fisher information $F_{N_0}(t_2)$ optimized in the subspace with $N_0$ atoms 
at time $t_2$ 
as a function of the loss rate
$\gamma_1$ (in units of $T^{-1}$).
Solid line: symmetric losses ($\gamma_1=\gamma_2$). 
Dashed line: asymmetric losses ($\gamma_2=0$).  
The Husimi functions are plotted in the insets
 for some specific choices of loss rates indicated by circles on the two curves.
Other parameters as in Fig.~\ref{fig4}. 
}
\label{fig5}
\end{figure}

We start by determining the conditional state $\hat{\rho}_{N_0}(t)$ with the initial number of atoms  $N_0$ at time $t$. In the quantum jump 
approach this corresponds to  the contribution of quantum trajectories with no jump in the time interval
$[0,t]$, which are given by (see Sec.~\ref{eq-general_QJ})
\begin{equation}
\ket{\widetilde{\psi}_{0} (t)} = e^{-\I t (\hat{H}_0 -\I \hat{D} )} \ket{N_0;\phi=0}
\,.
\end{equation}
The unnormalized conditional state is 
$\widetilde{\rho}_{N_0}^{\rm{(no\,loss)}}  (t)  = \ketbra{\widetilde{\psi}_{0} (t) }{\widetilde{\psi}_{0} (t)}$.
In the Fock basis diagonalizing both 
$\hat{H}_0$ and $\hat{D}$, it takes the form
\begin{eqnarray} \label{eq-state_no_jump}
& & 
\nn
\bra{n_1,n_2}  \widetilde{\rho}_{N_0}^{\rm{(no\,loss)}} (t)  \ket{n_1',n_2'} = 
\\
& &  \hspace*{0.6cm}
 e^{-t [d_{N_0} (n_1)+d_{N_0} (n_1')]}
 \bra{n_1,n_2} \hat{\rho}^{(0)}_{N_0} (t) \ket{n_1',n_2'}\;,
\end{eqnarray}
where $\hat{\rho}^{(0)}_{N_0} (t)$ is the state in the absence of losses (see Sec.~\ref{sec-dynamics_no_loss}),
 $n_2 = N_0-n_1$, $n_2'=N_0-n_1'$, and
$d_{N_0}(n_1)=\bra{n_1,n_2} \hat{D} \ket{n_1,n_2}$.
The  probability to find $N_0$ atoms at time $t$ is found with the help of
Eqs.(\ref{eq-density_matrix_no_losses}) and (\ref{eq-state_no_jump}). One finds
\begin{equation} \label{eq-proba_w_N}
w_{N_0} (t) = \tr \widetilde{\rho}_{N_0}^{\rm{(no\,loss)}}  (t) = 
\frac{1}{2^{N_0}} 
\sum_{n_1=0}^{N_0} 
\left( \begin{array}{c} N_0 \\ n_1 \end{array}\right)
e^{-2 t d_{N_0} (n_1)} \,.
\end{equation}

In the following we restrict our attention to {\it symmetric three-body losses}  $\kappa_1=\kappa_2$ and $\kappa_{12}= \kappa_{21}$.
The asymmetric three-body loss case is treated in Appendix~\ref{app-rho_N_0_strong_losses}.
Let us set $\kappa=(3 \kappa_1-\kappa_{12})/2$ and
\begin{equation} \label{eq=def_damping_a}
a  =  \frac{1}{2} \bigl( \gamma_1 + \gamma_2 -\gamma_{12} \bigr) + (N_0-2) \kappa \, .
\end{equation}
If $a \not=0$, the damping factor  in Eq.(\ref{eq-state_no_jump}) (\ie the exponential factor in the right-hand side) is Gaussian. Actually,
by using Eqs.(\ref{eq-K_2_body}), (\ref{eq-D_1body}), and (\ref{eq-D_3body}), one obtains 
\begin{equation} \label{eq-d(n_1)}
d_{N_0} (n_1) 
= a  (n_1-\overline{n}_1 )^2 + c\;,
\end{equation}
where $c$ is an irrelevant $n_1$-independent constant
which can be absorbed in the normalization of the density matrix, and
\begin{equation} \label{eq-overline_n_1}
\overline{n}_1
 =  
\frac{1}{4 a} 
 \Bigl(
  \Delta \alpha - \Delta \gamma  + N_0 (2\gamma_2 - \gamma_{12}) + 2 N_0 (N_0-2)  \kappa 
 \Bigr)
\end{equation}
with $\Delta \alpha = \alpha_2 - \alpha_1$ and $\Delta \gamma = \gamma_2 - \gamma_1$.

In order to estimate the QCs in $\hat{\rho}_{N_0} (t)$  and the typical loss rates at which this state is 
affected by the Gaussian damping, we now focus on three particular cases.

(i) {\it Symmetric loss rates} $\gamma_1=\gamma_2$ and $\alpha_1=\alpha_2$.
In this case $a=\gamma/2 + (N_0-2) \kappa$ with $\gamma = 2 \gamma_1 - \gamma_{12}$ and
 $\overline{n}_1=N_0/2$. Hence 
the damping factor in Eq.(\ref{eq-state_no_jump}) is a Gaussian centered 
at $(n_1,n_1') = (N_0/2,N_0/2)$. This center coincides with the peak of the  
matrix elements
in the absence of losses, which have a width $\approx \sqrt{N_0}$, see Eq.(\ref{eq-density_matrix_no_losses_asymptotic}).
Thus the effect of the Gaussian damping begins to set in for times $t$ such that $a t  \approx 1/N_0$.  In particular, the
macroscopic superposition at time $t_q = \pi/( \chi q)$ is noticeably affected by damping for
$a \gtrsim \chi q/N_0$.    
It  is shown in Appendix~\ref{app-rho_N_0_strong_losses}  that
$\hat{\rho}_{N_0} (t_q)$ converges at large loss rates $a \gg \chi q$ 
to the pure Fock state $\ket{N_0/2,N_0/2}$ with equal numbers of atoms in each mode if $N_0$ is even, as
  it could have been expected from the symmetry of the losses. This convergence is illustrated in 
the upper panels in Fig.~\ref{fig4}, which represent the density matrix (\ref{eq-state_no_jump}) at time $t=t_3$  
for increasing symmetric two-body loss rates and vanishing one-body, three-body, and inter-mode rates. 
The Fisher information $F_{N_0}(t_2)$ in the subspace with $N_0$ atoms at time $t_2$  is displayed in Fig.~\ref{fig5}.
For $\gamma_1 =\gamma_2 \gtrsim 5/T$, it is close to the Fisher  information $F_{N_0}(\infty) = N_0 (N_0/2+1)$ 
of the Fock state $\ket{N_0/2,N_0/2}$.
Let us, however,  stress  that  at such loss rates 
 $\hat{\rho}_{N_0} (t_2)$ has a negligible contribution to the total density matrix (\ref{eq-block-structure}) and
is very unlikely to show up in a single-run experiment,
because the no-jump probability $w_{N_0} (t_2)$ is very small.
  Thus, the large value of $F_{N_0}(t_2)$ for strong symmetric losses does  not 
mean that  the total atomic state $\hat{\rho}(t)$ has a large amount of QCs.  
The Husimi distributions of the conditional state $\hat{\rho}_{N_0} (t_2)$ are
shown in the upper insets in Fig.~\ref{fig5} for various rates $\gamma_1$. 
The two peaks at $(\theta,\phi)=(\pi/2,0)$ and 
$(\pi/2,\pi)$ of the two-component superposition in the absence of losses
are progressively washed out
at increasing $\gamma_1$, until one reaches the $\phi$-independent distribution of the Fock state 
$\ket{N_0/2,N_0/2}$.

(ii) {\it Completely asymmetric two-body losses} and  {\it no three-body losses},
$\gamma_2=\gamma_{12}=\kappa=0$. Then $a=\gamma_1/2$ and
$\overline{n}_1=\Delta \alpha/(2 \gamma_1) + 1/2$.
The onset of the damping on the $q$-component superposition is at the loss rate 
$\gamma_1 \approx \chi q/N_0^2$, which is smaller
by a factor   of $N_0$ compared with the symmetric case, except for strongly asymmetric one-body loss rates 
 satisfying $\Delta \alpha \approx \gamma_1 N_0$. In the last case, this onset occurs when 
$\gamma_1 \approx \chi q/N_0$ as in case (i). Therefore, if
$\Delta \alpha$ is not of the order of $\gamma_1 N_0$, the Gaussian damping   
 affects more strongly the macroscopic superpositions than in the symmetric case.  
The lower panels in Fig. \ref{fig4}  represent the
matrix elements of $\hat{\rho}_{N_0}(t_3)$ in the Fock basis and the  dashed curve in  Fig.~\ref{fig5} displays the Fisher information 
$F_{N_0}(t_2)$ for
$\Delta \alpha = \kappa = 0$.
Except at small values of $\gamma_1$, $F_{N_0}(t_2)$ is much smaller than for symmetric losses.
This can be explained from the results of Appendix~\ref{app-rho_N_0_strong_losses}, which show that
$\hat{\rho}_{N_0} ( t_q)$ converges  in the strong loss limit $\gamma_1 \gg \chi q$
to  the Fock state  $\ket{0,N_0}$ if $\alpha_2 < \alpha_1$ and to  a superposition of
Fock states with $n_1=0$ or $1$ atoms in the first mode if $\alpha_1 = \alpha_2$.
These pure states  have Fisher informations of the order of $ N_0$,   which are smaller by a factor of $N_0$ than those obtained
for strong symmetric losses. 
Because the aforementioned  Fock states are  
localized near the south pole of the Bloch sphere ($\theta=0$), 
the two peaks in the Husimi functions (lower insets in Fig.~\ref{fig5}) 
move  to values of $\theta$ smaller than $\pi/2$ when increasing $\gamma_1$.
Note that this picture is drastically modified when $\alpha_2 = \gamma_1 N_0 + \alpha_1$: then
$\hat{\rho}_{N_0}(t_2)$ converges to a superposition of the Fock states $\ket{N_0/2,N_0/2}$ and
$\ket{N_0/2+1,N_0/2+1}$ for even $N_0$ (see Appendix~\ref{app-rho_N_0_strong_losses}), and thus
 $F_{N_0}(t_2)$ behaves like in the case (i).  

(iii) {\it Strong inter-mode two-body losses} 
$\gamma_{12} > \gamma_1 + \gamma_2+2 (N_0-2) \kappa$, \ie $a < 0$. 
Then the onset of damping at time $t_q$ occurs for $|a| \approx \chi q/N_0^2$,   except when 
$\Delta \gamma \approx - \Delta \alpha/(N_0-1)$, in which case it occurs 
for $|a| \approx \chi q/N_0$. As shown in the Appendix~\ref{app-rho_N_0_strong_losses},
$\hat{\rho}_{N_0} ( t_q)$ converges at strong losses either to the Fock state with 
$n_1=0$ or $n_1=N_0$ atoms in the first mode, which has a Fisher information $F_{N_0}(\infty)= N_0$, or,
 if $\Delta \gamma  = -\Delta \alpha/(N_0-1)$, 
to the so-called NOON state, which has the highest possible Fisher information 
$F_{N_0}(\infty)=N_0^2$.

\subsection{Conditional states with $N<N_0$ atoms} \label{sec-subspace_N_0-2_atoms}

We study in this subsection the contribution to the total atomic density matrix
$\hat{\rho}(t)$ of
quantum trajectories  having $J\geq 1$ jumps in the time interval $[0,t]$. 

\subsubsection{General results} 

We first fix some notation.
Let $t \mapsto \ket{\psi_J (t)}$ be a trajectory subject to $J$ loss events,   
occurring at times $0 \leq s_1 \leq \cdots \leq s_J \leq t$. 
As in Sec.~\ref{eq-general_QJ} we denote each type of loss by the pair $m=(m_1,m_2)  \in \{1,2,3\}^2$, where $m_1$ and $m_2$
are the number   of atoms lost in the first and second  modes, respectively. 
The associated 
jump operator is $\hat{M}_m = \hat{a}_1^{m_1} \hat{a}_2^{m_2}$.
We use the vector notation $\sv = (s_1, \ldots, s_J)$ for the
sequence of  loss times $s_\nu$
and $\mv = (m_1,\ldots ,m_J )$ for the sequence of loss types $m_\nu$. 
Here $m_\nu= (m_{\nu,1}, m_{\nu,2})$ with $m_{\nu,i}$ the number of atoms lost in mode $i$ during the $\nu$th loss process.
Finally, let  $|\mv |= \sum_{\nu=1}^{J} (m_{\nu,1}+m_{\nu,2})$ be
the total number of atoms ejected from the condensate between times $0$ and $t$. 

It is easy to see that each jump (\ref{eq-jump_dyn}) transforms a CS $\ket{N_0;\theta ,\phi}$
into a CS  $\ket{N_0-r;\theta ,\phi}$, where $r=1,2,3$ is the number of atoms lost during the jump.
This CS is rotated on the Bloch sphere by the 
evolution between jumps driven by the nonlinear effective Hamiltonian $\Heff$. This rotation is due to the different numbers of atoms  in the BJJ
in the time intervals $[0,s_1]$, $[s_1,s_2], \cdots , [s_J, t]$, leading to different interaction energies. 
More precisely, it is shown in Appendix~\ref{eq-app-D} that for three-body loss rates satisfying
\begin{equation} \label{eq-cond_on_3_body_losses}
\kappa_{i} , \kappa_{ij} \ll ( N_0 t)^{-1}\;, i,j=1,2, i \not= j \;,
\end{equation} 
the wave function $\ket{\psi_J (t)}$ is up to a normalization factor given by
 \begin{equation} 
\label{eq-wave function_one_jump_bis_bis}
\ket{{\psi}_J (t)}
\propto   e^{-\I t \Heff }  \ket{N_0-|\mv| ; \theta_{\mv} (\sv) , \phi_{\mv} (\sv) } \;,
\end{equation}
where $\theta_{\mv} (\sv)$ and $\phi_{\mv} (\sv)$ are random angles depending on the loss types 
and loss times. These angles are given by
\begin{eqnarray} \label{eq-random_phases_J>1}
\nn
\theta_{\mv} (\sv)  &  =  & 
2 \arctan \Bigl( \exp \Bigl\{ -\sum_{\nu=1}^J \frac{s_\nu}{2} 
 \bigl( \delta_1 m_{\nu,1}+ \delta_2 m_{\nu,2}\bigr)  \Bigr\} \Bigr)
\\ 
\phi_{\mv} (\sv)  & = &  \sum_{\nu=1}^J s_\nu (\chi_1 m_{\nu,1} + \chi_2 m_{\nu,2} )\;,
\end{eqnarray}
where we have introduced the interaction energies 
\begin{equation} \label{eq-def_chi2}
 \chi_1  =  U_1-U_{12} \quad , \quad \chi_2 = - (U_2 - U_{12}) \;,
\end{equation}
and the loss rate differences
\begin{eqnarray}\label{eq-def_delta}
\non
 \delta_{1} & = & 2 \gamma_1 -  \gamma_{12} + (3 \kappa_1 - \kappa_{21}) N_0 \;,
\\
\delta_{2} & = & -( 2 \gamma_2 - \gamma_{12} + (3 \kappa_2 - \kappa_{12}) N_0 )
\;.
\end{eqnarray}
Equation (\ref{eq-wave function_one_jump_bis_bis}) means that, apart from damping effects 
due to  the effective Hamiltonian  $\Heff$, atom losses can be accounted for  
by random fluctuations of the two  phases $\theta$ and $\phi$ of the CS.
For a single loss event ($J=1$), these fluctuations have magnitude  
\begin{equation} \label{eq-phase_fluctuations}
\delta \theta_{m}  \simeq  \frac{1}{2} \delta s_{m}   \bigl| \sum_{i=1,2} \delta_i m_i  \bigr|
\quad , \quad 
\delta \phi_m   =  \delta s_{m}  \bigl| \sum_{i=1,2} \chi_i m_i  \bigr|
\end{equation}
(we assume here $\delta \theta_{m} \ll 1$), where
$\delta s_{m}$ is the fluctuation of the loss time $s$, whose 
distribution is given by Eq.~(\ref{eq-distribution_jump_timebis}) in Appendix~\ref{eq-app-D}. 
This  analogy between atom losses and $\phi$-noise is already known in the literature 
in the large $N_0$ regime~\cite{sinatra2012}. 
In this regime the $\theta$-noise is negligible (see below).

The conditional states $\hat{\rho}_{N} (t)$ with $N< N_0$ atoms
turn out to be simply related  to the 
(unnormalized) density matrix conditioned to no loss event 
for an  initial CS with $N$ atoms, defined as follows
\begin{equation} \label{eq-rho_no_jump_N_0-2}
\widetilde{\rho}_{N}^{\rm{(no\,loss)}} (t) 
 =  e^{-\I t \Heff} \ketbra{N;\phi= 0}{N;\phi=0}  e^{\I t \Heff^\dagger}\, .
\end{equation}
The matrix elements of  $\widetilde{\rho}_{N}^{\rm{(no\,loss)}} (t)$ in the Fock basis are
given by Eq.(\ref{eq-state_no_jump}) upon the replacement $N_0 \rightarrow N$.
It is demonstrated in Appendix~\ref{eq-app-D}
that if $N < N_0$, the matrix $\hat{\rho}_{N}(t)$ is given in this basis by
\begin{eqnarray} \label{eq-density_matrix_N_0-2_sector_in_Fock_basis_J>1}
\nn
& & 
\bra{n_1 ,n_2}  \hat{\rho}_{N} (t ) \ket{n_1',n_2'} 
\propto  
\\
& & \hspace*{7mm}  
{\mathcal{E}}_N ( t ; n_1, n_1')  
\bra{n_1 ,n_2}  \widetilde{\rho}_{N}^{\rm{(no\,loss)}} (t ) \ket{n_1',n_2'} \;,
\end{eqnarray}
where  ${\mathcal{E}}_{N} ( t; n_1,n'_1)$ is an envelope depending
on time and on the matrix entries $n_1$ and $n_1'$. This envelope is determined explicitly
in  Appendix~\ref{eq-app-D}. If a  single $r$-body loss event occurs between times $0$ and $t$, 
it is denoted by ${\mathcal{E}}^{\rm{(1-jump)} }_{N_0-r}  ( t; n,n')$ and is given by Eq.(\ref{eq-envelope_1_jump}). 
According to Eq.(\ref{eq-density_matrix_N_0-2_sector_in_Fock_basis_J>1}), 
$\hat{\rho}_{N} (t )$ is given in the Fock basis by the lossless density matrix $\hat{\rho}^{(0)}_N (t)$ for an
initial CS with $N$ atoms modulated  by the envelope ${\mathcal{E}}_{N}$ and by
 the  damping factor of Eq.(\ref{eq-state_no_jump}).

Let us assume that, in addition to the above condition (\ref{eq-cond_on_3_body_losses}) on three-body losses, 
the two-body loss rates  satisfy $\gamma_i ,\gamma_{12} \ll t^{-1}$. Furthermore, let
the total number $| \mv | = N_0 - N$ of atoms lost between times $0$ and $t$ be much smaller than $N_0$. Then
one finds (see  Appendix~\ref{eq-app-D})
\begin{eqnarray} \label{eq-enveloppe_J>1}
\nn 
{\mathcal{E}}_{N} ( t; n,n')
& = & \sum_{J_1,J_2,J_3 \geq 0, J_1+2 J_2+3J_3 = N_0-N} 
\frac{1}{J_1! J_2! J_3!} 
\\
& &
\prod_{r=1}^3
\Bigl[ {\mathcal{E}}^{\rm{(1-jump)} }_{N_0-r}  ( t; n,n') \Bigr]^{J_r}\, .
\end{eqnarray}
Therefore, the envelope for several jumps is obtained  by multiplying together the single-jump envelopes 
raised to the power $J_r$, and by summing over all the numbers
 $J_r$ of $r$-body losses in the time interval $[0,t]$ such that 
$N=N_0-J_1 - 2 J_2 - 3 J_3$.

Equations (\ref{eq-wave function_one_jump_bis_bis}), (\ref{eq-density_matrix_N_0-2_sector_in_Fock_basis_J>1}), 
and (\ref{eq-enveloppe_J>1}) are our main analytical results from the quantum trajectory approach. 
They are valid provided that $t \Gamma_m \ll N_0^{2-|m|}$ for 
all two-body ($|m|=2$) and three-body ($|m|=3$) loss rates $\Gamma_m$.
This is not a strong restriction since  for large $N_0$ the mean number $\langle \hat{N} \rangle_t$ of atoms 
in the BJJ at time $t$ when the BJJ is subject to   two-body  (respectively, three-body)  losses is of the order of
 $N_0 \gamma_i t  $ (respectively, $N_0^2 \kappa_i t$)\,
\footnote{If the BJJ is subject to symmetric two-body (respectively three-body) losses only,  the 
phenomenological rate equations give
${\langle} \hat{N} {\rangle }_t \simeq N_0 ( \gamma_1 N_0 t + 1)^{-1}$ 
(respectively ${\langle} \hat{N} {\rangle }_t \simeq N_0 ( 2 \kappa_1 N_0^2 t + 1)^{-1/2}$) for
$N_0 \gg 1$.}.
Hence the aforementioned condition  is still fulfilled if a large fraction (e.g. $50\%$)
of the initial atoms
 are lost between times $0$ and $t$. 
For that reason, we will say that the  loss rates $\Gamma_m$ such that  $( tN_0^{|m|-1} )^{-1}  \lesssim \Gamma_m \ll ( t N_0^{|m|-2} )^{-1}$
pertain to the intermediate loss rate regime.

\begin{figure*}
\begin{minipage}[t]{0.2\textwidth}
\centering
$\gamma_1 T=\gamma_2 T=0.025$
\includegraphics[width=\textwidth]{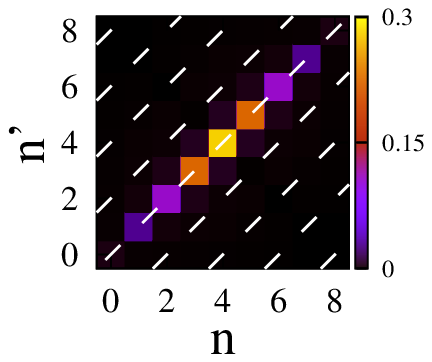}
\end{minipage}
\hspace{0.01\textwidth}
\begin{minipage}[t]{0.2\textwidth}
\centering
$\gamma_1 T=\gamma_2 T=0.5$
\includegraphics[width=\textwidth]{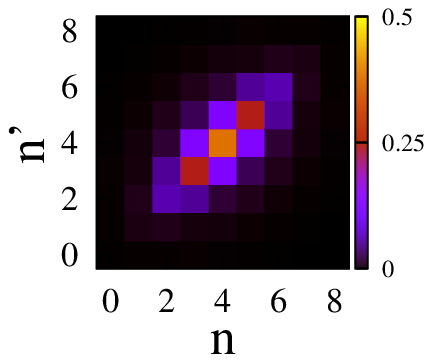}
\end{minipage}
\hspace{0.01\textwidth}
\begin{minipage}[t]{0.2\textwidth}
\centering
$\gamma_1 T=\gamma_2 T=2.0$
\includegraphics[width=\textwidth]{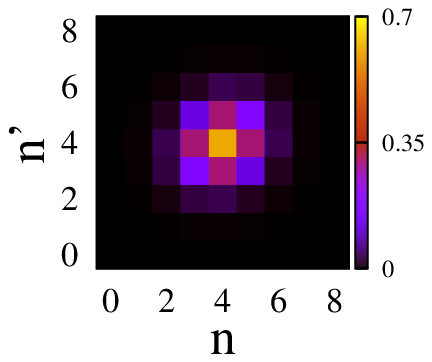}
\end{minipage}
\hspace{0.01\textwidth}
\begin{minipage}[t]{0.2\textwidth}
\centering
 $\gamma_1 T=\gamma_2 T=10$
\includegraphics[width=\textwidth]{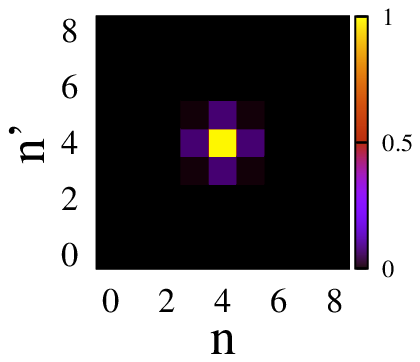}
\end{minipage}


\begin{minipage}[t]{0.2\textwidth}
\centering
$\gamma_1 T=0.05$, $\gamma_2=0$
\includegraphics[width=\textwidth]{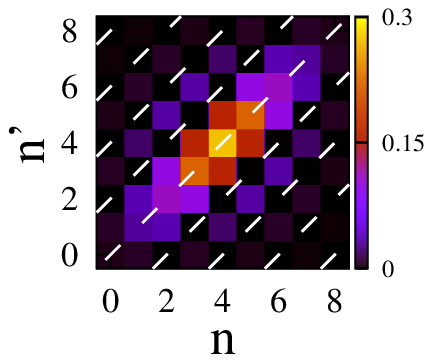}
\end{minipage}
\hspace{0.01\textwidth}
\begin{minipage}[t]{0.2\textwidth}
\centering
 $\gamma_1 T=1$, $\gamma_2=0$
\includegraphics[width=\textwidth]{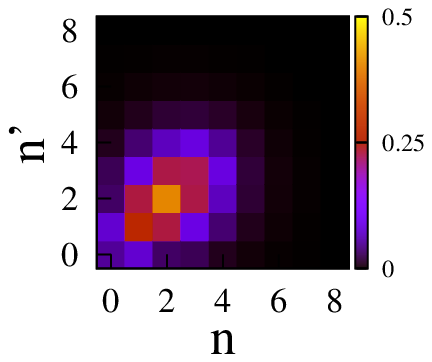}
\end{minipage}
\hspace{0.01\textwidth}
\begin{minipage}[t]{0.2\textwidth}
\centering
$\gamma_1 T=4.0$, $\gamma_2=0$
\includegraphics[width=\textwidth]{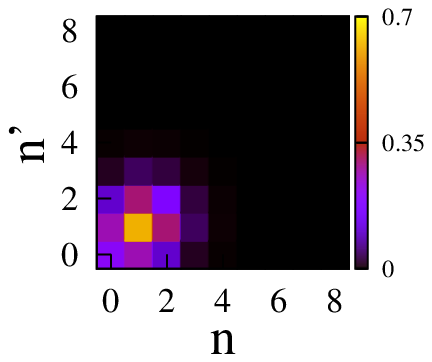}
\end{minipage}
\hspace{0.01\textwidth}
\begin{minipage}[t]{0.2\textwidth}
\centering
$\gamma_1 T=20$, $\gamma_2=0$
\includegraphics[width=\textwidth]{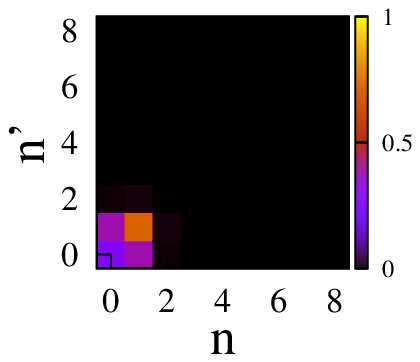}
\end{minipage}


\begin{minipage}[t]{0.2\textwidth}
\centering
 $\gamma_1=0$, $\gamma_2 T=0.05$
\includegraphics[width=\textwidth]{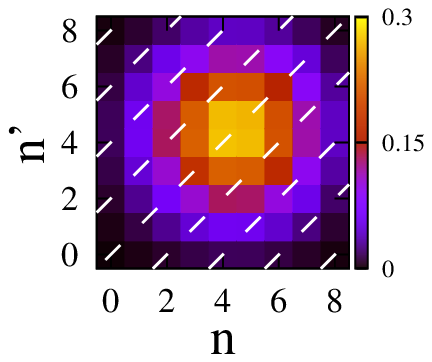}
\end{minipage}
\hspace{0.01\textwidth}
\begin{minipage}[t]{0.2\textwidth}
\centering
$\gamma_1=0$, $\gamma_2 T=1$
\includegraphics[width=\textwidth]{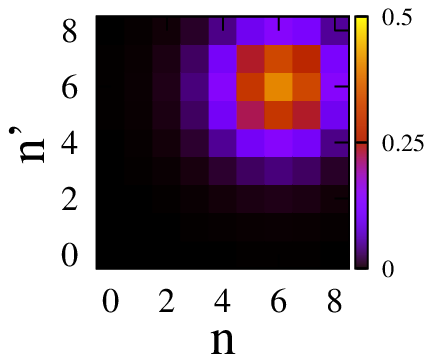}
\end{minipage}
\hspace{0.01\textwidth}
\begin{minipage}[t]{0.2\textwidth}
\centering
$\gamma_1=0$, $\gamma_2 T=4.0$
\includegraphics[width=\textwidth]{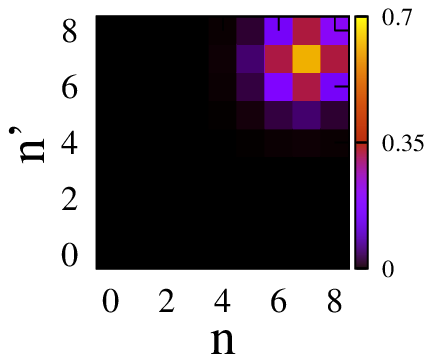}
\end{minipage}
\hspace{0.01\textwidth}
\begin{minipage}[t]{0.2\textwidth}
\centering
 $\gamma_1=0$, $\gamma_2 T=20$
\includegraphics[width=\textwidth]{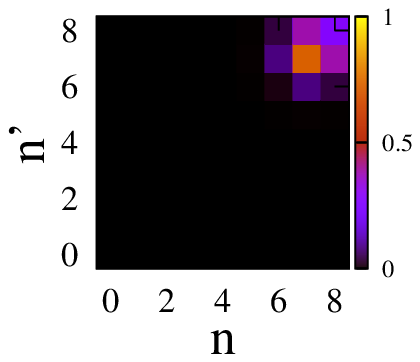}
\end{minipage}

\caption{(Color online) Moduli $| \bra{n,N_1-n} \hat{\rho}_{N_1} (t_2) \ket{n',N_1 -n'}|$ of the density matrix  in the 
Fock basis at time $t_2=T/4$ in the subspace with $N_1=N_0-2$ atoms
for increasing two-body losses rates (from left to right), from exact diagonalization.
The upper panels correspond to 
symmetric losses  and  energies ($\gamma_1=\gamma_2$ and $U_1 = U_2$), the middle panels to asymmetric 
losses and  
symmetric energies  ($\gamma_2=0$ and $U_1=U_2$), and the bottom panels to asymmetric losses 
and energies ($\gamma_1=0$ and $U_2=U_{12}$).
The revival time $T$ is the same in all cases. White dashed lines are marking the 
values of $(n,n')$ for which the matrix elements of the 
diagonal part   
$[\hat{\rho}^{(0)}(t_2)]_{\rm d}$ of the two-component superposition do not vanish.
Other parameters as in Fig.~\ref{fig4}. 
\label{fig6}}
\end{figure*}

\subsubsection{Small loss regime}

For small loss rates satisfying
\begin{equation} \label{eq-small_rate_condition}
\Gamma_m \ll  N_0^{1-r} t^{-1} \quad , \quad r = |m|=1,2,3 \;,
\end{equation}
the general results of Appendix~\ref{eq-app-D} take a simpler form which we discuss here.
For concreteness we restrict our attention to the times $t=t_q$. 
Let us first observe that  one can neglect the $\theta$-noise. In fact, 
$\delta \theta_m$ is much smaller than the quantum fluctuations  
in the CSs forming the components of the superposition 
(\ref{eq-q-component_cat_state}) (the latter are of the order of $1/\sqrt{N_0}$).
In contrast, due to the large fluctuations 
$\delta s_m \approx t_q$ of the loss time -- which has an almost flat distribution between $0$ and $t_q$, 
the $\phi$-fluctuations  are quite large. Indeed, if $\chi_1 m_1 + \chi_2 m_2 \approx \chi$ then one finds from Eq.(\ref{eq-phase_fluctuations}) that
$\delta \phi_m$  is of the order of the   phase separation $\phi_{k+1,q}-\phi_{k,q}=2\pi/q$ between the CSs.

According to Eqs.(\ref{eq-state_no_jump})  and (\ref{eq-density_matrix_N_0-2_sector_in_Fock_basis_J>1}),  for a single $r$-body loss process 
the conditional state $\hat{\rho}_{N_1}^{\rm{(1-jump)}} (t_q)$ with $N_1=N_0-r$ atoms  is obtained in the Fock basis
 by multiplying the matrix elements of the superposition $\rho^{(0)}_{N_1} ( t_q)$ in the absence of losses by the
damping factor 
\begin{equation} \label{eq-damping_factor}
D_q (n_1,n_1') = \exp \Bigl\{ - \frac{\pi}{\chi q} \bigl( d_{N_1} (n_1) + d_{N_1} (n_1') \bigr) \Bigr\}
\end{equation}
and by the envelope (see Eq.(\ref{eq-envelope_1_jump})  in  Appendix~\ref{eq-app-D})
\begin{equation} \label{eq-enveloppe_t_q}
{\cal{E}}_{q,r} (n_1,n_1') 
= \frac{q \chi}{\pi} \sum_{|m|=r} \Gamma_m C_m ( t_q; n_1,n_1')\;.
\end{equation}
In the last formula,  $C_m ( t_q ;n,n')$ is given by Eq.(\ref{eq-C_m(t)}) and the factor in front of the sum is put for convenience (then
 ${\cal{E}}_{q,r}(n_1,n_1)= \sum_{|m|=r} \Gamma_m$) and disappears in the state normalization.
In the limit 
(\ref{eq-small_rate_condition}), $C_m ( t_q ;n,n')$  can be approximated 
for symmetric energies $U_1=U_2$ (\ie  $\chi_1=-\chi_2=\chi$) by 
\begin{equation} \label{eq-approx_C_m} 
C_m (t_q; n ,n') 
 \simeq   
  \frac{1 - \exp \bigl\{ - \I \frac{\pi}{q} (m_1-m_2)  (n-n') \bigr\} }{\I \chi (m_1-m_2) ( n- n')}\,.
\end{equation}
%

\subsection{Channel effects and protection of quantum correlations against phase noise}
\label{sec-reduced_phase_noise}

In this subsection we study the conditional density matrix $\hat{\rho}_{N_1} (t_q)$ with $N_1=N_0-r$ atoms
for a BJJ subject to a single $r$-body loss event (with $r=1,2$, or $3$), focusing on the times of formation $t=t_q$ of the macroscopic superpositions. 
The more complex case of combined loss processes and
several loss events will be discussed in the next subsection.
We first single out the peculiar behavior
of the Fisher information $F_{N_1}(t_q)$ in the subspace with  $N_1$ atoms as one varies   the loss rates and interaction energies, by
relying on the exact diagonalization method. This behavior is  
then interpreted in the light of the analytical results of Sec.~\ref{sec-subspace_N_0-2_atoms}.
We identify several physical  effects explaining  the different decoherence 
scenarios discussed in Sec.~\ref{quantum_correlations}
for symmetric and asymmetric loss rates and energies.

\subsubsection{Density matrix and quantum Fisher information in the subspace with $N_0-2$ atoms} \label{sec-result_exact_diag}

\begin{figure}
\begin{minipage}[t]{8.6cm}
a) symmetric loss rates ($\gamma_1=\gamma_2$)

\includegraphics[width=\textwidth]{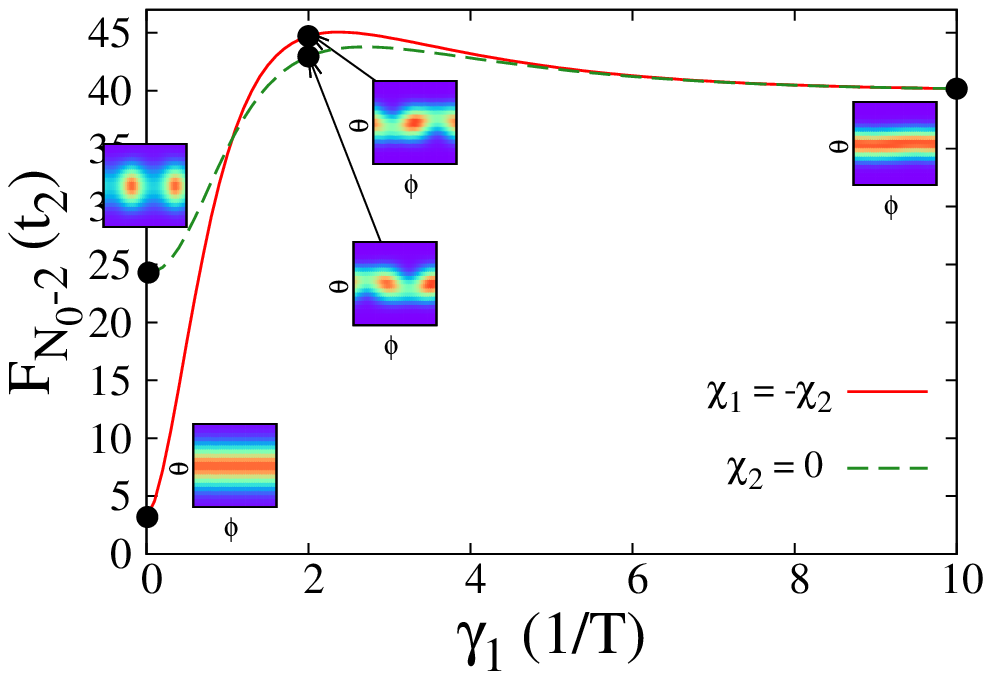} 
\end{minipage}
\vspace{2mm}

\begin{minipage}[b]{8.6cm}
b) asymmetric loss rates ($\gamma_2=0$)

\includegraphics[width=\textwidth]{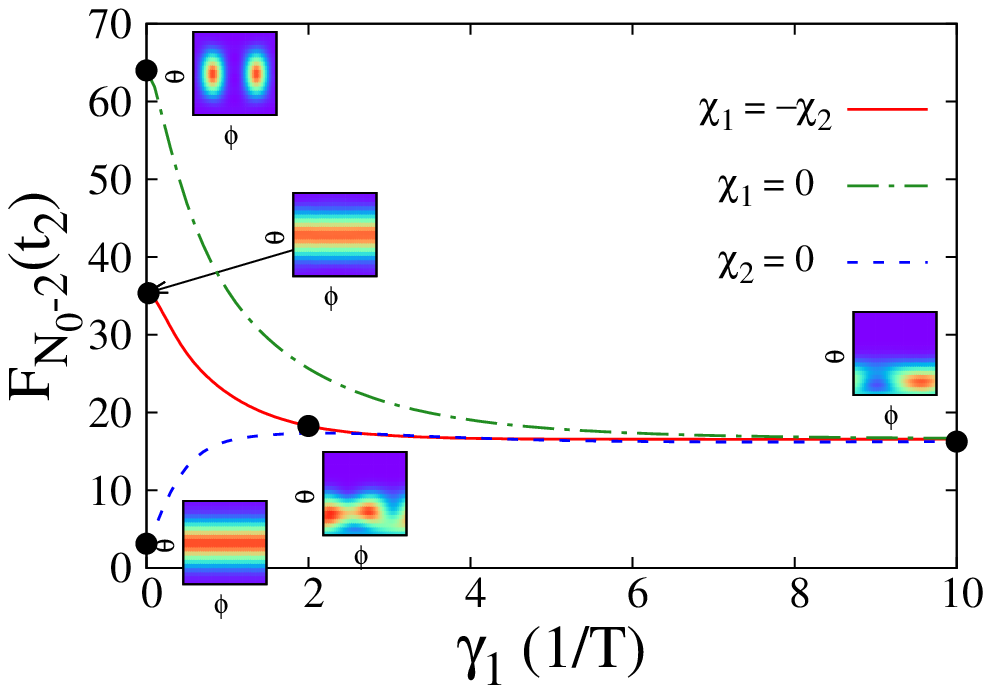} 
\end{minipage}

\caption{(Color online) Quantum  Fisher information optimized in the subspace with
$(N_0-2)$ atoms at time $t_2$
as a function of the two-body loss rate (in units of $T^{-1}$) for 
(a) symmetric losses $\gamma_1=\gamma_2$ with $U_1=U_2$ (red solid line)
 and $U_2=U_{12}$ (green dashed line);
(b) completely asymmetric losses with $\gamma_2=0$ and
 $U_1=U_2$ (red solid line), $U_1 = U_{12}$
(green  dot dashed line), and $U_2=U_{12}$ (blue  dashed line).
The $U_i$ are chosen in such a way that the revival time $T$ does not change.
Insets: plots of the Husimi functions for 
some specific choices of loss rates (indicated by circles and arrows).
Other parameters as in Fig.~\ref{fig4}.   All results are obtained from the exact diagonalization method. 
}
\label{fig7}
\end{figure}

Let us start by presenting the amount of QCs in the subspace with $N_1$ atoms calculated from the exact
diagonalization method. We restrict ourselves here to two-body losses, assuming no
one-body, three-body, and inter-mode losses (\ie $\alpha_i = \gamma_{12}=\kappa_i = \kappa_{ij} = 0$).
The density matrix $\hat{\rho}_{N_1}(t_2)$  in the Fock basis  is shown  in Fig.~\ref{fig6}.
If the interaction energies in the two modes are equal, \ie $U_1=U_2$,
we observe that $\hat{\rho}_{N_1}(t_2)$ is almost diagonal in the Fock basis for weak symmetric loss rates 
$\gamma_1=\gamma_2 \lesssim \chi /N_0$ (upper left panel), while
it has non-vanishing off-diagonal elements for odd values of $n_1'-n_1$   
for completely asymmetric rates $\gamma_{2}=0$ (middle left panel).
Moreover, if one  takes 
$\gamma_1=0$ in the first mode and tunes the energies such that $U_2=U_{12}$, keeping $\chi=(U_1+U_2-2 U_{12})/2$ fixed, 
the density matrix
has the same structure as that of a two-component superposition with $N_1= N_0-2$ atoms (lower left panel). 
This is confirmed by looking at the  Fisher information $F_{N_1}(t_2)$, which is
displayed in Fig.~\ref{fig7}. We stress that, unlike in Fig.~\ref{fig2},
 this Fisher information is not multiplied by the one-jump probability $w_{N_1}(t_q)$ and the optimization over the interferometer direction $\vec{n}$ 
is done in the $N_1$-atom sector, independently of the other sectors.
If one of the modes does not lose atoms and $U_i=U_{12}$ in the other mode (Fig.~\ref{fig2}b), 
$F_{N_1}(t_2)$ is approximately equal for $\gamma_1 \ll \chi/N_0$ 
to the Fisher information $N_1^2$ of a two-component superposition.  At stronger loss rates $\gamma_1 \approx 1/T =\chi/(2\pi)$, $F_{N_1}(t_2)$
decreases  to much lower values.
In contrast, for symmetric losses and energies, $F_{N_1}(t_2)$ starts  below the shot-noise limit 
at weak losses and increases with $\gamma_1$ to reach a maximum when $\gamma_1 \simeq 2/T$.
As we will see below, these  different behaviors of the Fisher information for symmetric and asymmetric loss rates and energies
occur in all subspaces with $N<N_0$ atoms and for all types of losses, thereby explaining the differences in 
the total Fisher information presented in Sec.~\ref{quantum_correlations}.

We now turn to the quantum jump approach.
By analyzing the form of 
the envelope (\ref{eq-enveloppe_t_q}) in the small loss regime, we argue below
that the aforementioned behavior of $F_{N_1} ( t_q)$ comes mainly from the combination of two effects: 
 a channel effect for $U_1=U_2$ and $q=2,3$, and the suppression of phase noise in the $i$th loss 
channel when $U_i = U_{12}$. Before discussing these two effects, we show that for $U_1=U_2$ the phase noise 
always induces  a complete phase relaxation  in the weak loss regime. However, we emphasize that this phase relaxation is not relevant for the Fisher information.    

\subsubsection{Complete phase relaxation for $U_1=U_2$} \label{sec-(1)}

Let us study the impact of the phase noise on $\hat{\rho}_{N_1} (t_q)$ for
{\it symmetric interaction energies} $U_1=U_2$ (\ie $\chi_1=-\chi_2 = \chi$) and small loss rates satisfying  (\ref{eq-small_rate_condition}).
Then $\delta s_m \approx t_q$ and Eq.(\ref{eq-phase_fluctuations}) gives 
 $\delta \phi_{m} \approx |m_1-m_2| \pi/q$. This implies that phase noise due to losses with unequal numbers of atoms $m_i$ 
in each mode $i$ blurs out the phases of the CSs, whereas 
a loss of one atom in each mode does not modify the state. 

Before showing this explicitly, let us mention some results established in~\cite{ferrini2010a, ferrini2010} 
concerning the effect of phase noise on macroscopic superpositions in BJJs.
We recall that  one can decompose 
 the density matrix as a sum of its diagonal and off-diagonal parts defined in Eq.(\ref{eq-density_matrix_no_losses_1}). 
This can also be done in the presence of noise. 
Phase noise flattens the Husimi distribution of the superposition  
in the $\phi$ direction (phase relaxation), which manifests itself  
by the convergence for strong noise of the diagonal part
of the density matrix  to a 
statistical mixture of Fock states with completely undefined phases.
A second effect of phase noise  
is the loss of the coherences between the CSs of the superposition, 
leading  to a convergence of the off-diagonal part to zero at strong noises. 
This off-diagonal part, albeit it does not influence the Husimi distribution,
contains the QCs useful for interferometry.
It was pointed out in~\cite{ferrini2010a, ferrini2010} that for intermediate phase noise one may have almost complete
phase relaxation while some QCs remain (weak decoherence).
 
In our case, the action of phase noise on the diagonal 
and off-diagonal parts of $\hat{\rho}_{N_1} (t_q)$ can be evaluated exactly. 
From Eqs.(\ref{eq-diag_and_off_diag_rho}), (\ref{eq-state_no_jump}), and (\ref{eq-density_matrix_N_0-2_sector_in_Fock_basis_J>1}), the matrix elements of 
$[ \hat{\rho}_{N_1} (t_q)]_{\rm d}$ in the Fock basis vanish for $n_1'\not= n_1$ modulo $q$. 
We may thus restrict our attention to $n_1' = n_1+pq$ for integer $p$'s.
Due to Eqs.(\ref{eq-enveloppe_t_q}) and (\ref{eq-approx_C_m}), if $p \not=0$ then the envelope reads 
\begin{equation}
{\cal{E}}_{q,r}(n_1,n_1+pq ) \simeq 
\begin{cases} 
- \I \Delta \alpha  \frac{1 - (-1)^p}{\pi p}   & \text{for $r=1$}
\\
\gamma_{12} & \text{for $r=2$}
\\
-\I ( \Delta \kappa + 3 \Delta \kappa_{12}) \frac{1 - (-1)^p}{3 \pi p} &  \text{for $r=3$}
\end{cases}
\end{equation}
with $\Delta \alpha = \alpha_2 - \alpha_1$, $\Delta \kappa = \kappa_2 - \kappa_1$, and $\Delta \kappa_{12} = \kappa_{21}- \kappa_{12}$.
Therefore, for weak two-body losses and in the absence of 
inter-mode losses  the diagonal part is  equal  to a statistical mixture of Fock states, 
\begin{eqnarray} \label{eq-added_formula}
\nn
& & \bra{n_1,N_1-n_1} \,[ \hat{\rho}_{N_1} ( t_q) ]_{\rm d}  \ket{n_1',N_1-n_1'}
\\
& & \hspace*{2cm} 
 \propto  \; \delta_{n_1,n_1'}
   \left( \begin{array}{c} N_1 \\ n_1 \end{array} \right) 
\E^{-\frac{2 \pi}{\chi q} d_{N_1} (n_1)} \;,
\end{eqnarray} 
where we have used Eqs.(\ref{eq-density_matrix_no_losses}), (\ref{eq-damping_factor}), and (\ref{eq-enveloppe_t_q}). 
This is confirmed  in the upper and middle left panels in Fig.~\ref{fig6},
where one observes vanishing matrix elements 
along the dashed lines $n_1'=n_1\pm 2$, $n_1'=n_1\pm 4, \ldots$.
We thus find that  the loss of two atoms in the same mode leads to complete phase relaxation. This  explains    
the $\phi$-independent profile of the Husimi distributions 
in the insets in Fig.~\ref{fig7} corresponding to $\chi_1 = - \chi_2$ and $\gamma_1 \ll 1/(T N_0)$.
In contrast, no phase relaxation occurs in the inter-mode channel $m=(1,1)$.

For one- and three-body  losses, 
complete phase relaxation occurs for symmetric losses ($\alpha_1=\alpha_2$,
$\kappa_1=\kappa_2$, and $\kappa_{12}=\kappa_{21}$) only.
This can be understood intuitively as follows. 
For weak losses the random phase $\phi_{1,0}=s \chi$ ($\phi_{0,1}= - s \chi$) produced  by
 the loss of one atom in the mode $i=1$ ($i=2$) is uniformly distributed 
 in $[0, \pi/q]$ ($[-\pi/q,0]$). Since the 
 components of the superposition have a phase separation of $2\pi/q$, it is clear that
  one needs equal loss probabilities in the two modes to wash out its phase content completely.
Note that here complete phase relaxation comes from an exact cancellation  when adding  the contributions 
of the two loss channels $m=(1,0)$ and $m=(0,1)$, which separately lead to non-diagonal
matrices $[ \hat{\rho}_{N_1} (t_q)]_{\rm d}$. A similar argument applies to three-body losses.

\subsubsection{Loss of quantum correlations when $U_1=U_2$ and  $q=2$ or $3$: channels effects} \label{sec-(2)}

As discussed above, the phenomenon of phase relaxation does not tell us anything about the  QCs useful for interferometry,
which can still be present in the atomic state even if one has complete phase relaxation.
Let us now study these QCs, contained in  the off-diagonal part $[ \hat{\rho}_{N_1} (t_q)]_{\rm od}$ of the conditional state.
We still assume symmetric energies $U_1=U_2$ and small losses satisfying  (\ref{eq-small_rate_condition}).
The off-diagonal part corresponds to the matrix elements of $\hat{\rho}_{N_1} (t_q)$ in the Fock basis
such that $n_1' \not= n_1$ modulo $q$ (see Eq.(\ref{eq-diag_and_off_diag_rho})).
In view of Eq.(\ref{eq-approx_C_m}), the main effect of phase noise is to multiply 
the matrix elements in the absence of noise by a factor of $(n_1-n_1')^{-1}$. This factor decays to zero as one moves away
from the diagonal but does not modify substantially the elements close to the diagonal.  
This explains the presence of off-diagonal matrix elements
for $n_1'=n_1 \pm 1$ and  $n_1'=n_1\pm 3$  in  
Fig.~\ref{fig6} when $\gamma_1 \ll \chi/N_0$, $\gamma_2=0$, and $U_1=U_2$ (left middle panel), 
as well as the relatively high value of the Fisher information $F_{N_0-2}(t_2)$  in Fig.~\ref{fig7}(b). 
For such loss rates and energies we are in the noise regime of the aforementioned weak decoherence, \ie phase noise is more efficient
 in washing out the phase content 
of each component of the superposition than in destroying the coherences. 

\begin{figure}
\includegraphics[width=8.6cm]{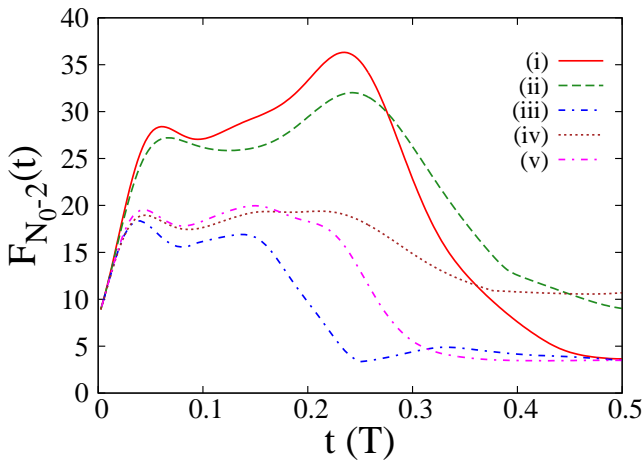} 
\caption{(Color online) Quantum Fisher information optimized in the subspace with $N_0-2$ atoms versus time for symmetric energies
$U_1=U_2$, from exact diagonalization. 
The red solid  and green dashed  lines (upper curves) correspond to
asymmetric two-body losses with (i) $\gamma_1 T=0.05$,   $\gamma_2=0$ and
(ii) $\gamma_1 T= 1.0$, $\gamma_2=0$. The
 blue dot-dashed, brown dotted and magenta dot-dashed lines 
(lower curves) correspond to symmetric two- and one-body 
losses with (iii)~$\gamma_1 = \gamma_2 =0.025/T$, $\alpha_i=0$, (iv)~$\gamma_1 =
\gamma_2 = 0.5/T$, $\alpha_i=0$, and (v) $\gamma_1 = \gamma_2  = 0.025/T$, 
$\alpha_1 =\alpha_2 = 0.5/T$.
Other parameters: $N_0=10$ and $\gamma_{12}=\kappa_i = \kappa_{ij} = 0$.}
\label{fig8}
\end{figure}

However, we see in Figs.~\ref{fig6} and \ref{fig7} that the situation is quite different for 
symmetric two-body losses $\gamma_1=\gamma_2$  and $\gamma_{12}=0$: 
then $[\hat{\rho}_{N_1} (t_2)]_{\rm od}$ 
vanishes completely and the Fisher information at weak losses is
smaller than $N_0$. This comes from a cancellation when adding the contributions of
the $m=(2,0)$ and $m=(0,2)$ loss channels, which 
occurs only at time $t_2$  and in the absence of  inter-mode losses.  
A similar cancellation occurs at time $t_3$ when the two modes are subject to three-body  
losses with symmetric rates $\kappa_1=\kappa_2$ and $\kappa_{12}=\kappa_{21}=0$.
In fact,  for such loss rates
Eqs.(\ref{eq-enveloppe_t_q}) and (\ref{eq-approx_C_m}) yield
${\cal{E}}_{r,r} ( n,n')  \simeq  2 \Gamma_{r,0} \delta_{n,n'}$ for $r=2,3$.
Hence $[\hat{\rho}_{N_1} (t_2)]_{\rm od}=0$, so that the whole density matrix
$\hat{\rho}_{N_1} (t_2)$ is diagonal in the Fock basis and given by Eq.(\ref{eq-added_formula}). This is clearly seen in 
the upper left panel in Fig.~\ref{fig6}. As a consequence of this channel effect, the two-component (three-component) superposition 
suffers   in the absence of inter-mode losses from a  {\it complete decoherence} in the $N_1$-atom subspace,
for arbitrary small symmetric two-body (three-body) loss rates. Note that this is not in contradiction with the fact that  
the total state $\hat{\rho}(t_2)$ converges  to 
$\hat{\rho}^{(0)} (t_2)$ when $\Gamma_m \rightarrow 0$, since the probability $w_{N_1} (t_2)$ converges to zero  
 in this limit and thus $\hat{\rho}_{N_1} (t_2)$ does not contribute to the total state. 
Such a channel effect does of course not occur for completely asymmetric losses involving only
one channel. It is illustrated in 
Fig.~\ref{fig8}, which displays the Fisher information $F_{N_1}(t)$
as a function of time. For asymmetric losses, $F_{N_1}(t)$ is 
maximum at time $t_2$ as in the lossless case. 
For symmetric two-body losses, instead, $F_{N_1}(t)$
is minimum at  $t_2$ due to the channel effect. 

We emphasize  that symmetric one-body losses $\alpha_1=\alpha_2$ and 
inter-mode three-body losses $\kappa_{12}=\kappa_{21}$
do not produce any channel effect. This means that 
these loss processes are less detrimental to the macroscopic superpositions than symmetric two-body losses. For instance, one has
${\cal{E}}_{q,1} ( n,n')= 2 \alpha_1 \sinc [ \pi (n-n')/q]$  for $\alpha_1=\alpha_2$
(we remind that we are treating for the moment the case  of symmetric interactions $U_1 = U_2$).
A striking consequence of this observation  
will be discussed in Sec.~\ref{sec-reduction_of_channel_effect} below.

\subsubsection{Protecting macroscopic superpositions by tuning the interaction energies}
\label{sec-(3)}

Let us now proceed to the case of {\it asymmetric interaction energies $U_1 \not= U_2$}.
In order to keep the  formation time $t_q= \pi/(\chi q)$ of the  superposition constant, we 
vary $U_{1}$ and $U_{2}$ while fixing  $2 \chi = \chi_{1}-\chi_{2}$. 
We still consider weak losses satisfying (\ref{eq-small_rate_condition}).
Then the phase relaxation described above is incomplete, as well as 
decoherence at times $t_2$  or $t_3$ for symmetric losses.
An interesting situation is $U_2=U_{12} < U_1$, \ie
$\chi_2=0$ and $\chi_1=2 \chi$. Then $\phi_{0,r}(s) =0$ by Eq.(\ref{eq-random_phases_J>1}), thus
the second mode is protected against phase noise,
whereas the first mode is subject to a strong noise with
fluctuations $\delta \phi_{r,0} \approx 2\pi r/q$. 
Taking for simplicity vanishing inter-mode rates, one gets from Eq.(\ref{eq-enveloppe_t_q}) and from Eq.(\ref{eq-C_m(t)}) in Appendix~\ref{eq-app-D}
\begin{equation} \label{eq-enveloppe_chi_2=0}
{\cal{E}}_{q,r} ( n,n') 
 =
  \Gamma_{0,r} + q \Gamma_{r,0} \frac{1- \exp \{  - \I \frac{2\pi r}{q} ( n-n') \}}{2\I \pi  r (n-n')}     \;.
\end{equation}
For symmetric rates $\Gamma_{0,r}=\Gamma_{r,0}$, the off-diagonal matrix elements of $\hat{\rho}_{N_1} (t_2)$ in the Fock basis coincide 
with those
of a two-component superposition up to a factor of the order of $1/2$. 
Loosely speaking, $\hat{\rho}_{N_1} (t_2)$ is a ``half macroscopic superposition''.
Such a state has a large Fisher information, as
shown in Fig.~\ref{fig7}(a). 
An even larger Fisher information is obtained for completely asymmetric losses with $\Gamma_{r,0}=0$, \ie if
atoms are lost in the protected mode $i=2$  only.
Then ${\cal{E}}_{q,r} ( n,n') = \Gamma_{0,r}$ and $\hat{\rho}_{N_1} (t_q) \propto  \widetilde{\rho}_{N_1}^{\rm{(no\,loss)}} (t_q)$, 
that is, the conditional state $\hat{\rho}_{N_1}(t_q)$  coincides with a superposition of $q$ CSs with $N_1$ atoms,
slightly modified by the damping factor (\ref{eq-damping_factor}).
This is in agreement with the convergence at weak losses and asymmetric energies  of
 the Fisher information $F_{N_0-2} (t_2)$ in Fig~\ref{fig7}(b)
to the highest possible value $(N_0-2)^2$, and to the presence of two well-pronounced 
peaks at $\phi=\pm \pi/2$ in the corresponding Husimi distributions.

In summary, by tuning the interaction energies $U_i$ such that
$\chi_1=0$ or $\chi_2=0$ 
one can protect one mode against phase noise, to the expense of enlarging
noise in the other mode, thereby limiting  decoherence effects 
on the conditional state with $N_1$ atoms.
This way of switching  phase noise off in one mode 
has been pointed out in~\cite{pawlowski2013} for two-body losses.   As a central result we find here that it applies to one- and three-body losses as well.
One can similarly switch the phase noise off in the two-body inter-mode loss channel $m=(1,1)$ by tuning the energies such that $\chi_2=-\chi_1$ 
(\ie by taking symmetric energies $U_1=U_2$) and  in the 
three-body inter-mode loss channels $m=(2,1)$ and $(1,2)$  by taking  $\chi_2 = - 2\chi_1$ and $\chi_1 = - 2 \chi_2$
 (\ie $U_1 = U_2 \mp 2 \chi/3$), respectively.  
We emphasize that it is impossible to suppress the noise in two different loss channels at the same time.
Therefore, the optimal energy tuning for protecting the macroscopic superpositions is to switch phase noise off in the  channel losing more atoms.

To  complete the description of Fig. \ref{fig7} we discuss in what follows three effects of atom losses occurring at
intermediate and strong loss rates.

\subsubsection{Increasing the loss rates reduces phase noise} \label{sec-(4)}

We first study the regime of intermediate loss rates. Surprisingly,
 the $\phi$-noise {\it decreases} if one  increases $\Gamma_m$. This results from the decrease of 
the loss time fluctuations $\delta s_{m}$ at increasing $\Gamma_m$,
leading to a decrease of the phase fluctuations $\delta \phi_{m}$ in Eq.(\ref{eq-phase_fluctuations}). 
In fact, while for  small rates the loss time is
uniformly distributed on the interval $[0, t_q ]$,
for larger rates the loss has more chance to occur at small times and 
$\delta s_m$ gets smaller.
For instance, it is easy to see by inspection of the
distribution (\ref{eq-distribution_jump_timebis}) in Appendix~\ref{eq-app-D}
that 
$\delta s_{m} \simeq G_m^{-1}$ when $G_m \gtrsim \chi q$ and  $N_0\gg 1$, $G_m$ being a non-decreasing 
function of the rates $\Gamma_m$ given by Eq.(\ref{eq-G_m}). 
This decreasing of the phase noise sets in for $G_m \approx \chi q $, that is, $\Gamma_m \approx \chi  q N_0^{1-|m|}$.
Therefore, by increasing the loss rates one protects the conditional state  with
$N_1$ atoms against phase noise and thus against decoherence. 
As seen in Fig.~\ref{fig2}, this  counter-intuitive effect does not manifest itself in the total Fisher information
$F_{\rm tot} (t)$. 
Indeed, when increasing $\gamma_1$ the subspaces contributing to $F_{\rm tot} (t)$ in Eq.(\ref{eq-total_Fisher})
have less atoms and hence are less quantum correlated, and the increase of $F_{N_0-2} (t)$ is counter-balanced by the decrease of the
probability $w_{N_0-2}(t)$.  
As a consequence, $F_{\rm tot} (t)$ is getting smaller.

\subsubsection{Effect of the $\theta$-noise} \label{sec-(5)}

The fact that the peaks of the Husimi distributions in  Fig.~\ref{fig7}(b) 
  at intermediate losses are centered at values of $\theta$ smaller than $\pi/2$  
is due to the $\theta$-noise. In fact, in this figure $\delta_1=2 \gamma_1 >0$ and $\delta_2 = 0$, so that
$\theta_{2,0} (s) < \pi/2$ and $\theta_{0,2} (s) = \pi/2$. For larger initial atom numbers $N_0$
the $\theta$-noise is always small, its fluctuations being of the order of $1/N_0$ when  $N_0\gg 1$.
Actually, when $G_m \gtrsim \chi q$ one finds  $\delta \theta_m \approx 1/N_0$ by replacing
$\delta s_m$ by $G_m^{-1}$ in Eq.(\ref{eq-phase_fluctuations}), whereas for $G_m < \chi q$
one has $\delta \theta_m < t_q r \max\{| \delta_1|, |\delta_2| \}/2$.

\subsubsection{Damping effects} \label{sec-(6)}
 
 Increasing further the  rates $\Gamma_m$, the  damping
due to the effective Hamiltonian begins to play the major role.
The combination of this damping with the reduced phase noise effect described above 
leads again to different behaviors of $F_{N_1}(t_q)$ as a function of the loss rates
for symmetric and asymmetric losses. 
Let us recall from  Sec.~\ref{eq-subspace_N_0_atoms} that the onset of  damping 
is for $\Gamma_m \approx \chi q N_0^{1-|m|}$ in the symmetric case 
and $\Gamma_m  \approx \chi q N_0^{-|m|}$ in the asymmetric case.   
On the other hand, we have seen above that phase noise reduction begins when 
$\Gamma_m \approx \chi q N_0^{1-|m|}$.
For symmetric losses, there  exists a small range  of loss rates $\Gamma_m$ on which phase noise is reduced
by increasing $\Gamma_m$ while the damping is still relatively small. 
This explains the increase of $F_{N_1}(t_2)$  with $\gamma_1$ seen in Fig. \ref{fig7}(a).
At the point where $F_{N_1}(t_2)$  
reaches a maximum,  two peaks are clearly visible in the Husimi distribution, as opposed to 
the flat distribution observed at weak losses. This nicely illustrates phase noise reduction.
In contrast, in the asymmetric case damping effects counter-balance 
phase noise reduction and the Fisher 
information decreases  when increasing $\gamma_1$ (even though some peaks  show up in the Husimi plots).
For  $\gamma_1=\gamma_2 \gg \chi$ and even initial atom numbers $N_0$,
$\hat{\rho}_{N_0-2}(t_q)$ converges  to
a Fock state with $(N_0-2)/2$ atoms in each mode, which has 
 a high Fisher information $(N_0-2)N_0/2$ (see Appendix~\ref{app-rho_N_0_strong_losses}).
For asymmetric losses, instead, $\hat{\rho}_{N_0-2}(t_q)$
 converges  to a superposition of Fock states with $n_1=0$ or $1$ atom in the first mode,
and $F_{N_0-2}(t_2) \approx N_0$, as seen in  Fig.~\ref{fig7}(b).
Similarly, the comparison of the two first rows in Fig.~\ref{fig6} shows that
an increase of $\gamma_{1}=\gamma_2$ 
makes non-vanishing off-diagonal matrix elements to appear, as a consequence of phase noise reduction, 
while for $\gamma_2=0$ the same operation moves the peak in the density matrix towards 
$n_1=n_1'=0$, as a consequence of damping.  

Let us stress again that these effects  on the conditional state with $N_1$ atoms   at strong losses do not affect the total density
matrix because of the small probability to have a single loss event between $t=0$ and $t=t_q$. 
Note also that  the approximations made in Sec.~\ref{sec-subspace_N_0-2_atoms} break down for strong losses, namely,
Eq.(\ref{eq-density_matrix_N_0-2_sector_in_Fock_basis_J>1}) is still valid but the envelope
${\cal E}_N$ has a more complex expression than that given in Eq.(\ref{eq-enveloppe_J>1})
(see Appendix~\ref{eq-app-D}). 
Since for $\Gamma_m \gg \chi N_0^{1-|m|}$ the most important effect is damping, the precise form of 
${\cal E}_N$ does, however, not matter.

\subsection{Conditional states for several loss events} \label{sec-many_losses}

We can now extend the previous results to the case of several loss events. 
In view of Eq.(\ref{eq-enveloppe_J>1}) the physical effects discussed above are  present in all subspaces with $N$ atoms, provided that
$0 < N_0 - N \ll N_0$. However,  some of our conclusions for a single loss event
must be modified  because of the combination of the different loss processes.

\subsubsection{Suppression of the channel effect due to one-body losses} \label{sec-reduction_of_channel_effect}

The channel effect leading to complete decoherence at times $t_2$ or $t_3$ for weak
symmetric  losses   is suppressed if, 
in addition to two- or three-body losses, also one-body losses are present.
In fact, the
density matrix in the subspace with $N_0-r$ atoms, $r=2$ or $3$, is given by 
Eq.(\ref{eq-density_matrix_N_0-2_sector_in_Fock_basis_J>1}) with
an envelope 
${\mathcal{E}}_{N_0-r}  (t_r;n,n') \propto {\mathcal{E}}_{r,r}  (n,n')
+ [ {\mathcal{E}}_{r,1}  (n,n')]^r/r!$, see Eq.(\ref{eq-enveloppe_J>1}).
As pointed out in  Sec.\ \ref{sec-(2)}, 
${\mathcal{E}}_{r,r}  (n,n')$ vanishes for $n \not= n'$ (channel effect), 
but this is not the case for the envelope ${\mathcal{E}}_{r,1}  (n,n')$ coming from
one-body losses. We thus find  that
 by adding one-body losses one can reduce decoherence  on the two-component or the three-component superpositions. We have checked that  off-diagonal elements
 indeed  appear in the density matrix $\hat{\rho}_{N_0-2}(t_2)$  
displayed in  the upper left panel in  Fig.~\ref{fig6}  when one adds one-body losses. 
A surprising consequence of these off-diagonal elements is shown in
Fig.~\ref{fig8}: if one-body losses are added to the two-body losses,  
the Fisher information $F_{N_0-2} (t_2)$ increases in spite of the larger amount of losses.
This does, however, not affect the total Fisher
information. Indeed, we have always  observed numerically a decrease of $F_{\rm tot} (t)$ when one-body losses
are added.

\subsubsection{Tuning the interaction energies}

For strongly asymmetric loss rates,  as in the case of a single loss event  it is possible 
to protect the QCs in the atomic state 
by tuning the interaction
energies while keeping fixed the  energy
$\chi$ governing the lossless dynamics. However, if the loss
rates are symmetric, decoherence effects are strong when  many loss events occur between times $0$ and $t_q$, whatever
the choice of the energies. The argument goes as follows.
If all losses occur mostly in the same mode $i$,
 one can switch phase noise off  in that mode $i$ by tuning the atomic 
energies in such a way 
that $\chi_i=0$ (see Sec.~\ref{sec-(3)}).
Then each conditional state $\hat{\rho}_N ( t_q)$, with  $N_0-N \ll N_0$, 
is close to a $q$-component superposition with $N$ atoms, apart from small damping effects.
This comes from the product form (\ref{eq-enveloppe_J>1}) of the envelope ${\cal{E}}_N$ 
in Eq.(\ref{eq-density_matrix_N_0-2_sector_in_Fock_basis_J>1}) and from our previous results on  
the  envelope ${\cal{E}}_{q,r} (n,n')$ for single loss events, which is almost constant  for weak loss rates $\Gamma_m \ll q \chi N_0^{1-|m|}$.
The same statement holds if losses occur mainly via inter-mode two-body processes 
(\ie $\gamma_{12} \gg \gamma_i , \alpha_i/N_0, \kappa_i N_0, \kappa_{ij} N_0$); then 
one must tune the energies such that $U_1=U_2$.
For symmetric losses the situation is different.
As soon as the number $N_0-N$ of lost atoms becomes large,
the tuning of the energies $U_i$ is inefficient to keep the 
coherences of the superposition, because the probability that all losses occur in the same mode 
decreases exponentially with the
number of loss events. Actually,  we have seen above that
only one mode can be protected against phase noise if $\chi$ is kept constant. 
Therefore, when a large number of atoms leave the BJJ, 
the loss rates must be strongly asymmetric 
in order to be able to  protect efficiently the superpositions from 
decoherence by tuning the $U_i$. 
These results provide a good explanation of the effects described in Figs.~\ref{fig2} and \ref{fig3}.

In order to fully explain the high values of the total Fisher information at time $t_q$ for strongly asymmetric losses found in Sec.~\ref{quantum_correlations},
we must also show that the interferometer direction $\vec{n}$ optimizing the Fisher information
$F (\hat{\rho}_N (t_q), \hat{J}_{\vec{n}})$   in the $N$-atom subspace is almost the same for all $N$
(otherwise one could not take advantage of the QCs to improve the phase precision of the interferometer
when the number of atoms at time $t_q$ is unknown). 
To see that this is indeed the case, let us note that the optimal directions roughly coincide with one of the phases $\phi_{k,q}$ of the
superposition (\ref{eq-q-component_cat_state}), which  are given by (we assume here $E_1=E_2$)
\begin{equation}
\phi_{k,q}=\Bigl[2 k+\epsilon_q -N+  \chi^{-1}(N-1) \frac{U_2-U_1}{2} \Bigr]\frac{\pi}{q}
\end{equation}
with $\epsilon_q=0$ if $q$ is even and $1$ otherwise. When $U_2=U_{12}$ (or
$U_1=U_{12}$) we obtain $\phi_{k,q}=[2 k+\epsilon_q -2 N+ 1]\pi/q$
(or $\phi_{k,q}=[2 k+\epsilon_q -1]\pi/q$). Thus
the components of the superpositions are transformed one into another by changing $N$ (for $q$ fixed) and the optimal directions are the same.
One deduces from this argument that
the total Fisher information $F_{\rm tot} (t_q)$ in Eq.(\ref{eq-total_Fisher}) is close to
the Fisher information  of a $q$-component superposition with $\langle \hat{N}\rangle_{t_q}$ atoms, and thus scale like $\langle N \rangle^2_{t_q}$,
where $\langle \hat{N} \rangle_{t_q}$ is the average number of
atoms at time $t_q$.

\subsection{Dependence of the Fisher information on the initial atom number $N_0$}

Let us briefly discuss the effect of an increase of the initial number of  atoms  $N_0$ on the QCs in the macroscopic superpositions, focusing on the
completely asymmetric loss  case $\alpha_1=\gamma_1=\kappa_1=0$ and no inter-mode losses. 
The increase of $N_0$ leads to a rapid increase of the 
probability for losing atoms. 
Indeed, for large $N_0$ the mean number $\langle N \rangle_{t_q}$ of atoms in the BJJ at time $t_q$ behaves like 
$N_0 ( \gamma_{2} N_0  t_q +1)^{-1}$
for two-body losses and $N_0 ( 2 \kappa_2 N_0^2  t_q+1)^{-1/2}$ for three-body losses, as follows from the phenomenological rate equations. 

We first 
consider the case of asymmetric interaction energies $U_2=U_{12}< U_1$. 
If $N_0$ is small enough so that the BJJ remains in the weak loss regime $\alpha_2 \ll \chi q$, $\gamma_{2} N_0 \ll \chi q$, and $\kappa_{2}  N_0^2 \ll \chi q$,  
it has been argued above that
the total Fisher information $F_{\rm tot}(t_q)$ should scale like  $\langle N \rangle^2_{t_q} \approx N_0^2$.  
However, when $\gamma_2 N_0^2$ and $\kappa_2 N_0^3$ become of the order of $\chi q$ one expects a less pronounced increase of $F_{\rm tot}(t_q)$ with $N_0$ 
because of the damping effects discussed in Secs.~\ref{eq-subspace_N_0_atoms} and~\ref{sec-(6)},
 as confirmed  by a comparison of Fig. \ref{fig2}(a) and Fig.~\ref{fig3}. 
Increasing further $N_0$, one enters into the intermediate loss rate regime with $\gamma_{2} N_0 \approx \chi q$ and $\kappa_2 N^2_0 \approx \chi q$,
characterized by a non negligible fraction of lost atoms, strong damping, and much stronger decoherence effects. 

Let us now take symmetric energies $U_1=U_2$. 
In the weak loss regime the envelope in Eq.(\ref{eq-density_matrix_N_0-2_sector_in_Fock_basis_J>1})
decays like $(n_1 - n_1')^{-J_1 - J_2 - J_3}$ as one moves away from the diagonal $n_1=n_1'$, 
see (\ref{eq-enveloppe_J>1}),  (\ref{eq-enveloppe_t_q}),  and (\ref{eq-approx_C_m}).
Thus, as the number $N_0-N=J_1+ 2 J_2 + J_3 $ of lost atoms  is getting larger the conditional states  $\hat{\rho}_N (t_q)$
become more diagonal in the Fock basis, in contrast with  what happens for asymmetric energies.
By increasing $N_0$, the mean number of lost atoms increases and 
the total atomic state $\hat{\rho} ( t_q)$ gets closer to  a statistical mixture of Fock states,  leading to a disappearance of  
the QCs in the macroscopic superposition. 
It should be noted, however, that  for fixed numbers of jumps $J_1$, $J_2$, and $J_3$, the decay  of the off-diagonal elements 
of the conditional states
$\hat{\rho}_{N} ( t_q)$  in the Fock basis is the same
for all $N_0$. This is related to the fact that in a BJJ subject to phase noise
the decoherence time is independent of the number of atoms~\cite{ferrini2010a}. 
Hence some QCs remain in the conditional states $\hat{\rho}_N (t_q)$ even for large $N_0$. The degradation of the QCs in the total state results from
the decay of the probabilities   $w_N (t_q)$ to have $N$ atoms  at time $t_q$. 
The behavior of the Fisher information $F_{\rm tot}(t_q)$  as a function of  $N_0$ strongly depends on the behavior 
of these probabilities.

\section{Summary and concluding remarks} \label{conclusions}

We have studied in detail the decoherence induced by  one-, two- and
three-body atom losses on the superpositions of coherent states dynamically
generated in BJJs.  For all loss types and at weak losses, 
  the degradation of the superposition is mainly due to 
a strong effective phase noise  
and to a  channel effect. The last effect gives rise to enhanced decoherence on the two-component (three-component) superposition
after summing over  the two loss channels when the two-body (three-body) loss rates and interaction energies are the same
in  the two modes   and there are no  inter-mode losses. Conversely, if all losses occur mostly in one mode, 
we have shown that it is possible to partially  prevent this degradation by adjusting the
interaction energy $U_i$ of each mode, keeping  their sum  fixed and  
exploiting the experimental
tunability of $U_i$. For instance, in the absence of inter-mode losses the effective phase noise
can be suppressed  in
the mode loosing more atoms by choosing an interaction
energy in this mode equal to the inter-mode interaction $U_{12}$.  For internal BJJs with Rubidium atoms as used in Ref.~\cite{gross2010}, this
could be done by  reducing the
scattering length $a_1$  in the  mode $i=1$ loosing less atoms.
Then, because $a_{2}$  and 
$a_{12}$ are almost equal, one has $U_2 \simeq U_{12}$, whereas $|U_1-U_{12} |$ can be large. 
For experimentally relevant loss rates
and initial atom  numbers, we have found that the amount of coherence left at the time of
formation of the  two-component superposition can be made
in this way  substantially  higher, provided that the system  has strongly
asymmetric losses (see Fig.~\ref{fig3}).  In  the experiment of
Ref.~\cite{gross2010}, this condition is met for two-body losses but not for one-body losses,
which are symmetric in the two modes. 
As a consequence, in the range of parameters corresponding to the experimental situation that we have studied,
 we predict that one-body loss processes lead to much stronger decoherence effects
on the macroscopic superposition than the asymmetric two-body   processes.

\acknowledgments

We are grateful to K. Rz\k a\.zewski, F.W.J. Hekking, P.~Treutlein, and M.K. Oberthaler for 
helpful discussions.
K.P. acknowledges support by the Polish Government Funds no. N202 174239 for the years
2010-2012 and financial support of the project  ``Decoherence in long range interacting
quantum systems and devices'' sponsored by the Baden-W\"urttemberg Stiftung. 
D.S acknowledges support from the ANR project no.\,ANR-09-BLAN-0098-01 and A.M. from the ERC ``Handy-Q'' grant
no. 258608. D.S. and A.M. acknowledge support from the ANR project no. ANR-13-JS01-0005-01

\appendix

\section{Solution of the master equation by the exact diagonalization method}
\label{app:exact_diagonalization}

We present in this appendix the exact solution of the master equation \eqref{eq-master_equation} 
with Lindblad generators  \eqref{eq-generator_2_body} in the  absence of inter-mode losses.
In the notation of Sec.~\ref{sec-no_loss} this
 means  $\kappa_{21} = \kappa_{12} = \gamma_{12} =0$.
The loss rates in the first and second modes are denoted  as in Sec.~\ref{eq-general_QJ} by 
$\Gamma_{r,0}$ and $\Gamma_{0,r}$, respectively,
with $r=1$, $2$, and $3$ for one-, two-, and three-body losses.

Let us first note that the inter-mode interaction energy $U_{12} \nopone \noptwo$ in the 
Bose-Hubbard Hamiltonian (\ref{eq-H_0})
can be absorbed in the intra-mode interactions  up to a term depending on the total number operator
$\hat{N}$ only, yielding
\begin{eqnarray}
\non
\hat{H}_0 & = & \hat{H}^{(1)} + \hat{H}^{(2)}  + \frac{U_{12}}{2} \hat{N} ( \hat{N}-1)
\\
\hat{H}^{(i)} & = & E_i \nopi + \frac{U_i-U_{12}}{2} \nopi ( \nopi -1)\;,\;i=1,2 \;.
\end{eqnarray}
As neither the initial density operator nor the dynamics 
couple subspaces with distinct total atom numbers, 
one can ignore the term depending on $\hat{N}$. Then $\hat{H}_0$ reduces to a sum of
two single-mode Hamiltonians $\hat{H}^{(1)}$ and $\hat{H}^{(2)}$. 
Since we assumed no inter-mode losses, the Lindblad generators can also be  expressed as sums of generators acting
on single modes.
Thus the two modes are not coupled in the master equation~\eqref{eq-master_equation} 
and the dynamics of the two-mode BEC can be deduced from that 
of two independent single-mode BECs, which are only coupled in the initial state 
$\hat{\rho} (0)=\ketbra{\psi(0)}{\psi(0)}$ given by Eq.~\eqref{eq-initial_state}.

Let us first focus on the single-mode master equation:
\begin{equation} \label{app:eq-master_equation}
		\frac{\D \hat{\rho}}{\D t}  =  - \I \bigl[ \hat{H} , \hat{\rho} (t) \bigr] 
   + \sum_{r=1}^3 {\Ll}_{\text{r-body}} (\hat{\rho} (t) ) 
\end{equation}
with $\hat{H} = \frac{U}{2} \bbr{\hat{a}^\dagger}^2\hat{a}^2$
and 
\begin{equation} \label{app:lindblad}
{\Ll}_{\text{r-body}}(\hat{\rho}) = \Gamma_r\, \hat{a}^r \hat{\rho}   \bbr{\hat{a}^\dagger}^r
   - \frac{\Gamma_r}{2} \bigl\{ \bbr{\hat{a}^\dagger}^r \hat{a}^r  , \hat{\rho}  \bigr\}\;,
\end{equation}
where  we 
denote the energies $U_1-U_{12}$ or $U_2-U_{12}$ collectively by $U$ and the loss rates  $\Gamma_{r,0}$ and $\Gamma_{0,r}$ 
collectively by $\Gamma_r$. 
Hereafter we use the notation
\begin{equation}
		\rho_{k,l+j}^{k+j,l}=\langle k,l+j|\hat{\rho}|k+j,l\rangle 
\end{equation}
for the matrix elements of the two-mode density operator in the Fock basis, and similarly
\begin{equation}
		\rho_{k}^{k+j}=\langle k|\hat{\rho}|k+j\rangle 
\end{equation} 
in the single-mode case. This unusual indexing will turn out to be convenient later. 
The master equation  \eqref{app:eq-master_equation} takes the following form in the Fock basis:
\begin{equation} \label{app:eq-DWful}
\frac{\D}{\D t} {\rho}_{k}^{k+j}  =  \lambda_{j,k}\, \rho_{k}^{k+j} (t) +  \sum_{r=1}^3 u_{j,k+r}^{(r)} \rho_{k+r}^{k+r+j} (t)\;,
\end{equation}
where we have set $\rho_l^{l+j} (t)=0$ for $l>N_0$ or $l+j > N_0$, 
\begin{eqnarray} 	\label{eq:diagonalcoeff}
	\lambda_{j,k} & = & -\frac{\I U}{2}\Bigl(k(k-1)-(k+j)(k+j-1)\Bigr) \\
	 &  & -\sum_{r=1}^3 \frac{\Gamma_{r}}{2} \bbr{   \prod_{w=0}^{r-1} ( k-w )+\prod_{w=0}^{r-1} (k+j-w ) } \nn \; ,
\end{eqnarray}
 and
\begin{equation}
		u_{j,k}^{(r)} = \Gamma_{r}\prod_{w=0}^{r} \sqrt{  (k-w) (k+j-w)}\,.
	\label{app:eq-r-coeff}
\end{equation}
From \eqref{app:eq-DWful} we conclude that  the master equation \eqref{app:eq-master_equation} couples only the matrix elements 
which are in the same distance $j$  from the diagonal. 
Therefore, the set of differential equations \eqref{app:eq-DWful} for all $j$ and $k$ can be grouped into families of 
equations with a fixed $j$, which can be solved independently from each other.

For a given $j$ and $N=j,\ldots, N_0$, we solve Eq.\eqref{app:eq-DWful}  with the initial condition
\begin{equation} \label{app:eqn-initial-condition}
\rho_{k}^{k+j} (0) =  \delta_{k,N-j} \quad , \quad k = 0,\ldots, N-j \,.
\end{equation}
Then $\rho_k^{k+j} (t)=0$ at all times $t$ when $k > N-j$.
It is convenient to   collect the matrix elements together into a vector having its
 $k$th component equal to $\rho_{k}^{k+j} (t)$,
\begin{equation}
		 \vv_{N,j} (t) = \left( \rho_{0}^{j} (t), \rho_{1}^{1+j} (t), \ldots, \rho_{N-j}^{N} (t)\right)\;.
\end{equation}
 Then the equations \eqref{app:eq-DWful} for different $k$ but fixed $j$ can be combined into a single 
equation for the vector $\vv_{N,j}$:
\begin{equation} \label{eq-diff_eq-for_vv}
 \frac{\D \vv_{N,j} }{\D t} = A_{N,j}  \vv_{N,j} (t) \, ,
\end{equation}
where $A_{N,j}$ is a time-independent $(N-j+1) \times (N-j+1)$ triangular superior matrix with coefficients determined by \eqref{app:eq-DWful}. 
The solution of Eq.\eqref{eq-diff_eq-for_vv} has the form
\begin{equation}\label{app:gen_sol}
\vv_{N,j} (t) = \exp\bbr{A_{N,j} t}\vv_{N,j} (0) .
\end{equation}
To determine the exponential in the right-hand side one has to  diagonalize $A_{N,j}$.

As this matrix is triangular, its eigenvalues are given by its diagonal elements
$\lambda_{j,n}  $, defined in Eq.\eqref{eq:diagonalcoeff}.
Let $\boldsymbol{l}_{j,n}$ and $\boldsymbol{p}_{j,n}$ be   the left and right  eigenvectors
of $A_{N, j}$ with eigenvalue $\lambda_{j,n}  $.
In the general case we have not been able to find explicit   expressions for these eigenvectors.
However,   by Eq.\eqref{app:eq-DWful} their components $p^k_{j,n} $ and $l^k_{j,n}$, $k=0, 1,\ldots, N-j$,
can be obtained recursively thanks to the formulas
\begin{equation} \label{app:eqn-eigen-right}
		p^k_{j,n} \! = \!\! \left\{
		  \begin{array}{lr}
		    0 & \quad \!\!\!\!\!\!\! \text{if $k>n$}\\
		    1 & \quad \!\!\!\!\!\!\!\!  \text{if $k=n$}\\
		\dss \frac{u_{j,k+1}^{(1)}p_{j,n}^{k+1}+u_{j,k+2}^{(2)}p_{j,n}^{k+2}+u_{j,k+3}^{(3)}p_{j,n}^{k+3} }{\lambda_{j,n}-\lambda_{j,k}} & 
\!\!\!\!\!\!\!\! \text{if $k<n$}
		  \end{array} \right.
\end{equation}
and 
\begin{equation}\label{app:eqn-eigen-left}
		l^k_{j,n} = \left\{
		  \begin{array}{lr}
		   \dss \frac{u_{j,k}^{(1)}l_{j,n}^{k-1}+u_{j,k}^{(2)}l_{j,n}^{k-2}+u_{j,k}^{(3)}l_{j,n}^{k-3}}{\lambda_{j,n}-\lambda_{j,k}} & \text{if $k>n$}\\
		   1 & \quad \text{if $k=n$}\\
		    0 & \quad \text{if $k<n$}
		  \end{array} \right. \; .
\end{equation}
Using these eigenvectors and eigenvalues, the solution  \eqref{app:gen_sol} with initial condition (\ref{app:eqn-initial-condition})
takes the form
\begin{equation}\label{app:eqn-sol-1mode}
 \vv_{N,j} ( t)  = \sum_{n=0}^{N-j}l_{j,n}^{N-j}\,\boldsymbol{p}_{j,n}e^{\lambda_{j,n} t} \; .
\end{equation}

Having the solution in a single mode, we can find the solution of the two-mode master equation \eqref{eq-master_equation}   in the main text, 
\begin{eqnarray} \label{app:eqn-sol-2mode}
& &\nonumber\rho^{k+j, l}_{k, l+j} (t) 
= \sum_{N=k+j}^{N_0-l} \rho_{N-j,N_0-N+j}^{N,N_0-N} (0) 
		\\
		 & & \hspace*{5mm} \times [\vv_{N,j}^{(1)} (t)]_{k} [ \vv_{N_0-N+j,j}^{(2)}( t) ]^{\ast}_{l} \, ,
		\label{app:eqn-solutionDW}
\end{eqnarray}
where we added superscripts    referring to the modes $1$ and $2$ to the vectors $\vv_{N,j} (t)$,
to stress that  the loss rates and interaction energies differ between modes.

The solutions \eqref{app:eqn-sol-1mode} and \eqref{app:eqn-sol-2mode} can be substantially simplified 
if only one-body losses are present.
For instance, Eq.\eqref{app:eqn-sol-1mode} takes the form
\begin{eqnarray}\label{app:eqn-rho_1bodyWRONG}
& & 		[\vv_{N,j} (t)]_{k}  =  \\
& &	\hspace*{5mm} 	\nonumber =  e^{z_j t} e^{ k x_j  t} \binom{N-j}{ k}^{\frac{1}{2}} \binom{N}{k+j}^{\frac{1}{2}}  
\left( \frac{e^{x_j t} -1 }{x_j/\Gamma_1}\right)^{N-k-j} 
\end{eqnarray}
with $z_j=  \I U (j^2-j)/2 -\Gamma_1 j/2$ and $x_j =   \I U j- \Gamma_1$.

Although the formula   \eqref{app:eqn-sol-2mode} for the density matrix elements  is a bit cumbersome, one can use it 
to derive simple expressions for the correlation functions
characterizing the state. 
To give an example, the first-order correlation function ($j=1$) reads 
(for simplicity we take symmetric loss rates $\Gamma_{1,0}=\Gamma_{0,1} = \alpha$ and 
energies $U_1  =U_2=U_{12}+\chi$):  
\begin{eqnarray*}
& & \text{Re}\left\{ \meanv{\hat{a}^{\dagger}_2\hat{a}_1}_t \right\} =  \text{Re} \Bigl\{ \sum_{k,l} \rho^{k+1, l}_{k, l+1} (t) \sqrt{k+1}\sqrt{l+1} \Bigr\} 
\\
& & \hspace*{1mm}		 
= \bbr{ \frac{ \alpha^2 + \chi^2e^{-\alpha t}\cos\bbr{\chi t}  + \alpha \chi e^{-\alpha t}\sin \bbr{\chi t}  }{\chi^2+\alpha^2} }^{N_0-1} 
\\
& & \hspace*{6mm} 
\times  \frac{N_0 e^{-\alpha t}}{2} \;.
\\
\end{eqnarray*}
The latter formula agrees with the results obtained with the help of the quantum trajectory method~\cite{yun2009} and  generating functions~\cite{pawlowskiBackground}.

In the case of two- and three-body losses, the eigenvectors  $\boldsymbol{p}_{j,n}$ and  $\boldsymbol{l}_{j,n}$ are
evaluated numerically using the recurrence  formulas \eqref{app:eqn-eigen-right} and \eqref{app:eqn-eigen-left}.

\section{Extraction of experimentally relevant parameters}
\label{app-parameters}

We choose a  symmetric trap with frequency $\omega = 2\pi\times500$\,Hz
 and initial number of atoms $N_0 = 100$ as in Ref.~\cite{sinatra1998}.  We compute the condensate
 wave function $\psi(\rv )$ with the help of the Gross-Pitaevskii equation, assuming no inter-species interaction, i.e. $a_{12}=0$,
and neglecting the interactions
in one of the two modes, namely $a_2 = 0$, where $a_{12}$ and $a_2$ are the    scattering lengths. 
Then $U_2 = U_{12}=0$. We estimate the interaction energy in the   first mode with
the formula $U_1 = \frac{4 \pi a_1}{M}\int |\psi|^4({\bf r})  \D^3 r$, $M$ being the atomic mass.
To compute the loss rates we use the constants for the atomic species used in Refs.~\cite{riedel2010,yun2009}.
 For such parameters the probability of three-body collisions is relatively small compared with the probability of two-body processes.
Moreover, the two-body processes are highly asymmetric for the internal states used in the experiment. Conversely, 
one-body losses appear to act symmetrically in the two modes.
For the results shown in Fig. \ref{fig3} we  additionally neglect  inter-mode losses (i.e. $\gamma_{12}=0$).

\section{Conditional state with $N_0$ atoms in the strong loss regime}
\label{app-rho_N_0_strong_losses}

In the strong loss regime,   the probability $w_{N_0} (t)$ that the BJJ does not lose any atom
in the time interval $[0,t]$ 
is very small (see Eq.(\ref{eq-proba_w_N})). As a consequence, the contribution  to the total
density matrix $\hat{\rho}(t)$ of the conditional state $\hat{\rho}_{N_0} (t)$ with $N_0$ atoms is negligible. However, one can 
gain some insight on QCs at intermediate loss rates
by investigating this strong loss regime. This study is performed in this appendix.
We also discuss the form of the damping factor  in Eq.(\ref{eq-state_no_jump}) for asymmetric three-body loss rates.
  We use the same notation as in Sec.~\ref{eq-subspace_N_0_atoms}.

Let us first consider {\it symmetric three-body losses} $\kappa_1=\kappa_2$ and 
$\kappa_{12}=\kappa_{21}$ and assume that  Eq.(\ref{eq=def_damping_a}) defines an effective loss rate $a>0$. 
The strong loss regime then corresponds to $a t  \gg 1$. The 
damping factor in Eq.(\ref{eq-state_no_jump}) is Gaussian and given by Eq.(\ref{eq-d(n_1)}). 
After renormalization by $w_{N_0}(t)$, this damping factor send 
all the matrix elements $\bra{n_1,N_0-n_1} \hat{\rho}_{N_0}(t)\ket{n_1',N_0-n_1'}$ of $\hat{\rho}_{N_0}(t)$ to zero
except those for which $n_1$ and $n_1'$ are the closest integer(s)  to $\overline{n}_1$. 
If $\overline{n}_1$ is an half integer, the four matrix elements with $n_1, n_1' = \overline{n}_1 \pm 1/2$ are damped
by exactly the same factor. Therefore, $\hat{\rho}_{N_0}(t)$ converges to a pure state,
\begin{equation}
\hat{\rho}_{N_0} (t) \rightarrow 
\ketbra{\psi^{(\infty)}_{0} (t)}{\psi^{(\infty)}_{0} (t)} \;.
\end{equation}
This state is either a Fock state
$\ket{\psi^{(\infty)}_{0}} = \ket{ E(\overline{n}_1),N_0-E(\overline{n}_1)}$
if $\overline{n}_1$ is not half integer (here $E(\overline{n}_1)$ denotes the closest integer to $\overline{n}_1$) or, if $\overline{n}_1$ is half integer, a superposition of two Fock states  
\begin{eqnarray} \label{eq-state_N_0_infinite_losses}
\nn
 \ket{\psi^{(\infty)}_{0} (t)}
& \propto &
\sum_{\pm} \left( \begin{array}{c} N_0\\ \overline{n}_1\pm \frac{1}{2} \end{array} \right)^{1/2}  
e^{\I    t \varphi_{\pm}} \\
& & \ket{\overline{n}_1 \pm \frac{1}{2}, N_0 -\overline{n}_1 \mp \frac{1}{2}} 
\end{eqnarray}
with 
$\varphi_+ =E_2 + (N_0- \overline{n}_1-\frac{1}{2}) U_2 + (2 \overline{n}_1 - N_0)  U_{12}$ and
$\varphi_{-}=E_1 + ( \overline{n}_1-\frac{1}{2}) U_1$.
In particular, if $\gamma_1=\gamma_2$ and $\alpha_1=\alpha_2$ (case (i) in  Sec.~\ref{eq-subspace_N_0_atoms}),
$\hat{\rho}_{N_0} ( t)$ converges to the Fock state $\ket{\psi_0^{(\infty)}}= \ket{\frac{N_0}{2},\frac{N_0}{2}}$ if $N_0$ is even
and to a superposition of the Fock states  $\ket{\frac{N_0\pm 1}{2},\frac{N_0 \mp 1}{2}}$  if $N_0$ is odd (since $\overline{n}_1 = N_0/2$).  
Similarly, if $\gamma_2=\gamma_{12}=\kappa=0$ (case (ii) in Sec.~\ref{eq-subspace_N_0_atoms}),
$\hat{\rho}_{N_0} (t )$  converges  to the Fock state $\ket{0,N_0}$ if $\alpha_2 < \alpha_1$ 
(since then $\overline{n}_1 < 1/2$, see Eq.(\ref{eq-overline_n_1})) and 
to a superposition of
Fock states with $n_1=0$ or $1$ atoms in the first mode if $\alpha_1=\alpha_2$ (since then $\overline{n}_1 = 1/2$).
Ignoring one-body losses, this can be explained as follows.
If one detects the same number of atoms initially and at time $t \gg 1/\gamma_1$,
the atomic state must have zero or one atom in the first mode suffering from two-body losses, since
otherwise    the BJJ would  have lost atoms in the  time interval $[0,t]$.

Let us turn to the case $a < 0$, \ie $\gamma_{12} > \gamma_1 + \gamma_2 + 2 (N_0-2) \kappa$ 
(case (iii) in  Sec.~\ref{eq-subspace_N_0_atoms}). We still assume symmetric three-body losses. 
It is easy to show from Eq.(\ref{eq-overline_n_1}) 
that $\overline{n}_1 > N_0/2$ if and only if $\Delta \gamma < - \Delta \alpha/ (N_0-1)$.
Therefore, $\hat{\rho}_{N_0} ( t)$ converges in the strong loss limit $| a| t \gg 1$  to
the Fock state  with   $n_{1,2}=0$  if 
$\pm \Delta \gamma < \mp  \Delta \alpha/(N_0-1)$, whereas
it converges 
 to the so-called NOON state
\begin{eqnarray} \label{eq-noon_state}
\non
\ket{\psi_{0}^{(\infty)} (t)} 
& = & 
\frac{1}{\sqrt{2}} \Bigl( e^{-\I t N_0 [ E_1 + U_1 (N_0-1)/2 ]} \ket{N_0,0} 
\\
& & 
+ e^{-\I t N_0 [ E_2 + U_2 (N_0-1)/2 ]} \ket{0,N_0} \Bigr)
\end{eqnarray}
 if $\Delta \gamma  = -\Delta \alpha/(N_0-1)$.
The latter state arises because if one knows that
the BJJ has not  lost any atom at time $t\gg |a|^{-1}$ one can be confident that it has either 
$n_1=0$ or $n_1=N_0$ atoms in the first mode, in such a way that no inter-mode collision
is possible. Since one cannot decide among the two possibilities, the state of the BJJ
is the superposition (\ref{eq-noon_state}). 

If $a=0$, \ie $\gamma_{12} = \gamma_1 + \gamma_2 + 2 (N_0-2) \kappa$, then
\begin{equation}
d_{N_0} (n_1) - d_{N_0} (0)= b n_1
\end{equation}
varies linearly with $n_1$, where  $b=- ( \Delta \alpha + (N_0-1) \Delta \gamma )/2$.
One easily finds that in the strong loss limit $|b | t \gg 1$, $\hat{\rho}_{N_0}(t)$ converges to the same states as in the previous
case $a<0$. Note that for $\Delta \gamma = - \Delta \alpha/(N_0-1)$ one has no damping, \ie $\hat{\rho}_{N_0} (t)$ 
coincides with the lossless density matrix.

  For completeness, let us now investigate the {\it asymmetric three-body loss} case
$K = 3 (\kappa_1-\kappa_2+\kappa_{21}-\kappa_{12})/2 \not=0$.   We do not assume anymore strong losses and
take $K >0$ (the case $K<0$ is treated by permuting the two modes).
The damping factor in Eq.(\ref{eq-state_no_jump}) is cubic in $n_1$,
\begin{equation} \label{eq-d(n_1)_asym}
d_{N_0} (n_1) 
= \frac{1}{3} K n_1^3
 +a  n_1^2 + b n_1 + c \;,
\end{equation}
where $c$ an irrelevant $n_1$-independent constant and
\begin{eqnarray}
\nn
a 
& = & \frac{1}{2} \bigl[ \gamma_1 + \gamma_2 -\gamma_{12} - 3 \kappa_1 + 3 (N_0-1)\kappa_2
\\ 
& &  + (N_0+1)\kappa_{12} - (2 N_0-1) \kappa_{21} \bigr]
\\ \nn
b
& = & 
\frac{1}{2} \bigl[ -\Delta \alpha + \Delta \gamma  - N_0 (2\gamma_2 - \gamma_{12}) - 2 \Delta \kappa 
\\
& & 
+ N_0 \bigl( - 3 (N_0-2) \kappa_2 - \kappa_{12} + (N_0-1) \kappa_{21} \bigr) \bigr] .
\end{eqnarray}
In the  last expression we have set $\Delta \kappa = \kappa_2 - \kappa_1$.
The minimum of $d_{N_0} (n_1)$ over all integers $n_1$ between $0$ and $N_0$ is reached 
either for $n_1=0$ or for $n_1 = \overline{n}_1 = (\sqrt{a^2 -  b K }-  a )/K$.
The effect of damping 
 at time $t_q$ on the lossless density matrix sets   in when $|K| N_0^3$, $|a| N_0^2$, or $|b| N_0$ are of the order of $\chi q$ or larger,
that is, for loss rates 
$\kappa \gtrsim \chi q/N_0^3$, $\gamma \gtrsim \chi q/N_0^2$,
or $\alpha \gtrsim \chi q/N_0$.   
For completely asymmetric losses of all kinds 
(\ie all rates vanish save for $\alpha_1$, $\gamma_1$, and $\kappa_1$)
one finds $\overline{n}_1 = 1+1/\sqrt{3}$ when only $\kappa_1$ is nonzero and
$\overline{n}_1 \simeq (-\gamma_1 + \sqrt{\gamma_1^2 - 3 \kappa_1 \alpha_1})/(3 \kappa_1) < 0$
when $N_0 \gg 1$ and $\alpha_1$, $N_0 \gamma_1$, and $N_0^2 \kappa_1$ have the same orders of magnitude.
In both cases, $\hat{\rho}_{N_0} (t)$ converges at strong losses to the Fock state $\ket{0,N_0}$, in 
analogy with what happens for two-body losses.

\section{Determination of the conditional states with $N<N_0$ atoms} \label{eq-app-D}

In this appendix we justify the formulas (\ref{eq-wave function_one_jump_bis_bis}), (\ref{eq-density_matrix_N_0-2_sector_in_Fock_basis_J>1}), and
(\ref{eq-enveloppe_J>1}) of Sec.~\ref{sec-subspace_N_0-2_atoms}.

\subsection{Contribution of trajectories with a single loss event} \label{app-density_matrix_N_0-2}

We first  determine the quantum trajectories  having exactly one jump in the time interval $[0,t]$ and the 
corresponding conditional  state $\hat{\rho}_{N_1} (t)$ with $N_1=N_0-r$ atoms, $r$ being the number of atoms lost during the jump process. 

Let $t \mapsto \ket{\psi_1 (t)}$ be such a trajectory subject to a single loss process,   
occurring at time $s \in [0,t]$ and of type  $m=(m_1,m_2)  \in \{1,2,3\}^2$, with 
$r=m_1+m_2$.
As a preliminary calculation, we take
a Fock state  $\ket{n_1,n_2}$  as initial state. This state is an eigenstate of $\Heff$ with eigenvalue $H_{\rm eff} (n_1,n_2)$. 
According to Eq.(\ref{eq-no_jump}) and given that $\hat{M}_m = \aopone^{m_1} \aoptwo^{m_2}$, the corresponding unnormalized wave function  $\ket{\widetilde{\psi}_1 (t)}$
at time $t$ is
(up to a prefactor) a Fock state with $n_i'=n_i-m_i \geq 0$ atoms in the mode $i=1,2$,
\begin{eqnarray} \label{eq-wavefunction_one_jump}
\ket{\widetilde{\psi}_1 (t)}
& = &   e^{-\I (t - s) \Heff}  \hat{M}_m  e^{-\I s \Heff} \ket{n_1,n_2}
\\ \nn
&  = & \sqrt{\frac{n_1! n_2!}{n_1' ! n_2' !}} e^{-\I \Phi_{m,s} (n_1',n_2')}  e^{-\I t \Heff} 
\ket{n_1',n_2'}\, ,
\end{eqnarray}
where
\begin{equation}
\Phi_{m, s} (n_1',n_2') =  s ( H_{\rm eff} (n_1,n_2) - H_{\rm eff} (n_1',n_2') )
\end{equation}
is a complex dynamical phase. 
The real part  of $\Phi_{m,s}$ 
is the dynamical phase associated to the change in the atomic interaction energy
 because of the reduction of particles at time $s$.
Since the Hamiltonian  (\ref{eq-H_0}) is quadratic in 
the number operators $\nopi$, this real part is
linear in $n_1'$ and $n_2'$. Setting $n_2'=N_1 - n_1'$, one finds
\begin{equation} \label{eq-real_part_Phi}
\re \Phi_{m, s} (n_1',n_2') 
=
\phi_{m} (s) n_1' + c_{m}\;,
\end{equation}
where  $c_{m}$ is an irrelevant $n_1'$-independent phase and
\begin{equation} \label{eq-random_phases1}
\phi_{m} (s) =   s \bigl( \chi_1 m_1 + \chi_2 m_2 \bigr)
\end{equation}
with $\chi_1 = U_1 - U_{12}$ and $\chi_2 = - ( U_2 - U_{12})$.

The imaginary part of $\Phi_{m, s}$ is associated to a change in the damping
due to the reduction of particles at time $s$.
It is quadratic in $n_i'$ because of the presence of the cubic damping operator 
$\hat{D}_{\rm{3-body}}$ (see Eq.(\ref{eq-D_3body})),  but we will see below
that one can neglect the quadratic term provided that the three-body loss rates satisfy $\kappa_i, \kappa_{ij} \ll (N_0 t )^{-1}$.   
In fact, by  neglecting all terms of the order of $s N_0\kappa_{i}$ and $s N_0\kappa_{ij}$
and keeping in mind that $n_1'$ and $n_2'=N_1-n_1'$ are at most of the order of $N_0$, one gets 
\begin{eqnarray} \label{eq-imaginary_part_Phi}
\nn
& & \im \Phi_{m, s} (n_1',n_2') 
 =  - \frac{s}{2} \biggl[ \sum_{i=1,2} \sum_{j\not= i}  ( 3 \kappa_i - 2 \kappa_{ij} + \kappa_{ji} ) m_i  \times
\\
& &
\Bigl( n_1' - \frac{N_1}{2} \Bigr)^2 
 + \bigl( \delta_1 m_1 + \delta_2 m_2 \bigr) \Bigl( n_1' - \frac{N_1}{2} \Bigr)
 + G_{m}   \biggr] 
\end{eqnarray}
with 
\begin{eqnarray} \label{eq-G_m}
\nn
G_{m} 
& = & \gamma_{12} \Bigl( \frac{r N_1}{2}+ m_1 m_2 \Bigr) +
\sum_{i=1,2} \Bigl( \alpha_i  + \gamma_i (N_1 -1 + m_i )  
\\
& & + \sum_{j\not= i} (3 \kappa_i + \kappa_{ji} + 2 \kappa_{ij} ) \frac{N_0^2}{4}  \Bigr) m_i  \; .
\end{eqnarray}
Here, we have set $\delta_1= 2 \gamma_1 - \gamma_{12} + ( 3 \kappa_1 - \kappa_{21} ) N_0$ and
$\delta_2= - ( 2 \gamma_2 - \gamma_{12} + ( 3 \kappa_2 - \kappa_{12} ) N_0)$ as in the main text.

We now take as initial state the CS $\ket{N_0;\phi=0}$. 
The corresponding unnormalized wave function is obtained from Eq.(\ref{eq-wavefunction_one_jump}) by using the 
Fock state expansion (\ref{eq-def_coherent_state}) for this CS.
This yields 
\begin{eqnarray} \label{eq-expansion_psi_bis}
\nn
\ket{\widetilde{\psi}_1 (t)} 
& = &
  \frac{1}{2^{N_0/2}} \sqrt{\frac{N_0!}{ N_1!}} \sum_{n_1' = 0}^{N_1} 
   \lt \begin{array}{c} N_1 \\ n_1' \end{array} \rt^{1/2} \E^{- \I t \hat{H}_{\rm eff}} \times
\\
& & 
  \E^{-\I \Phi_{m,s} (n_1',N_1-n_1')} \ket{n_1',N_1-n_1'}\,.
\end{eqnarray}
Note that only the terms with $|n_1'-N_1/2| \lesssim \sqrt{N_1}$ contribute significantly to the last sum. 
Thus one can neglect the quadratic term in the  dynamical phase (\ref{eq-imaginary_part_Phi})  
in the limit $\kappa_i,\kappa_{ij}  \ll (N_0 t)^{-1}$. 
Plugging Eqs.(\ref{eq-real_part_Phi}) and (\ref{eq-imaginary_part_Phi}) into
Eq.(\ref{eq-expansion_psi_bis}), one recognizes the Fock state expansion of a CS
with $N_1$ atoms. We get
\begin{eqnarray} \label{eq-expansion_psi_1}
\nn
\ket{\widetilde{\psi}_1 (t)} 
& = & 2^{-\frac{r}{2}} \sqrt{\frac{N_0!}{N_1!}} e^{-s G_{m} /2} 
\Bigl[ \cosh \Bigl( \frac{s}{2} \sum_i \delta_i m_i \Bigr) \Bigr]^{\frac{N_1}{2}} \times
\\
& & 
e^{-\I t \Heff} \ket{N_1; \theta_{m} (s), \phi_{m} (s)}
\end{eqnarray}
with
\begin{equation} \label{eq-random_phases2}
\theta_{m} (s)    =   
2 \arctan \Bigl( \exp \Bigl\{ -\frac{s}{2} \bigl( \delta_1 m_1 + \delta_2 m_2 \bigr) \Bigr\} \Bigr)\;.
\end{equation} 
This justifies Eq.(\ref{eq-wave function_one_jump_bis_bis}) for $J=1$, namely
\begin{equation} 
\label{eq-app_wave function_one_jump_bis_bis}
\ket{{\psi}_{1} (t)}
\propto   e^{-\I t \Heff }  \ket{N_0-r; \theta_{m} (s) , \phi_{m} (s) } \;.
\end{equation}
Moreover, from  Eqs.(\ref{eq-distrib_jump_times}) and (\ref{eq-expansion_psi_1}) we find the 
probability $\D p_{m}^{(t)} (s;1)$ that a loss event of type $m$ occurs in the time 
interval $[s, s+\D s]$ and that no other loss occur in $[0,t]$,
\begin{eqnarray} \label{eq-distribution_jump_time}
\nn
& & \D p_{m}^{(t)} (s;1) 
 =   \widetilde{p}_{m}^{(t)} ( s; 1)  \| e^{-\I t \Heff} \ket{N_1; \theta_{m} (s) , \phi_{m} (s) } \|^2 \D s
\\
& & 
\widetilde{p}_{m}^{(t)} ( s ) 
 =     \frac{\Gamma_m}{2^r}  \frac{N_0!}{N_1!}
e^{-s G_{m} }
  \cosh^{N_1} \Bigl( \frac{s}{2} \sum_{i=1,2} \delta_i m_i \Bigr) \,.
\end{eqnarray}

Let us assume that the BJJ is subject to $r$-body losses only, with  $r=1,2$, or $3$ fixed. 
According to Eq.(\ref{eq-density_matrix_subspace_N_0-j}),
the state in the subspace with $N_1=N_0-r$ atoms is
\begin{eqnarray} 
\label{eq-density_matrix_one_jump}
& & \hat{\rho}_{N_1} (t) \; \propto \; \widetilde{\rho}_{N_1}^{\text{(1-jump)}} (t)  =  
\sum_{m,|m|=r} \int_0^{t} \D s\, \widetilde{p}_{m}^{(t)} (s ) 
\times
\\ 
\nn
& & 
\hspace*{3mm}
\E^{-\I t \Heff}  \ket{N_1;\theta_{m}(s), \phi_{m} (s)} \bra{N_1;\theta_{m} (s),\phi_{m} (s) } \,\E^{\I t \Heff^{\dagger}}
\end{eqnarray}
and the probability to have $N_1$ atoms in the BJJ at time $t$ is 
$w_{N_1} ( t) = \tr  \widetilde{\rho}_{N_1}^{\text{(1-jump)}} (t)$.
Equation (\ref{eq-density_matrix_one_jump}) means that by conditioning to a single loss
event one obtains the same state as if there were no atom loss, one had  initially $N_1$ atoms,  
and the BJJ was subject to some external noises $\theta$ and $\phi$ rotating the state 
around the Bloch sphere. 
More precisely, with the help of the commutation of $\Heff$ with the angular momentum 
$\hat{J}_z = (\nopone-\noptwo)/2$ and the identity
$\ket{N;\theta,\phi}
= (e^{\I \phi} \cosh u )^{-N/2} e^{(-\I \phi+ u)  \hat{J}_z} \ket{N; \phi=0}$ with $u=\ln (\tan (\theta/2))$, 
one can rewrite (\ref{eq-density_matrix_one_jump})  as
\begin{eqnarray} \label{eq-density_matrix_one_jump_bis}
& & \widetilde{\rho}_{N_1}^{\text{(1-jump)}} (t)   
\\ \nn
& & \propto    
\sum_{m, |m|=r} \Bigl\langle U_m^{\rm{eff}} (s) \ketbra{N_1;\phi= 0}{N_1;\phi = 0} U_m^{\rm{eff}} (t)^\dagger \Bigr\rangle_s \;,
\end{eqnarray}
where
$U_m^{\rm eff} (s)= e^{-\I  (\phi_{m} (s)+ \I \ln ( \tan (\theta_{m}(s)/2)) \hat{J}_z} e^{- \I t \Heff}$ is a 
non-unitary random evolution operator
and the brackets denote the average with respect to the exponential distribution  
$h_{m} (s ) \propto \Theta ( t-s)  e^{-s G_{m}}$ 
of the loss time $s$ 
(here $\Theta $ denotes the Heaviside step function).
Hence the impact of atom losses on the conditional state 
can be fully described by introducing the effective noises
$\theta$ and $\phi$, in addition to the damping coming from the non self-adjoint Hamiltonian $\Heff$. 
These noises have fluctuations given by Eq.(\ref{eq-phase_fluctuations}) in the main text, where $\delta s_m$ is the fluctuation of the loss time 
with respect to the distribution
\begin{equation} \label{eq-distribution_jump_timebis}
f_m (s) = \frac{\widetilde{p}_{m}^{(t)} (s)\Theta (t-s)}{\int_0^t \D s \,\widetilde{p}_{m}^{(t)} (s)}\;.
\end{equation}

We now proceed to evaluate 
the density matrix (\ref{eq-density_matrix_one_jump})  explicitly
in the Fock basis. It reads
\begin{eqnarray} \label{eq-density_matrix_N_0-2_sector_in_Fock_basis}
& & 
\bra{n_1 ,n_2}  \widetilde{\rho}_{N_1}^{\text{(1-jump)}} (t ) \ket{n_1',n_2'} 
\propto  
\\
\nn
& & \hspace*{5mm}  
\Ee_{N_1}^{\text{(1-jump)}}   (t; n_1, n_1') 
\bra{n_1 ,n_2}  \widetilde{\rho}_{N_1}^{\rm{(no\,loss)}} (t ) \ket{n_1',n_2'} \;,
\end{eqnarray}
where $\widetilde{\rho}_{N_1}^{\rm{(no\,loss)}} (t ) $ is the  unnormalized density matrix conditioned to no loss event between times  $0$ and $t$
for an  
initial phase state with $N_1$ atoms (see Eq.(\ref{eq-rho_no_jump_N_0-2})) and 
\begin{equation} \label{eq-envelope_1_jump}
\Ee_{N_1}^{\text{(1-jump)}}   (t; n_1, n_1')  = \sum_{m,|m|=r} \Gamma_m C_{m} (t; n_1, n_1')  
\end{equation}
with
\begin{widetext}
\begin{equation} \label{eq-C_m(t)}
C_{m} ( t; n,n' ) 
 = \frac{1-e^{- t [G_m +(\delta_1 m_1+\delta_2 m_2)(n+n'-N_1)/2 + \I (\chi_1 m_1 + \chi_2 m_2) (n-n') ]}}
{G_m +(\delta_1 m_1+\delta_2 m_2)(n+n'-N_1)/2 + \I (\chi_1 m_1 + \chi_2 m_2) (n-n')} \,.
\end{equation}
%
\end{widetext}
Note that in the presence of both one- and two-body losses,
 to get the state $\hat{\rho}_{N_1}(t)$ in the subspace  with $N_1=N_0-2$ atoms one must add
to $\hat{\rho}_{N_1}^{\text{(1-jump)}}(t)$ the contribution of trajectories having two one-body loss events, which we now proceed to
evaluate.

\subsection{Contribution of trajectories with several loss events}
\label{app-several_jumps}

The extension to $J>1$ loss events of the previous calculation  does not present any difficulty.
We denote by $\ket{\psi_J (t)}$ the wave function 
after $J$ jumps
of types $m_1, \ldots, m_J$ occurring at times $0 \leq s_1 \leq \cdots \leq s_J \leq t$.
One easily finds  that if $\kappa_{i}   , \kappa_{ij}  \ll (N_0 t)^{-1}$ then
$\ket{\psi_J (t)}$ is the time-evolved CS
defined in Eq.(\ref{eq-wave function_one_jump_bis_bis}), with phases $\phi_{\mv} (\sv)$ 
and $\theta_{\mv} (\sv)$ given by Eq.(\ref{eq-random_phases_J>1}).
As  in the case $J=1$, $\phi_{\mv} (\sv)$ and $\I \ln ( \tan (\theta_{\mv} (\sv) /2))$ 
are the real and imaginary dynamical phases per atom in the first mode associated to 
the variations in the interaction energy and damping subsequent to the losses.

We are now ready to determine
the conditional states $\hat{\rho}_N(t)$ for all $N$ when the BJJ is subject simultaneously to 
one-, two-, and three-body losses. To this end 
one needs not only the wave function $\ket{{\psi}_J (t)}$
but also the norm of $\ket{\widetilde{\psi}_J (t)}$ giving the probability (\ref{eq-distrib_jump_times}). 
Using Eq.(\ref{eq-no_jump}) and the vector notation of Sec.~\ref{sec-subspace_N_0-2_atoms},
a simple (but somehow tedious) generalization of the calculation leading to 
Eq.(\ref{eq-expansion_psi_1}) yields 
\begin{widetext}
\begin{equation} \label{eq-expansion_psi_J}
\ket{\widetilde{\psi}_J (t)} 
 =  2^{-\frac{|\mv|}{2}} \sqrt{\frac{N_0!}{N_J!}} \exp \biggl\{ - \frac{1}{2} \sum_{\nu=1}^J s_\nu G_{\mv,\nu} \biggr\}  
\biggl[ \cosh \biggl( \sum_{\nu=1}^J \frac{s_\nu}{2} \sum_{i=1,2} \delta_i m_{\nu,i} \biggr) \biggr]^{\frac{N_J}{2}}
e^{-\I t \Heff} \ket{N_J; \theta_{\mv} (\sv), \phi_{\mv} (\sv)} \;,
\end{equation}
where  $m_{\nu,i}$ is the number of atoms lost in mode $i$ in the $\nu$th event, 
$N_J = N_0 - |\mv|$ is the remaining number of atoms  in the BJJ after the $J$ jumps
(\ie $|\mv | = \sum_{\nu,i} m_{\nu,i}$),  and
\begin{eqnarray} \label{eq-G_m_J_jump}
\nn
G_{\mv,\nu} 
& = & 
 \sum_{i=1,2} \Bigl[ \alpha_i  + \gamma_i \bigl( N_J -1 + \mu_{\nu,i} + \mu_{\nu+1,i} \bigr) 
+ \sum_{j\not= i} (3 \kappa_i + \kappa_{ji} + 2 \kappa_{ij} ) \frac{N_0^2}{4}  \Bigr] m_{\nu,i}   
\\
& & + \gamma_{12} \Bigl( \frac{|m_\nu| N_J }{2}+ \mu_{\nu,1} \mu_{\nu,2} - \mu_{\nu+1,1} \mu_{\nu+1,2} \Bigr) 
\end{eqnarray}
with $\mu_{\nu,i}= \sum_{\nu'=\nu}^{J} m_{\nu',i}$ for $\nu=1,\ldots,J$ and $i=1,2$. 
Thanks to 
 Eqs.(\ref{eq-density_matrix_subspace_N_0-j})
and (\ref{eq-expansion_psi_J}), the matrix elements  in the Fock basis  
of the unnormalized conditional state $\widetilde{\rho}_{N}(t)$ with $N$ atoms  are
\begin{eqnarray} \label{eq-density_matrix_N_0-2_sector_in_Fock_basis_J>1_app}
\nn
\bra{n_1 ,n_2}  \widetilde{\rho}_{N} (t ) \ket{n_1',n_2'} 
& = & \sum_{J=1}^{N_0} \sum_{\mv, N_0 - |\mv|=N} \Gamma_{m_1} \ldots \Gamma_{m_J} \int_{0\leq s_1 \leq \cdots \leq s_J \leq t} \!\!\!\!\!\! \D s_1 \ldots \D s_J \,
\braket{n_1,n_2}{\widetilde{\psi}_J (t)} \braket{\widetilde{\psi}_J (t)}{n_1',n_2'}
\\
& \propto  &
\hspace*{5mm} {\mathcal{E}}_N ( t ; n_1, n_1')  
\bra{n_1 ,n_2}  \widetilde{\rho}_{N}^{\rm{(no\,loss)}} (t ) \ket{n_1',n_2'} \;,
\end{eqnarray}
with
\begin{eqnarray} \label{eq-C_m(t)_J_jumps}
\nn
{\mathcal{E}}_N ( t; n,n' ) 
& = & 
\sum_{J=1}^{N_0}
\int_{0 \leq s_1 \leq \cdots \leq s_J \leq t} \D s_1 \ldots \D s_J   
\sum_{\mv, N_0 - | \mv | = N}
\Gamma_{m_1}\ldots \Gamma_{m_J} 
\exp \Bigl\{ - \I (n-n') \sum_{\nu=1}^J s_\nu \sum_{i} \chi_i m_{\nu,i}  \Bigr\}  \times
\\
& & 
\exp \Bigl\{ -\sum_{\nu=1}^J s_\nu G_{\mv,\nu}  
 -  (n+n'-N_J )  \sum_{\nu=1}^J \frac{s_\nu}{2} \sum_{i} \delta_i m_{\nu,i} \Bigr\}\,.
\end{eqnarray}

If, in addition to the above condition on three-body losses, 
the two-body loss rates  satisfy $ \gamma_i , \gamma_{12} \ll t^{-1} $ and
$|\mv| \ll N_0$,
the envelope  ${\mathcal{E}}_N (t;n,n')$ takes
the particularly simple form given by Eq.(\ref{eq-enveloppe_J>1}). 
Actually, in these limits the expression (\ref{eq-G_m_J_jump}) of $G_{\mv,\nu}$ 
reduces to the corresponding expression (\ref{eq-G_m}) for a single loss event
of type $m=m_\nu$,
\begin{equation}
G_{\mv,\nu} 
\simeq G_{m_\nu} \simeq  \gamma_{12} \frac{|\mv| N_0}{2}  + \sum_{i=1,2} \Bigl(  \alpha_i + \gamma_i N_0 
+\sum_{j\not=i} ( 3 \kappa_i + \kappa_{ji} + 2 \kappa_{ij}) \frac{N_0^2}{4} \Bigr)
 m_{\nu,i} \;.
\end{equation}
%
The integrand  
in Eq.(\ref{eq-C_m(t)_J_jumps}) is then symmetric  
under the exchange of the $s_\nu$'s, allowing us to replace the integration range by 
$[0,t]^J$ upon division by $J!$.
With the help of a simple counting argument, one obtains  Eq.(\ref{eq-enveloppe_J>1}) of  Sec.~\ref{sec-subspace_N_0-2_atoms}.
\end{widetext}
\bibliographystyle{apsrev4-1}
\bibliography{refs}

\begin{thebibliography}{10}%
\makeatletter
\providecommand \@ifxundefined [1]{%
 \ifx #1\undefined \expandafter \@firstoftwo
 \else \expandafter \@secondoftwo
\fi
}%
\providecommand \@ifnum [1]{%
 \ifnum #1\expandafter \@firstoftwo
 \else \expandafter \@secondoftwo
\fi
}%
\providecommand \enquote [1]{``#1''}%
\providecommand \bibnamefont  [1]{#1}%
\providecommand \bibfnamefont [1]{#1}%
\providecommand \citenamefont [1]{#1}%
\providecommand\href[0]{\@sanitize\@href}%
\providecommand\@href[1]{\endgroup\@@startlink{#1}\endgroup\@@href}%
\providecommand\@@href[1]{#1\@@endlink}%
\providecommand \@sanitize [0]{\begingroup\catcode`\&12\catcode`\#12\relax}%
\@ifxundefined \pdfoutput {\@firstoftwo}{%
 \@ifnum{\z@=\pdfoutput}{\@firstoftwo}{\@secondoftwo}%
}{%
 \providecommand\@@startlink[1]{\leavevmode\special{html:<a href="#1">}}%
 \providecommand\@@endlink[0]{\special{html:</a>}}%
}{%
 \providecommand\@@startlink[1]{%
  \leavevmode
  \pdfstartlink
   attr{/Border[0 0 1 ]/H/I/C[0 1 1]}%
   user{/Subtype/Link/A<</Type/Action/S/URI/URI(#1)>>}%
  \relax
 }%
 \providecommand\@@endlink[0]{\pdfendlink}%
}%
\providecommand \url  [0]{\begingroup\@sanitize \@url }%
\providecommand \@url [1]{\endgroup\@href {#1}{\urlprefix}}%
\providecommand \urlprefix [0]{URL }%
\providecommand \Eprint[0]{\href }%
\@ifxundefined \urlstyle {%
  \providecommand \doi [1]{doi:\discretionary{}{}{}#1}%
}{%
  \providecommand \doi [0]{doi:\discretionary{}{}{}\begingroup
  \urlstyle{rm}\Url }%
}%
\providecommand \doibase [0]{http://dx.doi.org/}%
\providecommand \Doi[1]{\href{\doibase#1}}%
\providecommand \bibAnnote [3]{%
  \BibitemShut{#1}%
  \begin{quotation}\noindent
    \textsc{Key:}\ #2\\\textsc{Annotation:}\ #3%
  \end{quotation}%
}%
\providecommand \bibAnnoteFile [2]{%
  \IfFileExists{#2}{\bibAnnote {#1} {#2} {\input{#2}}}{}%
}%
\providecommand \typeout [0]{\immediate \write \m@ne }%
\providecommand \selectlanguage [0]{\@gobble}%
\providecommand \bibinfo [0]{\@secondoftwo}%
\providecommand \bibfield [0]{\@secondoftwo}%
\providecommand \translation [1]{[#1]}%
\providecommand \BibitemOpen[0]{}%
\providecommand \bibitemStop [0]{}%
\providecommand \bibitemNoStop [0]{.\EOS\space}%
\providecommand \EOS [0]{\spacefactor3000\relax}%
\providecommand \BibitemShut [1]{\csname bibitem#1\endcsname}%
\bibitem{fano1961}%
  \BibitemOpen
  \bibfield{author}{%
  \bibinfo {author} {\bibfnamefont{U.}~\bibnamefont{Fano}},\ }%
  \bibfield{journal}{%
  \Doi{10.1103/PhysRev.124.1866}{\bibinfo {journal} {Phys. Rev.}}\ }%
  \textbf{\bibinfo {volume} {124}},\ \bibinfo {pages} {1866} (\bibinfo {month}
  {Dec}\ \bibinfo {year} {1961})%
  \bibAnnoteFile{NoStop}{fano1961}%
\bibitem{feshbach1958}%
  \BibitemOpen
  \bibfield{author}{%
  \bibinfo {author} {\bibfnamefont{H.}~\bibnamefont{Feshbach}},\ }%
  \bibfield{journal}{%
  \Doi{10.1016/0003-4916(58)90007-1}{\bibinfo {journal} {Annals of Physics}}\
  }%
  \textbf{\bibinfo {volume} {5}},\ \bibinfo {pages} {357} (\bibinfo {year}
  {1958}),\ ISSN \bibinfo {issn} {0003-4916}%
  \bibAnnoteFile{NoStop}{feshbach1958}%
\bibitem{Bloch2005}%
  \BibitemOpen
  \bibfield{author}{%
  \bibinfo {author} {\bibfnamefont{I.}~\bibnamefont{Bloch}},\ }%
  \bibfield{journal}{%
  \bibinfo {journal} {Nature Physics}\ }%
  \textbf{\bibinfo {volume} {1}},\ \bibinfo {pages} {23} (\bibinfo {year}
  {2005})%
  \bibAnnoteFile{NoStop}{Bloch2005}%
\bibitem{Bloch2008}%
  \BibitemOpen
  \bibfield{author}{%
  \bibinfo {author} {\bibfnamefont{I.}~\bibnamefont{Bloch}},\ }%
  \bibfield{journal}{%
  \bibinfo {journal} {Nature}\ }%
  \textbf{\bibinfo {volume} {453}},\ \bibinfo {pages} {1016} (\bibinfo {year}
  {2008})%
  \bibAnnoteFile{NoStop}{Bloch2008}%
\bibitem{Weitenberg2011}%
  \BibitemOpen
  \bibfield{author}{%
  \bibinfo {author} {\bibfnamefont{C.}~\bibnamefont{Weitenberg}}, \bibinfo
  {author} {\bibfnamefont{M.}~\bibnamefont{Endres}}, \bibinfo {author}
  {\bibfnamefont{J.~F.}\ \bibnamefont{Sherson}}, \bibinfo {author}
  {\bibfnamefont{M.}~\bibnamefont{Cheneau}}, \bibinfo {author}
  {\bibfnamefont{P.}~\bibnamefont{Schauss}}, \bibinfo {author}
  {\bibfnamefont{T.}~\bibnamefont{Fukuhara}}, \bibinfo {author}
  {\bibfnamefont{I.}~\bibnamefont{Bloch}},\ and\ \bibinfo {author}
  {\bibfnamefont{S.}~\bibnamefont{Kuhr}},\ }%
  \bibfield{journal}{%
  \bibinfo {journal} {Nature}\ }%
  \textbf{\bibinfo {volume} {471}},\ \bibinfo {pages} {319} (\bibinfo {year}
  {2011})%
  \bibAnnoteFile{NoStop}{Weitenberg2011}%
\bibitem{kitagawa1993}%
  \BibitemOpen
  \bibfield{author}{%
  \bibinfo {author} {\bibfnamefont{M.}~\bibnamefont{Kitagawa}}\ and\ \bibinfo
  {author} {\bibfnamefont{M.}~\bibnamefont{Ueda}},\ }%
  \bibfield{journal}{%
  \Doi{10.1103/PhysRevA.47.5138}{\bibinfo {journal} {Phys. Rev. A}}\ }%
  \textbf{\bibinfo {volume} {47}},\ \bibinfo {pages} {5138} (\bibinfo {month}
  {Jun}\ \bibinfo {year} {1993})%
  \bibAnnoteFile{NoStop}{kitagawa1993}%
\bibitem{soerensen2001a}%
  \BibitemOpen
  \bibfield{author}{%
  \bibinfo {author} {\bibfnamefont{A.~S.}\ \bibnamefont{S\o{}rensen}}\ and\
  \bibinfo {author} {\bibfnamefont{K.}~\bibnamefont{M\o{}lmer}},\ }%
  \bibfield{journal}{%
  \Doi{10.1103/PhysRevLett.86.4431}{\bibinfo {journal} {Phys. Rev. Lett.}}\ }%
  \textbf{\bibinfo {volume} {86}},\ \bibinfo {pages} {4431} (\bibinfo {month}
  {May}\ \bibinfo {year} {2001})%
  \bibAnnoteFile{NoStop}{soerensen2001a}%
\bibitem{soerensen2001}%
  \BibitemOpen
  \bibfield{author}{%
  \bibinfo {author} {\bibfnamefont{A.}~\bibnamefont{S\o{}rensen}}, \bibinfo
  {author} {\bibfnamefont{L.~M.}\ \bibnamefont{Duan}}, \bibinfo {author}
  {\bibfnamefont{J.~I.}\ \bibnamefont{Cirac}},\ and\ \bibinfo {author}
  {\bibfnamefont{P.}~\bibnamefont{Zoller}},\ }%
  \bibfield{journal}{%
  \bibinfo {journal} {Nature}\ }%
  \textbf{\bibinfo {volume} {409}},\ \bibinfo {pages} {63} (\bibinfo {year}
  {2001})%
  \bibAnnoteFile{NoStop}{soerensen2001}%
\bibitem{yurke1986}%
  \BibitemOpen
  \bibfield{author}{%
  \bibinfo {author} {\bibfnamefont{B.}~\bibnamefont{Yurke}}\ and\ \bibinfo
  {author} {\bibfnamefont{D.}~\bibnamefont{Stoler}},\ }%
  \bibfield{journal}{%
  \Doi{10.1103/PhysRevLett.57.13}{\bibinfo {journal} {Phys. Rev. Lett.}}\ }%
  \textbf{\bibinfo {volume} {57}},\ \bibinfo {pages} {13} (\bibinfo {month}
  {Jul}\ \bibinfo {year} {1986})%
  \bibAnnoteFile{NoStop}{yurke1986}%
\bibitem{stoler1971}%
  \BibitemOpen
  \bibfield{author}{%
  \bibinfo {author} {\bibfnamefont{D.}~\bibnamefont{Stoler}},\ }%
  \bibfield{journal}{%
  \Doi{10.1103/PhysRevD.4.2309}{\bibinfo {journal} {Phys. Rev. D}}\ }%
  \textbf{\bibinfo {volume} {4}},\ \bibinfo {pages} {2309} (\bibinfo {month}
  {Oct}\ \bibinfo {year} {1971})%
  \bibAnnoteFile{NoStop}{stoler1971}%
\bibitem{esteve2008}%
  \BibitemOpen
  \bibfield{author}{%
  \bibinfo {author} {\bibfnamefont{J.}~\bibnamefont{Esteve}}, \bibinfo {author}
  {\bibfnamefont{C.}~\bibnamefont{Gross}}, \bibinfo {author}
  {\bibfnamefont{A.}~\bibnamefont{Weller}}, \bibinfo {author}
  {\bibfnamefont{S.}~\bibnamefont{Giovanazzi}},\ and\ \bibinfo {author}
  {\bibfnamefont{M.~K.}\ \bibnamefont{Oberthaler}},\ }%
  \bibfield{journal}{%
  \bibinfo {journal} {Nature}\ }%
  \textbf{\bibinfo {volume} {455}},\ \bibinfo {pages} {1216} (\bibinfo {year}
  {2008})%
  \bibAnnoteFile{NoStop}{esteve2008}%
\bibitem{riedel2010}%
  \BibitemOpen
  \bibfield{author}{%
  \bibinfo {author} {\bibfnamefont{F.}~\bibnamefont{Riedel}}, \bibinfo {author}
  {\bibfnamefont{P.}~\bibnamefont{B\"ohi}}, \bibinfo {author}
  {\bibfnamefont{Y.}~\bibnamefont{Li}}, \bibinfo {author}
  {\bibfnamefont{T.~W.}\ \bibnamefont{H\"ansch}}, \bibinfo {author}
  {\bibfnamefont{A.}~\bibnamefont{Sinatra}},\ and\ \bibinfo {author}
  {\bibfnamefont{P.}~\bibnamefont{Treutlein}},\ }%
  \bibfield{journal}{%
  \bibinfo {journal} {Nature}\ }%
  \textbf{\bibinfo {volume} {464}},\ \bibinfo {pages} {1170} (\bibinfo {year}
  {2010})%
  \bibAnnoteFile{NoStop}{riedel2010}%
\bibitem{gross2010}%
  \BibitemOpen
  \bibfield{author}{%
  \bibinfo {author} {\bibfnamefont{C.}~\bibnamefont{Gross}}, \bibinfo {author}
  {\bibfnamefont{T.}~\bibnamefont{Zibold}}, \bibinfo {author}
  {\bibfnamefont{E.}~\bibnamefont{Nicklas}}, \bibinfo {author}
  {\bibfnamefont{J.}~\bibnamefont{Est\`eve}},\ and\ \bibinfo {author}
  {\bibfnamefont{M.~K.}\ \bibnamefont{Oberthaler}},\ }%
  \bibfield{journal}{%
  \bibinfo {journal} {Nature}\ }%
  \textbf{\bibinfo {volume} {464}},\ \bibinfo {pages} {1165} (\bibinfo {year}
  {2010})%
  \bibAnnoteFile{NoStop}{gross2010}%
\bibitem{sinatra1998}%
  \BibitemOpen
  \bibfield{author}{%
  \bibinfo {author} {\bibfnamefont{A.}~\bibnamefont{Sinatra}}\ and\ \bibinfo
  {author} {\bibfnamefont{Y.}~\bibnamefont{Castin}},\ }%
  \bibfield{journal}{%
  \bibinfo {journal} {Eur. Phys. J. D}\ }%
  \textbf{\bibinfo {volume} {4}},\ \bibinfo {pages} {247} (\bibinfo {year}
  {1998})%
  \bibAnnoteFile{NoStop}{sinatra1998}%
\bibitem{yun2008}%
  \BibitemOpen
  \bibfield{author}{%
  \bibinfo {author} {\bibfnamefont{Y.}~\bibnamefont{Li}}, \bibinfo {author}
  {\bibfnamefont{Y.}~\bibnamefont{Castin}},\ and\ \bibinfo {author}
  {\bibfnamefont{A.}~\bibnamefont{Sinatra}},\ }%
  \bibfield{journal}{%
  \Doi{10.1103/PhysRevLett.100.210401}{\bibinfo {journal} {Phys. Rev. Lett.}}\
  }%
  \textbf{\bibinfo {volume} {100}},\ \bibinfo {pages} {210401} (\bibinfo
  {month} {May}\ \bibinfo {year} {2008})%
  \bibAnnoteFile{NoStop}{yun2008}%
\bibitem{yun2009}%
  \BibitemOpen
  \bibfield{author}{%
  \bibinfo {author} {\bibfnamefont{L.}~\bibnamefont{Yun}}, \bibinfo {author}
  {\bibfnamefont{P.}~\bibnamefont{Treutlein}}, \bibinfo {author}
  {\bibfnamefont{J.}~\bibnamefont{Reichel}},\ and\ \bibinfo {author}
  {\bibfnamefont{A.}~\bibnamefont{Sinatra}},\ }%
  \bibfield{journal}{%
  \Doi{10.1140/epjb/e2008-00472-6}{\bibinfo {journal} {Eur. Phys. J. B}}\ }%
  \textbf{\bibinfo {volume} {68}},\ \bibinfo {pages} {365} (\bibinfo {year}
  {2009})%
  \bibAnnoteFile{NoStop}{yun2009}%
\bibitem{pawlowskiBackground}%
  \BibitemOpen
  \bibfield{author}{%
  \bibinfo {author} {\bibfnamefont{K.}~\bibnamefont{Paw\l{}owski}}\ and\
  \bibinfo {author} {\bibfnamefont{K.}~\bibnamefont{Rz\k{a}\.{z}ewski}},\ }%
  \bibfield{journal}{%
  \Doi{10.1103/PhysRevA.81.013620}{\bibinfo {journal} {Phys. Rev. A}}\ }%
  \textbf{\bibinfo {volume} {81}},\ \bibinfo {pages} {013620} (\bibinfo {month}
  {Jan}\ \bibinfo {year} {2010})%
  \bibAnnoteFile{NoStop}{pawlowskiBackground}%
\bibitem{ferrini2010}%
  \BibitemOpen
  \bibfield{author}{%
  \bibinfo {author} {\bibfnamefont{G.}~\bibnamefont{Ferrini}}, \bibinfo
  {author} {\bibfnamefont{D.}~\bibnamefont{Spehner}}, \bibinfo {author}
  {\bibfnamefont{A.}~\bibnamefont{Minguzzi}},\ and\ \bibinfo {author}
  {\bibfnamefont{F.~W.~J.}\ \bibnamefont{Hekking}},\ }%
  \bibfield{journal}{%
  \bibinfo {journal} {Phys. Rev. A}\ }%
  \textbf{\bibinfo {volume} {84}},\ \bibinfo {pages} {043628} (\bibinfo {year}
  {2011})%
  \bibAnnoteFile{NoStop}{ferrini2010}%
\bibitem{huang2006}%
  \BibitemOpen
  \bibfield{author}{%
  \bibinfo {author} {\bibfnamefont{Y.~P.}\ \bibnamefont{Huang}}\ and\ \bibinfo
  {author} {\bibfnamefont{M.~G.}\ \bibnamefont{Moore}},\ }%
  \bibfield{journal}{%
  \bibinfo {journal} {Phys. Rev. A}\ }%
  \textbf{\bibinfo {volume} {73}},\ \bibinfo {pages} {023606} (\bibinfo {year}
  {2006})%
  \bibAnnoteFile{NoStop}{huang2006}%
\bibitem{ferrini2010a}%
  \BibitemOpen
  \bibfield{author}{%
  \bibinfo {author} {\bibfnamefont{G.}~\bibnamefont{Ferrini}}, \bibinfo
  {author} {\bibfnamefont{D.}~\bibnamefont{Spehner}}, \bibinfo {author}
  {\bibfnamefont{A.}~\bibnamefont{Minguzzi}},\ and\ \bibinfo {author}
  {\bibfnamefont{F.~W.~J.}\ \bibnamefont{Hekking}},\ }%
  \bibfield{journal}{%
  \bibinfo {journal} {Phys. Rev. A}\ }%
  \textbf{\bibinfo {volume} {82}},\ \bibinfo {pages} {033621} (\bibinfo {year}
  {2010})%
  \bibAnnoteFile{NoStop}{ferrini2010a}%
\bibitem{pawlowski2013}%
  \BibitemOpen
  \bibfield{author}{%
  \bibinfo {author} {\bibfnamefont{K.}~\bibnamefont{Pawlowski}}, \bibinfo
  {author} {\bibfnamefont{D.}~\bibnamefont{Spehner}}, \bibinfo {author}
  {\bibfnamefont{A.}~\bibnamefont{Minguzzi}},\ and\ \bibinfo {author}
  {\bibfnamefont{G.}~\bibnamefont{Ferrini}},\ }%
  \bibfield{journal}{%
  \bibinfo {journal} {Phys. Rev. A}\ }%
  \textbf{\bibinfo {volume} {88}},\ \bibinfo {pages} {013606} (\bibinfo {year}
  {2013})%
  \bibAnnoteFile{NoStop}{pawlowski2013}%
\bibitem{sinatra2012}%
  \BibitemOpen
  \bibfield{author}{%
  \bibinfo {author} {\bibfnamefont{A.}~\bibnamefont{Sinatra}}, \bibinfo
  {author} {\bibfnamefont{J.-C.}\ \bibnamefont{Dornstetter}},\ and\ \bibinfo
  {author} {\bibfnamefont{Y.}~\bibnamefont{Castin}},\ }%
  \bibfield{journal}{%
  \bibinfo {journal} {Front. Phys}\ }%
  \textbf{\bibinfo {volume} {7}},\ \bibinfo {pages} {86} (\bibinfo {year}
  {2012})%
  \bibAnnoteFile{NoStop}{sinatra2012}%
\bibitem{braunstein1994}%
  \BibitemOpen
  \bibfield{author}{%
  \bibinfo {author} {\bibfnamefont{S.~L.}\ \bibnamefont{Braunstein}}\ and\
  \bibinfo {author} {\bibfnamefont{C.~M.}\ \bibnamefont{Caves}},\ }%
  \bibfield{journal}{%
  \Doi{10.1103/PhysRevLett.72.3439}{\bibinfo {journal} {Phys. Rev. Lett.}}\ }%
  \textbf{\bibinfo {volume} {72}},\ \bibinfo {pages} {3439} (\bibinfo {month}
  {May}\ \bibinfo {year} {1994})%
  \bibAnnoteFile{NoStop}{braunstein1994}%
\bibitem{Zhang1990}%
  \BibitemOpen
  \bibfield{author}{%
  \bibinfo {author} {\bibfnamefont{W.~M.}\ \bibnamefont{Zhang}}, \bibinfo
  {author} {\bibfnamefont{D.~H.}\ \bibnamefont{Feng}},\ and\ \bibinfo {author}
  {\bibfnamefont{R.}~\bibnamefont{Gilmore}},\ }%
  \bibfield{journal}{%
  \bibinfo {journal} {Rev. Mod. Phys.}\ }%
  \textbf{\bibinfo {volume} {62}},\ \bibinfo {pages} {867} (\bibinfo {year}
  {1990})%
  \bibAnnoteFile{NoStop}{Zhang1990}%
\bibitem{Milburn97}%
  \BibitemOpen
  \bibfield{author}{%
  \bibinfo {author} {\bibfnamefont{G.}~\bibnamefont{Milburn}}, \bibinfo
  {author} {\bibfnamefont{J.}~\bibnamefont{Corney}}, \bibinfo {author}
  {\bibfnamefont{E.}~\bibnamefont{Wright}},\ and\ \bibinfo {author}
  {\bibfnamefont{D.}~\bibnamefont{Walls}},\ }%
  \bibfield{journal}{%
  \bibinfo {journal} {Phys. Rev. A}\ }%
  \textbf{\bibinfo {volume} {55}},\ \bibinfo {pages} {4318} (\bibinfo {year}
  {1997})%
  \bibAnnoteFile{NoStop}{Milburn97}%
\bibitem{anglin1997}%
  \BibitemOpen
  \bibfield{author}{%
  \bibinfo {author} {\bibfnamefont{J.}~\bibnamefont{Anglin}},\ }%
  \bibfield{journal}{%
  \Doi{10.1103/PhysRevLett.79.6}{\bibinfo {journal} {Phys. Rev. Lett.}}\ }%
  \textbf{\bibinfo {volume} {79}},\ \bibinfo {pages} {6} (\bibinfo {month}
  {Jul}\ \bibinfo {year} {1997})%
  \bibAnnoteFile{NoStop}{anglin1997}%
\bibitem{jack2002}%
  \BibitemOpen
  \bibfield{author}{%
  \bibinfo {author} {\bibfnamefont{M.~W.}\ \bibnamefont{Jack}},\ }%
  \bibfield{journal}{%
  \Doi{10.1103/PhysRevLett.89.140402}{\bibinfo {journal} {Phys. Rev. Lett.}}\
  }%
  \textbf{\bibinfo {volume} {89}},\ \bibinfo {pages} {140402} (\bibinfo {month}
  {Sep}\ \bibinfo {year} {2002})%
  \bibAnnoteFile{NoStop}{jack2002}%
\bibitem{jack2003}%
  \BibitemOpen
  \bibfield{author}{%
  \bibinfo {author} {\bibfnamefont{M.~W.}\ \bibnamefont{Jack}},\ }%
  \bibfield{journal}{%
  \Doi{10.1103/PhysRevA.67.043612}{\bibinfo {journal} {Phys. Rev. A}}\ }%
  \textbf{\bibinfo {volume} {67}},\ \bibinfo {pages} {043612} (\bibinfo {month}
  {Apr}\ \bibinfo {year} {2003})%
  \bibAnnoteFile{NoStop}{jack2003}%
\bibitem{itah2010}%
  \BibitemOpen
  \bibfield{author}{%
  \bibinfo {author} {\bibfnamefont{A.}~\bibnamefont{Itah}}, \bibinfo {author}
  {\bibfnamefont{H.}~\bibnamefont{Veksler}}, \bibinfo {author}
  {\bibfnamefont{O.}~\bibnamefont{Lahav}}, \bibinfo {author}
  {\bibfnamefont{A.}~\bibnamefont{Blumkin}}, \bibinfo {author}
  {\bibfnamefont{C.}~\bibnamefont{Moreno}}, \bibinfo {author}
  {\bibfnamefont{C.}~\bibnamefont{Gordon}},\ and\ \bibinfo {author}
  {\bibfnamefont{J.}~\bibnamefont{Steinhauer}},\ }%
  \bibfield{journal}{%
  \Doi{10.1103/PhysRevLett.104.113001}{\bibinfo {journal} {Phys. Rev. Lett.}}\
  }%
  \textbf{\bibinfo {volume} {104}},\ \bibinfo {pages} {113001} (\bibinfo
  {month} {Mar}\ \bibinfo {year} {2010})%
  \bibAnnoteFile{NoStop}{itah2010}%
\bibitem{gross2011}%
  \BibitemOpen
  \bibfield{author}{%
  \bibinfo {author} {\bibfnamefont{C.}~\bibnamefont{Gross}}, \bibinfo {author}
  {\bibfnamefont{J.}~\bibnamefont{Est\`eve}}, \bibinfo {author}
  {\bibfnamefont{M.~K.}\ \bibnamefont{Oberthaler}}, \bibinfo {author}
  {\bibfnamefont{A.~D.}\ \bibnamefont{Martin}},\ and\ \bibinfo {author}
  {\bibfnamefont{J.}~\bibnamefont{Ruostekoski}},\ }%
  \bibfield{journal}{%
  \Doi{10.1103/PhysRevA.84.011609}{\bibinfo {journal} {Phys. Rev. A}}\ }%
  \textbf{\bibinfo {volume} {84}},\ \bibinfo {pages} {011609} (\bibinfo {month}
  {Jul}\ \bibinfo {year} {2011})%
  \bibAnnoteFile{NoStop}{gross2011}%
\bibitem{hume2013}%
  \BibitemOpen
  \bibfield{author}{%
  \bibinfo {author} {\bibfnamefont{D.~B.}\ \bibnamefont{Hume}}, \bibinfo
  {author} {\bibfnamefont{I.}~\bibnamefont{Stroescu}}, \bibinfo {author}
  {\bibfnamefont{M.}~\bibnamefont{Joos}}, \bibinfo {author}
  {\bibfnamefont{W.}~\bibnamefont{Muessel}}, \bibinfo {author}
  {\bibfnamefont{H.}~\bibnamefont{Strobel}},\ and\ \bibinfo {author}
  {\bibfnamefont{M.~K.}\ \bibnamefont{Oberthaler}},\ }%
  \bibfield{journal}{%
  \bibinfo {journal} {Phys. Rev. Lett.}\ }%
  \textbf{\bibinfo {volume} {111}},\ \bibinfo {pages} {253001} (\bibinfo {year}
  {2013})%
  \bibAnnoteFile{NoStop}{hume2013}%
\bibitem{Klauder1986}%
  \BibitemOpen
  \bibfield{author}{%
  \bibinfo {author} {\bibfnamefont{B.}~\bibnamefont{Yurke}}, \bibinfo {author}
  {\bibfnamefont{S.~L.}\ \bibnamefont{McCall}},\ and\ \bibinfo {author}
  {\bibfnamefont{J.~R.}\ \bibnamefont{Klauder}},\ }%
  \bibfield{journal}{%
  \bibinfo {journal} {Phys. Rev. A}\ }%
  \textbf{\bibinfo {volume} {33}},\ \bibinfo {pages} {4033} (\bibinfo {year}
  {1986})%
  \bibAnnoteFile{NoStop}{Klauder1986}%
\bibitem{pezze09}%
  \BibitemOpen
  \bibfield{author}{%
  \bibinfo {author} {\bibfnamefont{L.}~\bibnamefont{Pezze}}\ and\ \bibinfo
  {author} {\bibfnamefont{A.}~\bibnamefont{Smerzi}},\ }%
  \bibfield{journal}{%
  \bibinfo {journal} {Phys. Rev. Lett.}\ }%
  \textbf{\bibinfo {volume} {102}},\ \bibinfo {pages} {100401} (\bibinfo {year}
  {2009})%
  \bibAnnoteFile{NoStop}{pezze09}%
\bibitem{Grond11}%
  \BibitemOpen
  \bibfield{author}{%
  \bibinfo {author} {\bibfnamefont{J.}~\bibnamefont{Grond}}, \bibinfo {author}
  {\bibfnamefont{U.}~\bibnamefont{Hohenester}}, \bibinfo {author}
  {\bibfnamefont{J.}~\bibnamefont{Schmiedmayer}},\ and\ \bibinfo {author}
  {\bibfnamefont{A.}~\bibnamefont{Smerzi}},\ }%
  \bibfield{journal}{%
  \bibinfo {journal} {Phys. Rev. A}\ }%
  \textbf{\bibinfo {volume} {84}},\ \bibinfo {pages} {023619} (\bibinfo {year}
  {2011})%
  \bibAnnoteFile{NoStop}{Grond11}%
\bibitem{tikhonenkov10}%
  \BibitemOpen
  \bibfield{author}{%
  \bibinfo {author} {\bibfnamefont{I.}~\bibnamefont{Tikhonenkov}}, \bibinfo
  {author} {\bibfnamefont{M.}~\bibnamefont{Moore}},\ and\ \bibinfo {author}
  {\bibfnamefont{A.}~\bibnamefont{Vardi}},\ }%
  \bibfield{journal}{%
  \bibinfo {journal} {Phys. Rev. A}\ }%
  \textbf{\bibinfo {volume} {82}},\ \bibinfo {pages} {043624} (\bibinfo {year}
  {2010})%
  \bibAnnoteFile{NoStop}{tikhonenkov10}%
\bibitem{HyllusPRL2010}%
  \BibitemOpen
  \bibfield{author}{%
  \bibinfo {author} {\bibfnamefont{P.}~\bibnamefont{Hyllus}}, \bibinfo {author}
  {\bibfnamefont{L.}~\bibnamefont{Pezz\'e}},\ and\ \bibinfo {author}
  {\bibfnamefont{A.}~\bibnamefont{Smerzi}},\ }%
  \bibfield{journal}{%
  \Doi{10.1103/PhysRevLett.105.120501}{\bibinfo {journal} {Phys. Rev. Lett.}}\
  }%
  \textbf{\bibinfo {volume} {105}},\ \bibinfo {pages} {120501} (\bibinfo
  {month} {Sept}\ \bibinfo {year} {2010})%
  \bibAnnoteFile{NoStop}{HyllusPRL2010}%
\bibitem{Hyllus2010}%
  \BibitemOpen
  \bibfield{author}{%
  \bibinfo {author} {\bibfnamefont{P.}~\bibnamefont{Hyllus}}, \bibinfo {author}
  {\bibfnamefont{O.}~\bibnamefont{G\"uhne}},\ and\ \bibinfo {author}
  {\bibfnamefont{A.}~\bibnamefont{Smerzi}},\ }%
  \bibfield{journal}{%
  \Doi{10.1103/PhysRevA.82.012337}{\bibinfo {journal} {Phys. Rev. A}}\ }%
  \textbf{\bibinfo {volume} {82}},\ \bibinfo {pages} {012337} (\bibinfo {month}
  {Jul}\ \bibinfo {year} {2010})%
  \bibAnnoteFile{NoStop}{Hyllus2010}%
\bibitem{carmichael1991}%
  \BibitemOpen
  \bibfield{author}{%
  \bibinfo {author} {\bibfnamefont{H.}~\bibnamefont{Carmichael}},\ }%
  \emph{\bibinfo {title} {An {O}pen {S}ystem {A}pproach to {Q}uantum
  {O}ptics}}\ (\bibinfo {publisher} {Springer-Verlag},\ \bibinfo {address} {New
  York},\ \bibinfo {year} {1991})%
  \bibAnnoteFile{NoStop}{carmichael1991}%
\bibitem{moelmer1993}%
  \BibitemOpen
  \bibfield{author}{%
  \bibinfo {author} {\bibfnamefont{K.}~\bibnamefont{M\o{}lmer}}, \bibinfo
  {author} {\bibfnamefont{Y.}~\bibnamefont{Castin}},\ and\ \bibinfo {author}
  {\bibfnamefont{J.}~\bibnamefont{Dalibard}},\ }%
  \bibfield{journal}{%
  \bibinfo {journal} {J. Opt. Soc. Am. B}\ }%
  \textbf{\bibinfo {volume} {10}},\ \bibinfo {pages} {524} (\bibinfo {year}
  {1993})%
  \bibAnnoteFile{NoStop}{moelmer1993}%
\bibitem{belavkin1990}%
  \BibitemOpen
  \bibfield{author}{%
  \bibinfo {author} {\bibfnamefont{V.~P.}\ \bibnamefont{Belavkin}},\ }%
  \bibfield{journal}{%
  \bibinfo {journal} {Journal of Mathematical Physics}\ }%
  \textbf{\bibinfo {volume} {31}},\ \bibinfo {pages} {2930} (\bibinfo {year}
  {1990})%
  \bibAnnoteFile{NoStop}{belavkin1990}%
\bibitem{barchielli1991}%
  \BibitemOpen
  \bibfield{author}{%
  \bibinfo {author} {\bibfnamefont{A.}~\bibnamefont{Barchielli}}\ and\ \bibinfo
  {author} {\bibfnamefont{V.~P.}\ \bibnamefont{Belavkin}},\ }%
  \bibfield{journal}{%
  \bibinfo {journal} {Journal of Physics A: Mathematical and General}\ }%
  \textbf{\bibinfo {volume} {24}},\ \bibinfo {pages} {1495} (\bibinfo {year}
  {1991})%
  \bibAnnoteFile{NoStop}{barchielli1991}%
\bibitem{Knight98}%
  \BibitemOpen
  \bibfield{author}{%
  \bibinfo {author} {\bibfnamefont{M.}~\bibnamefont{Plenio}}\ and\ \bibinfo
  {author} {\bibfnamefont{P.}~\bibnamefont{Knight}},\ }%
  \bibfield{journal}{%
  \bibinfo {journal} {Rev. Mod. Phys.}\ }%
  \textbf{\bibinfo {volume} {70}},\ \bibinfo {pages} {101} (\bibinfo {year}
  {1998})%
  \bibAnnoteFile{NoStop}{Knight98}%
\bibitem{Haroche}%
  \BibitemOpen
  \bibfield{author}{%
  \bibinfo {author} {\bibfnamefont{S.}~\bibnamefont{Haroche}}\ and\ \bibinfo
  {author} {\bibfnamefont{J.-M.}\ \bibnamefont{Raimond}},\ }%
  \emph{\bibinfo {title} {Exploring the Quantum: Atoms, Cavities, and
  Photons}}\ (\bibinfo {publisher} {Oxford University Press},\ \bibinfo
  {address} {Oxford},\ \bibinfo {year} {2006})%
  \bibAnnoteFile{NoStop}{Haroche}%
\bibitem{gross2010dissJPB}%
  \BibitemOpen
  \bibfield{author}{%
  \bibinfo {author} {\bibfnamefont{C.}~\bibnamefont{Gross}},\ }%
  \bibfield{journal}{%
  \bibinfo {journal} {J. Phys. B}\ }%
  \textbf{\bibinfo {volume} {45}},\ \bibinfo {pages} {103001} (\bibinfo {year}
  {2012})%
  \bibAnnoteFile{NoStop}{gross2010dissJPB}%
\bibitem{Note1}%
  \BibitemOpen
  \bibinfo {note} {If the BJJ is subject to symmetric two-body (respectively
  three-body) losses only, the phenomenological rate equations give
  ${\delimiter "426830A } \protect \mathaccentV {hat}05E{N} {\delimiter
  "526930B }_t \simeq N_0 ( \gamma _1 N_0 t + 1)^{-1}$ (respectively
  ${\delimiter "426830A } \protect \mathaccentV {hat}05E{N} {\delimiter
  "526930B }_t \simeq N_0 ( 2 \kappa _1 N_0^2 t + 1)^{-1/2}$) for $N_0 \gg
  1$.}%
  \bibAnnoteFile{Stop}{Note1}%
\end{thebibliography}%

\end{document}